\documentclass[graybox,envcountchap,sectrefs,natbib]{svmono}
\usepackage{mathptmx} 
\usepackage{helvet}       
\usepackage{courier}    
\usepackage{type1cm}  
\usepackage{amsfonts}
\usepackage{amssymb}
\usepackage{xspace}
\usepackage{chapterbib}
\usepackage{makeidx}         
\usepackage{multicol}        
\usepackage[bottom]{footmisc}
\usepackage{graphicx}        

\begin{document}
\mainmatter
\setcounter{page}{0} 
\setcounter{chapter}{3} %

%
%
\let\jnl=\rmfamily
\def\refe@jnl#1{{\jnl#1}}%

\newcommand\aj{\refe@jnl{AJ}}%
\newcommand\actaa{\refe@jnl{Acta Astron.}}%
\newcommand\araa{\refe@jnl{ARA\&A}}%
\newcommand\apj{\refe@jnl{ApJ}}%
\newcommand\apjl{\refe@jnl{ApJ}}%
\newcommand\apjs{\refe@jnl{ApJS}}%
\newcommand\ao{\refe@jnl{Appl.~Opt.}}%
\newcommand\apss{\refe@jnl{Ap\&SS}}%
\newcommand\aap{\refe@jnl{A\&A}}%
\newcommand\aapr{\refe@jnl{A\&A~Rev.}}%
\newcommand\aaps{\refe@jnl{A\&AS}}%
\newcommand\azh{\refe@jnl{AZh}}%
\newcommand\gca{\refe@jnl{GeoCh.Act}}%
\newcommand\grl{\refe@jnl{Geo.Res.Lett.}}%
\newcommand\jgr{\refe@jnl{J.Geoph.Res.}}%
\newcommand\memras{\refe@jnl{MmRAS}}%
\newcommand\jrasc{\refe@jnl{J.RoySocCan}}%
\newcommand\mnras{\refe@jnl{MNRAS}}%
\newcommand\na{\refe@jnl{New A}}%
\newcommand\nar{\refe@jnl{New A Rev.}}%
\newcommand\pra{\refe@jnl{Phys.~Rev.~A}}%
\newcommand\prb{\refe@jnl{Phys.~Rev.~B}}%
\newcommand\prc{\refe@jnl{Phys.~Rev.~C}}%
\newcommand\prd{\refe@jnl{Phys.~Rev.~D}}%
\newcommand\pre{\refe@jnl{Phys.~Rev.~E}}%
\newcommand\prl{\refe@jnl{Phys.~Rev.~Lett.}}%
\newcommand\pasa{\refe@jnl{PASA}}%
\newcommand\pasp{\refe@jnl{PASP}}%
\newcommand\pasj{\refe@jnl{PASJ}}%
\newcommand\skytel{\refe@jnl{S\&T}}%
\newcommand\solphys{\refe@jnl{Sol.~Phys.}}%
\newcommand\sovast{\refe@jnl{Soviet~Ast.}}%
\newcommand\ssr{\refe@jnl{Space~Sci.~Rev.}}%
\newcommand\nat{\refe@jnl{Nature}}%
\newcommand\iaucirc{\refe@jnl{IAU~Circ.}}%
\newcommand\aplett{\refe@jnl{Astrophys.~Lett.}}%
\newcommand\apspr{\refe@jnl{Astrophys.~Space~Phys.~Res.}}%
\newcommand\nphysa{\refe@jnl{Nucl.~Phys.~A}}%
\newcommand\physrep{\refe@jnl{Phys.~Rep.}}%
\newcommand\procspie{\refe@jnl{Proc.~SPIE}}%

\newcommand{\Al}{$^{26}$Al\xspace}
\newcommand{\al}{$^{26}$Al\xspace}
\newcommand{\Be}{$^{7}$Be\xspace}
\newcommand{\be}{$^{7}$Be\xspace}
\newcommand{\bem}{$^{10}$Be\xspace}
\newcommand{\ca}{$^{44}$Ca\xspace}
\newcommand{\Ca}{$^{44}$Ca\xspace}
\newcommand{\cam}{$^{41}$Ca\xspace}
\newcommand{\Co}{$^{56}$Co\xspace}
\newcommand{\co}{$^{56}$Co\xspace}
\newcommand{\csm}{$^{135}$Cs\xspace}
\newcommand{\ct}{$^{13}$C\xspace}
\newcommand{\ci}{$^{57}$Co\xspace}
\newcommand{\Ci}{$^{57}$Co\xspace}
\newcommand{\ch}{$^{60}$Co\xspace}
\newcommand{\Ch}{$^{60}$Co\xspace}
\newcommand{\Cl}{$^{36}$Cl\xspace}
\newcommand{\li}{$^{7}$Li\xspace}
\newcommand{\Li}{$^{7}$Li\xspace}
\newcommand{\Fe}{$^{60}$Fe\xspace}
\newcommand{\fh}{$^{60}$Fe\xspace}
\newcommand{\fe}{$^{56}$Fe\xspace}
\newcommand{\Fr}{$^{57}$Fe\xspace}
\newcommand{\fr}{$^{57}$Fe\xspace}
\newcommand{\mg}{$^{26}$Mg\xspace}
\newcommand{\Mg}{$^{26}$Mg\xspace}
\newcommand{\mn}{$^{54}$Mn\xspace}
\newcommand{\Na}{$^{22}$Na\xspace}
\newcommand{\Ne}{$^{22}$Ne\xspace}
\newcommand{\Ni}{$^{56}$Ni\xspace}
\newcommand{\nh}{$^{60}$Ni\xspace}
\newcommand{\Nh}{$^{60}$Ni\xspace}
\newcommand\nuk[2]{$\rm ^{\rm #2} #1$}  
\newcommand{\pd}{$^{107}$Pd\xspace}
\newcommand{\pb}{$^{205}$Pb\xspace}
\newcommand{\tc}{$^{99}$Tc\xspace}
\newcommand{\Sc}{$^{44}$Sc\xspace}
\newcommand{\Ti}{$^{44}$Ti\xspace}
\newcommand{\ti}{$^{44}$Ti\xspace}
\def\aa{$\alpha$}
\newcommand{\about}{$\simeq$}
\newcommand{\cms}{cm\ensuremath{^{-2}} s\ensuremath{^{-1}}\xspace}
\newcommand{\degree}{$^{\circ}$}
\newcommand{\flux}{ph~cm\ensuremath{^{-2}} s\ensuremath{^{-1}}\xspace}
\newcommand{\fluxrad}{ph~cm$^{-2}$s$^{-1}$rad$^{-1}$\ }
\newcommand{\ga}{\ensuremath{\gamma}}
\newcommand{\gam}{\ensuremath{\gamma}}
\def\nn{$\nu$}
\def\ra{$\rightarrow$}
\newcommand{\Msol}{M\ensuremath{_\odot}\xspace}
\newcommand{\msol}{M\ensuremath{_\odot}\xspace}
\newcommand{\Msolppc}{M\ensuremath{_\odot} pc\ensuremath{^{-2}}{\xspace}}
\newcommand{\Msolpy}{M\ensuremath{_\odot} y\ensuremath{^{-1}}{\xspace}}
\newcommand{\msb}{M\ensuremath{_\odot}\xspace}
\newcommand{\Msun}{M\ensuremath{_\odot}\xspace}
\newcommand{\Rsun}{R\ensuremath{_\odot}\xspace}
\newcommand{\rsun}{R\ensuremath{_\odot}\xspace}
\newcommand{\Lsun}{L\ensuremath{_\odot}\xspace}
\newcommand{\lsun}{L\ensuremath{_\odot}\xspace}
\newcommand{\solar}{\ensuremath{_\odot}\xspace}
\newcommand{\zs}{Z\ensuremath{_\odot}\xspace}

\chapauthor{Friedrich-Karl Thielemann\footnote{University of Basel, 4056 Basel, Switzerland (f-k.thielemann@unibas.ch)}, Raphael Hirschi\footnote{University of Keele, Keele, ST5 5BG, United Kingdom}, Matthias Liebend\"orfer\footnote{University of Basel, 4056 Basel, Switzerland}, and Roland Diehl\footnote{Max Planck Institut f\"ur extraterrestrische Physik, 85748 Garching, Germany}}
\chapter{Massive Stars and their Supernovae}
\label{massive-stars}

\section{Cosmic Significance of Massive Stars}
\label{sec:4-1}

Our understanding of stellar evolution and the final explosive endpoints such as supernovae or
hypernovae or gamma-ray bursts relies on the combination of 
\begin{itemize}
\item[a)]
 {(magneto-)hydrodynamics}
\item[b)]
 {engergy generation due to nuclear reactions accomanying composition changes}
\item[c)]
{radiation transport}
\item[d)]
{thermodynamic properties (such as  the equation of state of stellar matter).} 
\end{itemize}
Hydrodynamics \index{hydrodynamics} is essentially embedded within the numerical schemes which implement the physics of  processes (b) to (d).
In early phases of stellar evolution, hydrodynamical processes can be approximated by
a hydro\emph{static} treatment. Nuclear energy production (b) includes all
nuclear reactions triggered during stellar evolution and explosive end stages, also among unstable isotopes produced on the way.
Radiation transport (c) covers atomic physics (e.g. opacities) for photon
transport, but also nuclear physics and neutrino nucleon/nucleus interactions \index{neutrino}
in late phases and core collapse. The \index{thermodynamics}  thermodynamical treatment (d) addresses the mixture of \emph{ideal gases}
of photons, electrons/positrons and nuclei/ions. These are fermions and 
bosons, in dilute media or at high temperatures their energies can
often be approximated by Maxwell-Boltzmann distributions. \index{Maxwell-Boltzmann distribution}
At very high densities, the \emph{nuclear} equation of state is 
required to relate pressure and density. \index{equation of state!nuclear} It
exhibits a complex behavior, with transitions from
individual nuclei to clusters of nucleons with a background neutron bath,
homogeneous phases of nucleons, the emergence of hyperons and pions up
to a possible hadron-quark phase transition. 
\index{stars!quark stars} \index{hyperons}

The detailed treatment of all these ingredients and their combined application
is discussed in more depth in textbooks 
\citep{1994sse..book.....K,2009pfer.book.....M,1996snai.book.....A,2007nps..book.....I}, 
and/or the preceding Chapter~(3),
where the evolution of low and intermediate mass stars is addressed.
That chapter also includes the stellar structure equations in spherical
symmetry and a discussion of opacities for photon transport. Ch.~8 and 9
(tools for modeling objects and their processes) go into more detail with 
regard to modeling hydro\emph{dynamics}, (convective) instabilities and energy 
transport as 
well as the energy generation due to nuclear reactions and the determination
of the latter. Here we want to focus on the 
astrophysical aspects, i.e. a description of the evolution of massive \index{stars!evolution} 
stars and their endpoints with a special emphasis on the composition of
their ejecta (in form of stellar winds during the evolution or of explosive ejecta). 
Low and intermediate mass stars end their evolution
as \index{stars!AGB} AGB stars, finally blowing off a planetary nebula via wind losses and
leaving a \index{stars!white dwarf} white dwarf with an unburned C and O composition. 
Massive stars evolve beyond this point and experience 
all stellar burning stages from H over He, C, Ne, O and Si-burning up to
core collapse and explosive endstages. In this chapter we want to discuss the
nucleosynthesis processes involved and the production of radioactive
nuclei\footnote{We focus especially on long-lived radioactivities which can be observed
with gamma-ray satellites, and refractory isotopes which can be observed in dust condensations included
in meteorites.}  in more detail. 
This includes
all hydrostatic nuclear-burning stages experienced by massive stars, and explosive burning
stages when a shock wave moves outward after a successful explosion
was initiated, but also final wind ejecta from the hot proto-neutron
star which emerged in the collapse and explosion phase. All these ejecta
will enter the interstellar medium in galaxies, initially appearing as gas and dust in
wind bubbles and supernova remnants, later determining 
the evolution of the larger-scale gas composition. 
The interstellar gas composition will evolve with time, and the composition 
of newly formed stars will witness this composition at the time of their formation. 

Massive stars play an important role as contributors to the gas composition of the interstellar medium 
via wind losses or explosions. In astronomical terms
they are the progenitors of blue supergiants (BSG), red
supergiants (RSG), Wolf-Rayet (WR) and luminous blue variable (LBV)
stars \index{stars!Wolf Rayet} \index{stars!blue supergiant} \index{stars!luminous blue variable}
\citep{2010RvMP...??..???B}.  
At the end of their life, they explode as 
core collapse supernovae (ccSNe), observed as 
SNe of type II or Ib,c \index{supernovae!core collapse} \index{supernovae!type II} \index{supernovae!type Ibc} 
\citep{2006ARA&A..44..507W} 
and also as long soft gamma-ray bursts (GRBs 
\citet{2004RvMP...76.1143P}). \index{gamma-ray burst}
After collapse, their cores become neutron stars or black holes. They
are one of the main sites for nucleosynthesis, which takes place during
both pre-SN (hydrostatic) burning stages and during explosive burning.
A weak $s$~process occurs during core He- (and C-) burning 
\citep{2007ApJ...655.1058T,2009SSRv..147....1E} and the
$r$~process probably occurs during the explosion 
\citep{1996ApJ...471..331Q}. 
These s(low) and r(apid) neutron \index{process!neutron capture}
capture processes are mainly responsible for the heavy nuclei beyong the
Fe-group.
Radioactive isotopes like $^{26}$Al and $^{60}$Fe detected by the INTEGRAL
\index{INTEGRAL}
satellite are produced by massive stars, plus many more radioactivities
from the final explosive ejecta (like e.g. $^{44}$Ti, $^{56}$Ni, $^{56}$Co
etc., see Sect.~4.4.2 and 4.5). Ch.~2 and 3 discussed also many long-lived 
heavy nuclei beyond Fe with half-lives larger than $10^7$ and up to $10^{11}$years.
As massive stars are probably not the origin of heavy $s$-process nuclei (see 
Ch.~3), we will address here those nuclei which are clearly 
identified with the $r$~process ( $^{232}$Th, $1.4\times 10^{10}$y,
$^{235}$U, $7\times 10^{8}$y,
$^{236}$U, $2.3\times 10^{7}$y,
$^{238}$U, $4.5\times 10^{9}$y,
$^{244}$Pu, $8\times 10^{7}$y,
$^{247}$Cm, $1.6\times 10^{7}$y) and where especially $^{232}$Th and $^{238}$U,
with half-lives comparable to the age of the Galaxy/Universe, can also serve
as chronometers. \index{process!r process} \index{nucleocosmochronology}

Massive stars, even though they are much less numerous than low mass
stars, contribute significantly (about two thirds) to the integrated 
luminosity of galaxies.
At high redshifts $z$, or low metallicities $Z$,
they are even more important drivers of characteristic phenomena and evolution. The first
stars formed are thought to be all massive or even very massive, and
to be the cause of the re-ionisation of the universe. \index{re-ionization}
As discussed above, if the final core collapse leads to a black hole\index{stars!black hole}, the
endpoint of this evolution can be the origin of the subset of
(long, soft) gamma ray bursts (GRBs).
GRBs are the new \emph{standard candles} for cosmology at high
redshifts. They are visible from higher redshifts than usual SNe
(of type I or II) are, and thus will  impose tighter constraints on
cosmological models. \index{cosmology}
Massive stars with their large energy output can be seen out to significant 
(cosmological) distances -- either directly through their thermal photospheric emission,
or indirectly through the impact on their surroundings (ionization, or heated dust). 
In their \emph{collapsar} and GRB extremes, emission is beamed into a jet, which makes
them visible even at greater distances. This can also give us information on the star \index{stars!formation}
formation history at a very early age of the universe ($z>$10) beyond the reach of galaxy observations.
Closer to home, recent surveys of metal poor halo
stars provide a rich variety of constraints for the early chemical evolution of our
Galaxy and thus the nucleosynthesis ejecta (\emph{astro-archeology}).

\section{Hydrostatic and Explosive Burning in Massive Stars}
\label{sec:4-2}

Following the motivation for studying massive stars 
in the previous section, we now discuss the ingredients for their modeling. 
Thermonuclear energy generation is one of the key aspects: \index{stars!structure} \index{stars!evolution}
It shapes the interior structure of the star, thus its evolutionary time scales, 
and the generation of new chemical elements and nuclei.
Without understanding these, the \emph{feedback} from massive stars \index{stars!feedback}
\emph{as it determines the evolution of galaxies} cannot be understood 
in astrophysical terms.\footnote{Empirical descriptions from observations of a multitude of galaxies are often utilized to substitute such astrophysical models in cosmological simulations.} 
Thermonuclear burning, nuclear energy generation and resulting nuclear 
abundances are determined by thermonuclear reactions and weak interactions. 
The treatment 
of the required nuclear/plasma physics, and a detailed technical description
of reaction rates, their determination and the essential features of 
composition changes and reaction networks is presented in Ch.~9. 
Here we want to discuss which types of reactions are involved specifically
in the evolution of massive stars and their catastrophic end stages.
Nuclear burning can in general be classified into two categories: 
\begin{itemize} 
\item[(1)]
hydrostatic \index{nucleosynthesis!hydrostatic} burning stages on timescales dictated by stellar energy loss
\item[(2)] explosive burning due to hydrodynamics of the specific
event.  \index{nucleosynthesis!explosive}
\end{itemize}
Massive stars (as opposed to low and intermediate mass stars) are the
ones which experience explosive burning (2) as a natural outcome at the end of their evolution
and they undergo more extended hydrostatic burning stages (1) than their
low- and intermediate-mass cousins. Therefore, we want to address some of these
features here in a general way, before describing the evolution and
explosion in more detail. 

The important ingredients for describing nuclear
burning and the resulting composition changes (i.e. nucleosynthesis) 
are (i) strong-interaction cross sections and photodisintegrations, 
(ii) weak interactions related to decay half-lives, electron or positron \index{decay!beta decay} \index{process!electron capture}
captures, and finally (iii) neutrino-induced reactions. They will now be discussed\footnote{A review of the sources for this microphysics input is given for (i) in
Ch.~9 and for (iii) in Ch.~8. We will review some of the required
weak interaction rates (ii) in the subsections on late phases of stellar
evolution / core collapse and the description of the explosion.}. 

\subsection{Nuclear Burning During Hydrostatic Stellar Evolution} 

Hydrostatic burning stages are characterized by temperature thresholds,
\index{nucleosynthesis!hydrostatic} 
permitting thermal Maxwell-Boltzmann distributions of (charged) particles
(nuclei) to penetrate increasingly larger Coulomb barriers of electrostatic
repulsion. These are (two body) reactions as discussed in Equ.~9.6 
and 9.9 of Ch.~9, representing terms of the type $_ir_j$ in the
network equation (9.1).
\index{process!H burning} \index{process!CNO} \index{process!pp}
H-burning converts $^1$H into $^4$He via pp-chains or the CNO-cycles. The
simplest pp-chain is initiated by $^1$H(p,$e^+\nu$)$^2$H(p,$\gamma$)$^3$He and
completed by $^3$He($^3$He,2p)$^4$He. The dominant CNO-cycle chain
$^{12}$C(p,$\gamma)^{13}$N($e^+\nu)^{13}$C(p,$\gamma)^{14}$N(p,$\gamma)^{15}
$O($e^+\nu)^{15}$N(p,$\alpha)^{12}$C is controlled by the \index{isotopes!14N}
slowest reaction $^{14}$N(p,$\gamma)^{15}$O. The major reactions in
He-burning are the  $3\alpha$- reaction $^4$He(2$\alpha,\gamma$)$^{12}$C and
$^{12}$C($\alpha, \gamma$)$^{16}$O. The  $3\alpha$- reaction, being essentially
a sequence of two two-body reactions with an extremely short-lived intermediate
\index{process! $3\alpha$-} \index{isotopes!8Be}
nucleus $^8$Be, is an example for the term $_i\hat{r}_j$ in Equ.~9.1,
 which includes the product of three abundances.
The H- and He-burning stages are also encountered
in low and intermediate mass stars, leaving white dwarfs as central
objects. They are discussed in much more detail
with all minor reaction pathways in Ch.~3. 
Massive stars, the subject of the present Chapter, undergo further burning 
stages up to those involving the production of 
Fe-group nuclei. Table \ref{4:burntime} lists these burning stages 
and their typical central densities and temperatures, their duration and the typical luminosity in photons 
(from \citet{1995ApJS..101..181W}), which involve the reaction types given below.
For further details see Sect.~\ref{sec:4-3}.

\begin{table}
\caption{Burning stages of a 20M$_\odot$ star}
\begin{center}
\begin{tabular}{r c c c c}
\hline
\hline
  \         &  $\rho_c$      &  $T_c$        &  $\tau$   &  $L_{phot}$\\
  Fuel     &  (g cm$^{-3}$) &  ($10^{9}$ K) &  (yr)   &  (erg s$^{-1}$)\\
\hline
  Hydrogen &  5.6(0)        &  0.04        &  1.0(7)   &  2.7(38)    \\
  Helium   &  9.4(2)        &  0.19         &  9.5(5)   &  5.3(38)    \\
  Carbon   &  2.7(5)        &  0.81         &  3.0(2)   &  4.3(38)    \\
  Neon     &  4.0(6)        &  1.70         &  3.8(-1)  &  4.4(38)    \\
  Oxygen   &  6.0(6)        &  2.10         &  5.0(-1)  &  4.4(38)    \\
  Silicon  &  4.9(7)        &  3.70         &  2~days   &  4.4(38)    \\
\hline
\end{tabular}
\end{center}
\label{4:burntime}
\end{table}

\begin{table}
\caption{Major Reactions in Carbon Burning}
\begin{center}
\begin{tabular}{l}
\hline
\hline
\\
(a) basic energy generation   \\
 $^{12}$C($^{12}$C,$\alpha)^{20}$Ne$~~~~^{12}$C($^{12}$C,p)$^{23}$Na \\
 $^{23}$Na(p,$\alpha)^{20}$Ne$~~~^{23}$Na(p,$\gamma)^{24}$Mg$~~~^{12}$C($\alpha,\gamma)^{16}$O \\
\\
(b) fluxes $>10^{-2}\times$(a) \\  
 $^{20}$Ne($\alpha,\gamma)^{24}$Mg$~~~~^{23}$Na($\alpha$,p)$^{26}$Mg(p,$\gamma)^{27}$Al \\
 $^{20}$Ne(n,$\gamma)^{21}$Ne(p,$\gamma)^{22}$Na
($e^+\nu)^{22}$Ne($\alpha$,n)$^{25}$Mg(n,$\gamma)^{26}$Mg \\
 $^{21}$Ne($\alpha$,n)$^{24}$Mg$~~~^{22}$Ne(p,$\gamma)^{23}$Na$~~~
^{25}$Mg(p,$\gamma)^{26}$Al($e^+\nu)^{26}$Mg \\
\\
(c) low temperature, high density burning    \\
 $^{12}$C(p,$\gamma)^{13}$N($e^+\nu)^{13}$C($\alpha$,n)$^{16}$O($\alpha,\gamma)^{20}$Ne \\
 $^{24}$Mg(p,$\gamma)^{25}$Al($e^+\nu)^{25}$Mg \\
 $^{21}$Ne(n,$\gamma)^{22}$Ne(n,$\gamma)^{23}$Ne($e^-\bar\nu)^{23}$Na(n,$\gamma)^{24}
$Na($e^-\nu)^{24}$Mg + $s$~processing\\
\hline
\end{tabular}
\end{center}
\label{4:carbon}
\end{table}

\begin{itemize}  
\item \emph{Heavy-ion fusion reactions:}
In
C-burning the reaction $^{12}$C($^{12}$C,$\alpha$)$^{20}$Ne dominates, in
\index{isotopes!20Ne} \index{isotopes!16O} \index{isotopes!28Si} \index{isotopes!12C}
O-burning it is $^{16}$O($^{16}$O,$\alpha$)$^{28}$Si.
The corresponding reaction rates $_ir_j$ (after integrating over a
Maxwell-Boltzmann distribution of targets and projectiles) have the
form given in Equ.~9.9 of Ch.~9 and contribute to the 
second term in Equ.~9.1.
Reactions going beyond these key reactions are provided in tables \ref{4:carbon}
and \ref{4:oxygen}.
Further features as well as the status of nuclear cross sections are
discussed in recent reviews on hydrostatic burning stages 
\citep{2006NuPhA.777..226H,2006NuPhA.777..254B,2009RPPh...72h6301C,2010ARNPS..60..175K} and Ch.~9.
\begin{table}
\caption{Major Reactions in Oxygen Burning} \index{process!O burning}
\begin{center}
\begin{tabular}{l}
\hline
\hline
\\
(a) basic energy generation   \\
 $^{16}$O($^{16}$O,$\alpha)^{28}$Si$~~~~^{16}$O($^{12}$O,p)$^{31}$P$~~~~
^{16}$O($^{16}$O,n)$^{31}$S($e^+\nu)^{31}$P \\
 $^{31}$P(p,$\alpha)^{28}$Si($\alpha,\gamma)^{32}$S\\
 $^{28}$Si($\gamma,\alpha)^{24}$Mg($\alpha$,p)$^{27}$Al($\alpha$,p)$^{30}$Si \\
 $^{32}$S(n,$\gamma)^{33}$S(n,$\alpha)^{30}$Si($\alpha,\gamma)^{34}$S \\
 $^{28}$Si(n,$\gamma)^{29}$Si($\alpha$,n)$^{32}$S($\alpha$,p)$^{35}$Cl \\
 $^{29}$Si(p,$\gamma)^{30}$P($e^+\nu)^{30}$Si \\
\\
electron captures \\
$^{33}$S($e^-,\nu)^{33}$P(p,n)$^{33}$S \\
$^{35}$Cl($e^-,\nu)^{35}$S(p,n)$^{35}$Cl \\
\\
(b) high temperature burning \\
 $^{32}$S($\alpha,\gamma)^{36}$Ar($\alpha$,p)$^{39}$K \\
 $^{36}$Ar(n,$\gamma)^{37}$Ar($e^+\nu)^{37}$Cl\\
 $^{35}$Cl($\gamma$,p)$^{34}$S($\alpha,\gamma)^{38}$Ar(p,$\gamma)^{39}$K(p,$\gamma)^{40}$Ca \\
 $^{35}$Cl($e^-,\nu)^{35}$S($\gamma$,p)$^{34}$S \\
 $^{38}$Ar($\alpha,\gamma)^{42}$Ca($\alpha,\gamma)^{46}$Ti \\
 $^{42}$Ca($\alpha$,p)$^{45}$Sc(p,$\gamma)^{46}$Ti \\
\\
(c) low temperature, high density burning    \\
 $^{31}$P($e^-\nu)^{31}$S$~~~~^{31}$P(n,$\gamma)^{32}$P \\
 $^{32}$S($e^-,\nu)^{32}$P(p,n)$^{32}$S \\
 $^{33}$P(p,$\alpha)^{30}$Si \\
\hline
\end{tabular}
\end{center}
\label{4:oxygen}
\index{isotopes!31P} \index{isotopes!32S} \index{isotopes!28Si} \index{isotopes!30Si}
\index{isotopes!24Mg} \index{isotopes!36Ar} \index{isotopes!40Ca} \index{isotopes!46Ti}
\end{table}
\item \emph{Photo-disintegrations:}
The alternative to fusion reactions are photodisintegrations which start to
play a role at sufficiently high temperatures $T$ when 30$kT\approx~Q$
(the Q-value \index{Q value}
or energy release of the inverse capture reaction). This ensures the existence
of photons with energies $>$$Q$ in the Planck distribution and
leads to Ne-Burning [$^{20}$Ne($\gamma,\alpha)^{16}$O,
$^{20}$Ne($\alpha,\gamma)^{24}$Mg] at $T$$>$$1.5\times 10^9$K (preceding
O-burning) due to a small Q-value of $\approx$4~MeV and Si-burning \index{process!Si burning}
at temperatures in excess of 3$\times$10$^9$K
[initiated like Ne-burning by photodisintegrations]. \index{process!Ne burning} \index{process!photo-disintegration}
Such photodisintegrations (after integrating over a thermal (Planck) distribution of 
photons at temperature $T$) have the form given in equation (9.4) of Ch.~9
and act similar to decays with a temperature-dependent decay constant, 
contributing (like decays) to the first term $_i\lambda_j$ in equation (9.1).
In table \ref{4:neon} we provide some of the main reactions of Ne-burning,
which is initiated by the photodisintegration of Ne.
\begin{table}
\caption{Major Reactions in Neon Burning}
\begin{center}
\begin{tabular}{l}
\hline
\hline
\\
(a) basic energy generation   \\
 $^{20}$Ne($\gamma,\alpha)^{16}$O$~~~~^{20}$Ne($\alpha,\gamma)^{24}$Mg($\alpha,\gamma)^{28}$Si\\
\\
(b) fluxes $>10^{-2}\times$(a) \\
\ $^{23}$Na(p,$\alpha)^{20}$Ne$~~~~^{23}$Na($\alpha$,p)$^{26}$Mg($\alpha$,n)$^{29}$Si \\
 $^{20}$Ne(n,$\gamma)^{21}$Ne($\alpha$,n)$^{24}$Mg(n,$\gamma)^{25}$Mg($\alpha$,n)$^{28}$Si \\
 $^{28}$Si(n,$\gamma)^{29}$Si(n,$\gamma)^{30}$Si \\
 $^{24}$Mg($\alpha$,p)$^{27}$Al($\alpha$,p)$^{30}$Si \\
 $^{26}$Mg(p,$\gamma)^{27}$Al(n,$\gamma)^{28}$Al($e^-\bar\nu)^{28}$Si\\
\\
(c) low temperature, high density burning    \\
 $^{22}$Ne($\alpha$,n)$^{25}$Mg(n,$\gamma)^{26}$Mg(n,$\gamma)^{27}$Mg($e^-\bar\nu)^{27}$Al \\
 $^{22}$Ne left from prior neutron-rich carbon burning  \\
\\
\hline
\end{tabular}
\end{center}
\index{isotopes!20Ne} \index{isotopes!23Na} \index{isotopes!29Si} \index{isotopes!22Ne}
\label{4:neon}
\end{table}
\item \emph{Electron capture reactions:}
Massive stellar cores eventually lead to electron-gas degeneracy, i.e. the \index{process!electron capture}
Pauli exclusion principle for fermions determines the population of energy 
states rather than the Boltzmann statistics, valid only for low densities /
high temperatures. 
The Fermi energy of electrons is \index{Fermi!energy}
\begin{equation} 
E_F={\hbar^2/2m_e}(3\pi^2)^{2/3}n_e^{2/3}
\end{equation}
Here $n_e$ is the density of the electron gas
$n_e=\rho N_AY_e$, $\rho$ denotes the matter density and
$N_A$ Avogadro's number. 
In late stages of O-burning, in Si-burning (and during the later collapse
stage) this Fermi energy of (degenerate) electrons, increases to the level of nuclear energies (MeV).  
In a neutral, completely
ionized plasma, the electron abundance $Y_e$ is equal to the total proton 
abundance $Y_e=\sum_i Z_i Y_i$ (summing over all abundances of nuclei, 
including protons/hydrogen) and limited by the 
extreme values 0 (only
neutrons) and 1 (only protons) with typical values during stellar evolution
close to 0.5 or slightly below.
Such conditions permit electron captures on protons and nuclei, if the
negative Q-value of the reaction can be overcome by the electron (Fermi) 
energy. The general features for typical conditions are presented in 
table \ref{4:electron}, example reactions were already given in table
\ref{4:oxygen}.
\begin{table}
\caption{Electron Capture}
\begin{center}
\begin{tabular}{c}
\hline
\hline
$ p + e^-  \rightarrow \nu_e + n$\ \ \ or \ \ \ p($e^-,\nu_e)$n\\ 
$ (A, Z) + e^- \rightarrow  \nu_e + (A, Z-1)$\ \ \ or \ \ \ $^A$Z($e^-,\nu_e)^A$Z-1\\
\\
$E_F(\rho Y_e=10^7$gcm$^{-3}$)=0.75~MeV\\
$E_F(\rho Y_e=10^9$gcm$^{-3}$)=4.70~MeV\\
\hline
\end{tabular}
\end{center}
\label{4:electron}
\end{table}
Thus, at sufficiently high densities, electron captures - which are 
energetically prohibited - \index{process!electron capture} \index{decay!beta decay}
can become possible and lead to an enhanced \emph{neutronization} of the
astrophysical plasma, in addition to the role of beta-decays and electron
captures with positive Q-values
\citep{1988PhR...163...13N}.
In degenerate Ne-O-Mg cores (after core C-burning of stars with 
$8<M/$M$_\odot<10$), electron captures on $^{20}$Ne and $^{24}$Mg
cause a loss of degeneracy pressure support and introduce a collapse
rather than only a contraction, which combines all further burning stages
on a short collapse time scale 
\citep{1987ApJ...322..206N}.
In Si-burning of more massive stars, electron capture on intermediate mass and
Fe-group nuclei becomes highly important and determines the neutronization
($Y_e$) of the central core. \index{process!Si burning}
As discussed in Ch.~9, such rates contribute to the one-body reaction
terms $_i\lambda_j$ in Equ.~9.1 with the effective decay constants
in Equ.~9.5 being a function of $T$ and $n_e=\rho N_A Y_e$, the 
electron number density. 
\begin{table}
\caption{Neutrino Reactions}
\begin{center}
\begin{tabular}{c}
\hline
\hline
$\nu_e + n \leftrightarrow p + e^-$\ \ \  or \ \ \ n($\nu_e,e^-$)p\\
$\bar{\nu}_e + p \leftrightarrow n+ e^+$\ \ \ or \ \ \ p(($\bar{\nu}_e,e^+)$n\\
$ \nu_e + (Z, A) \leftrightarrow (Z+1,A) + e^-$\ \ \  or \ \ \ $^A$Z($\nu_e,e^-)^A$Z+1\\
$\bar{\nu}_e +  (Z,A) \leftrightarrow (Z-1,A) + e^+$\ \ \  or \ \ \ $^A$Z($\bar{\nu}_e,e^+)^A$Z-1\\
$ (Z,A) +  \nu  \leftrightarrow \nu + (Z,A)^*$\\
\hline
\end{tabular}
\end{center}
\label{4:neutrinos}
\end{table}

Neutrino cross section on nucleons, nuclei and 
electrons are minute, by comparison to above reactions. 
It therefore requires high densities of the order $\rho >10^{12}$g~cm$^{-3}$  
that also the inverse process to electron/positron capture (neutrino capture) \index{process!electron capture} \index{process!positron capture} \index{process!neutrino capture} 
can occur on relevant timescales.
The same is true for other processes such as e.g. 
inelastic scattering, leaving a nucleus in an excited
state which can emit nucleons and alpha particles. 
Such neutrino-induced reactions can be expressed in a similar way as photon 
and electron \index{neutrino}
captures, integrating now over the corresponding neutrino distribution. The 
latter is, however, not necessarily in thermal equilibrium and not just
a function of temperature and neutrino densities. Neutrino distributions
are rather
determined by (neutrino) radiation transport calculations (see
Ch.~8, where also other neutrino scattering processes are discussed).
\end{itemize}

\noindent
All the reactions presented above and occurring at different times in the
sequence of burning stages, contribute to the three types of terms in 
the reaction network equation (Equ.9.1 in Ch.~9). If one is interested
to show how nuclear abundances $Y_i$ enter in this set of equations, it can
also be written in the form\footnote{The formal difference to Equ.~9.1 is that one does not sum here 
over the reactions but rather over all reaction partners (see also
the equation following Table 3.2 in Ch.~3). However, in total,
all the terms which appear are identical. Due to the different 
summation indices, the P's have a slightly different notation, $\lambda$'s denote 
decay rates called $L$ in Ch.~9, and $<j,k>$ correspond to $<\sigma^* v>$ of reactions
between nuclei $j$ and $k$, while $<j,k,l>$ includes a similar
expression for three-body reactions 
\citep{1985A&A...149..239N}.
A survey of computational methods to solve nuclear networks is given in 
\citet{1999JCoAM.109..321H,1999ApJS..124..241T}.
(Like for electron abundances $Y_e$, the abundances $Y_i$  in Eq.(\ref{4:network}) are related to number densities $n_i=\rho N_A Y_i$ and mass 
fractions of the corresponding nuclei via $X_i=A_i Y_i$, where $A_i$ is the
mass number of nucleus $i$ and $\sum X_i = 1$.)}

\begin{equation}
{{d Y_i} \over dt}  =  \sum_j P^i _j\ \lambda_j Y_j + \sum_{j,k} P^i _{j,k}\ 
\rho N_A <j,k> Y_j Y_k 
+ \sum_{j,k,l} P^i _{j,k,l}\ \rho^2 N_A^2 <j,k,l> Y_j Y_k Y_l. 
\label{4:network}
\end{equation}


Core Si-burning, the final burning stage during stellar evolution, which
is initiated by the photodisintegration $^{28}$Si($\gamma,\alpha)^{24}$Mg 
close to $3\times 10^9$~K - and
\index{process!photo-disintegration} \index{process!Si burning} \index{stars!evolution}
followed by a large number of fusion and photodisintegration reactions -
ends with nuclear reactions in a complete  \index{equilibrium!chemical} \index{process!nuclear statistical equilibrium} 
\emph{chemical equilibrium}\footnote{all strong (thermonuclear) and 
photodisintegration reactions are equilibrized, while weak interaction
reactions, changing $Y_e$, may occur on longer timescales.} 
(nuclear statistical equilibrium, NSE) and an abundance distribution 
centered around Fe
(as discussed in Ch.~9, Equ.~9.14 and 9.15).
These temperatures permit photodisintegrations with typical Q-values of 8-10~MeV
as well as the penetration of Coulomb barriers in capture reaction.
In such an NSE the abundance of each nucleus Y$_i$ is only dependent on
temperature $T$, density $\rho$, its nuclear binding energy $B_i$, \index{binding energy}
and via charge conservation on 
$\sum_i Z_i Y_i =Y_e$.
$Y_e$ is altered by weak interactions on longer timescales.
\index{process!quasiequilibrium}
\emph{A quasi-equilibrium (QSE)} can occur, if localized nuclear mass regions 
are in equilibrium with the background of free neutrons, protons and $\alpha$~particles,  
but offset from other regions of nuclei and thus their NSE values 
\citep{1996ApJ...460..869H,1999ApJ...511..862H,2007ApJ...667..476H}. 
Different quasi-equilibrium regions are usually
separated from each other by slow reactions with typically small Q-values.
Such boundaries between QSE groups, which are due to slow reactions, can be related
to neutron or proton shell closures, like e.g. $Z=N=20$, separating the
Si- and Fe-groups in early phases of Si-burning.

\begin{figure}[!tbp] 
\begin{center}
\includegraphics[width=\textwidth]{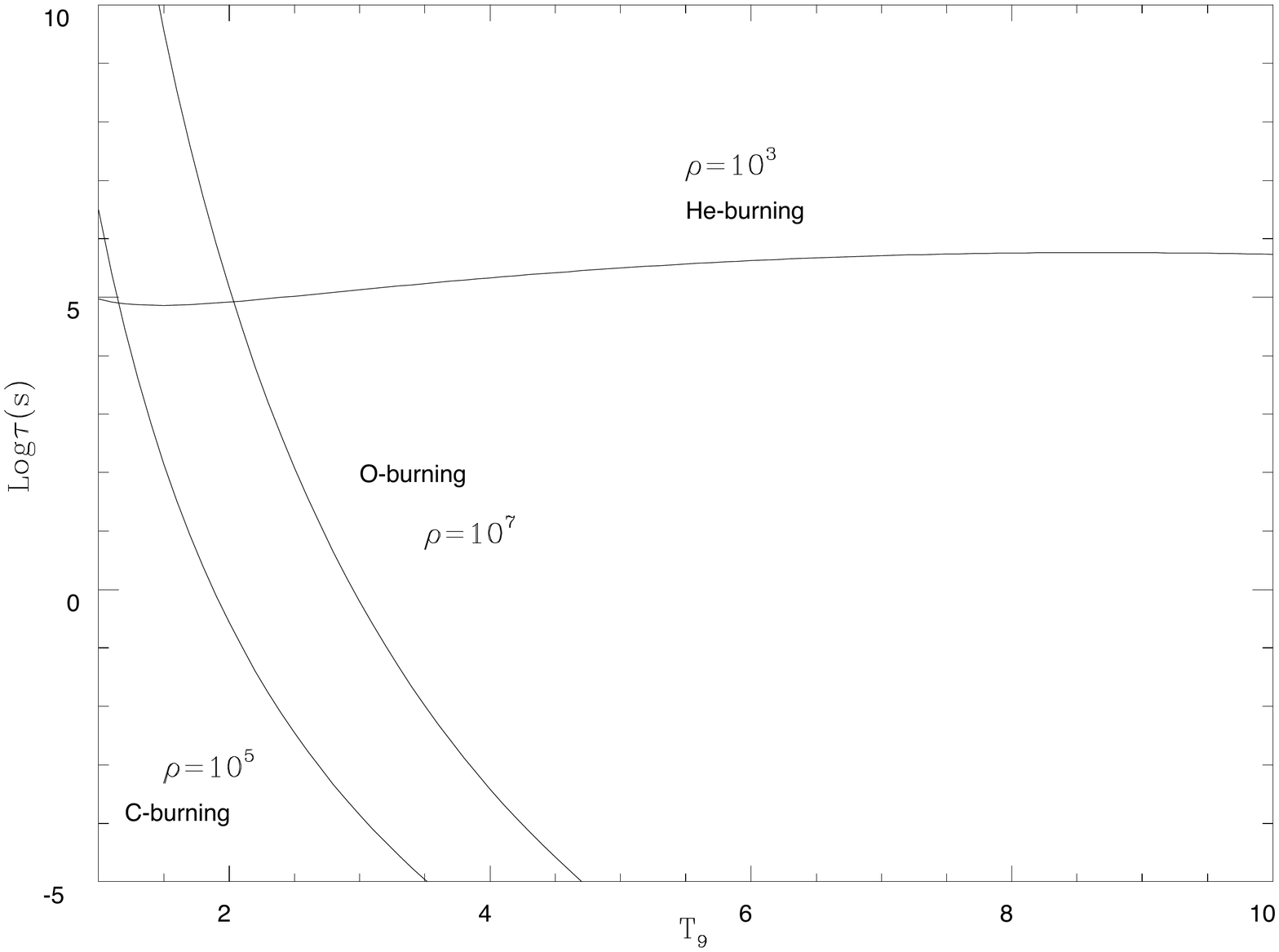}
\includegraphics[width=\textwidth]{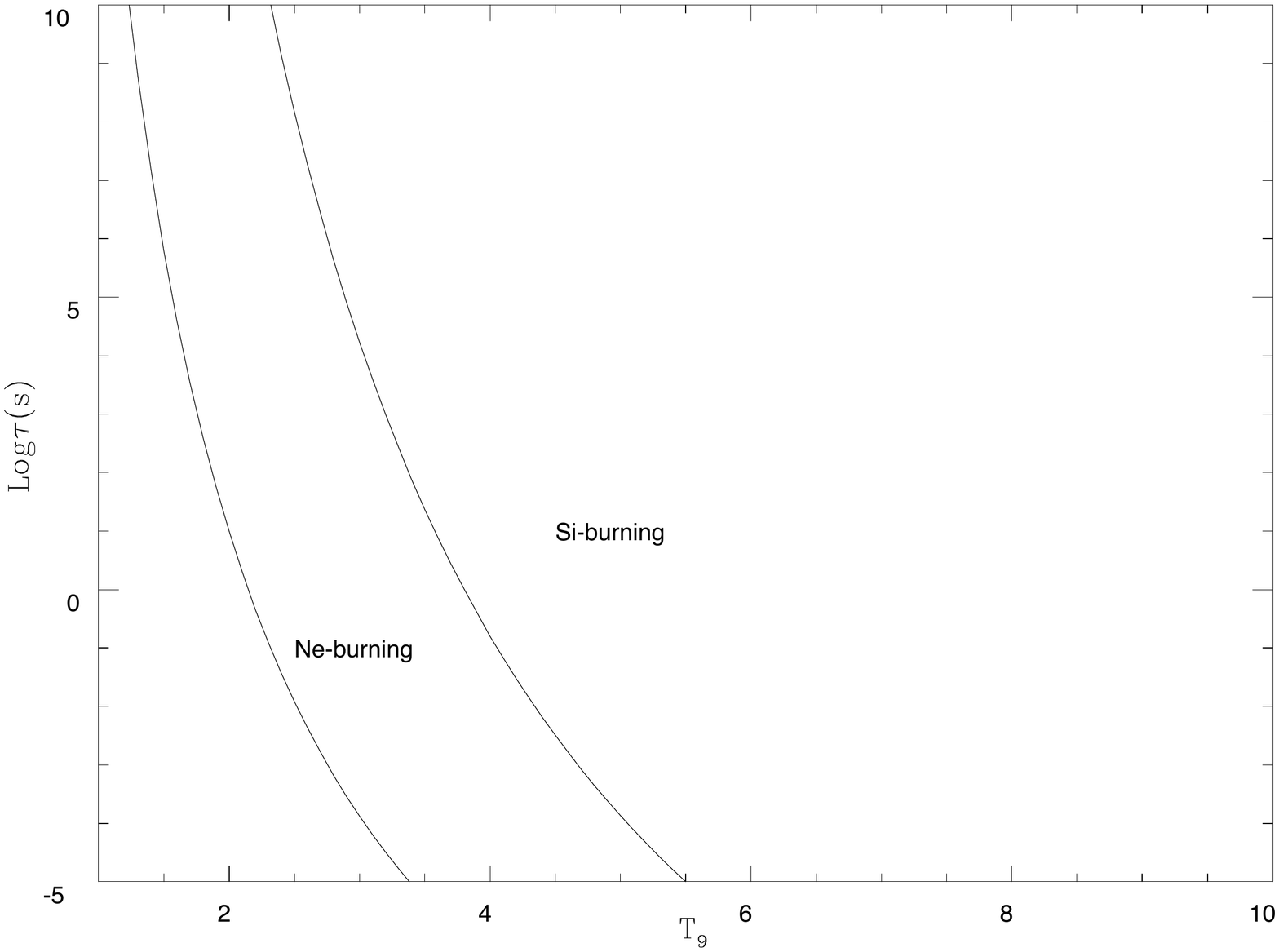}
\end{center}
\caption{Burning timescales in ($log_{10}$) seconds for fuel exhaustion of He-, C-, and
O-burning (top) and Ne- and Si-burning (buttom), 
as a function of temperature.
Density-dependent timescales are labeled with a
chosen typical density (in g~cm$^{-3}$). They scale with $1/\rho$ for C- and
O-burning and $1/\rho^2$ for He-burning. \index{process!H burning} \index{process!C burning} \index{process!He burning} \index{process!Ne burning} \index{process!Si burning}
Ne- and Si-burning, initiated by
photodisintegrations, are not density-dependent.
The almost constant
He-burning timescale beyond $T_9$=$T/10^9$K=1 permits efficient
destruction on explosive timescales only for high
densities.}  
\label{4:explo1}
\end{figure} 

All reactions discussed above, occurring during all stellar burning stages,
are essentially related to nuclei from H to the Fe-group, and not much
beyond. 

\begin{itemize}
\item \emph{Neutron capture processes:} \index{process!neutron capture}
Through neutron capture reactions, also during regular stellar evolution, there is a chance to
produce heavier nuclei. 
During core and shell He-burning specific  $\alpha$-induced reactions can
liberate neutrons which are responsible for the slow neutron capture process
($s$~process). A major neutron source is the reaction 
$^{22}$Ne($\alpha,n)^{25}$Mg, with $^{22}$Ne being produced via succesive
$\alpha$-captures on the H-burning CNO product 
$^{14}$N($\alpha,\gamma)^{18}$F($\beta^+)^{18}$O($\alpha,\gamma)^{22}$Ne.
If occurring, the mixing of $^{12}$C into H-burning shells can produce an even
stronger neutron source $^{13}$C($\alpha,n)^{16}$O  via 
$^{12}$C($p,\gamma)^{13}$N($\beta^+)^{13}$C. In massive, rotating, low 
metallicity stars, mixing can lead to the production of \emph{primary}
$^{14}$N and $^{22}$Ne, i.e. a neutron source which does not reflect the
initial metallicity of $^{14}$N in the CNO-cycle, and can thus be much 
stronger. Ch.~3  
discusses in full detail the strong $s$~process via a combination of
$^{13}$C and $^{22}$Ne in He-shell flashes of low and 
intermediate
mass stars. In a similar way mixing processes can also occur in massive
stars due to rotation or convective instabilities. Without such mixing 
processes only \emph{secondary} (metallicity-dependent) $^{22}$Ne is available 
for $^{22}$Ne($\alpha,n)^{25}$Mg and core He-burning 
as well as shell C-burning
lead to a weak $s$~process 
\citep{2007ApJ...655.1058T}. 
The $s$~process can in principle form elements up to Pb and Bi
through a series of neutron captures and $\beta^-$-decays,
starting on existing heavy nuclei around Fe 
\citep{2006NuPhA.777..291K}.
Weak $s$~processing, based on \emph{secondary} $^{22}$Ne, does not proceed beyond mass 
numbers of $A=80-90$.
The production of heavier nuclei is possible in massive stars if 
\emph{primary} $^{14}$N and $^{22}$Ne are available.
\end{itemize}

\subsection{Explosive Burning}

\begin{figure}[!b]  
\begin{center}
\includegraphics[width=\textwidth]{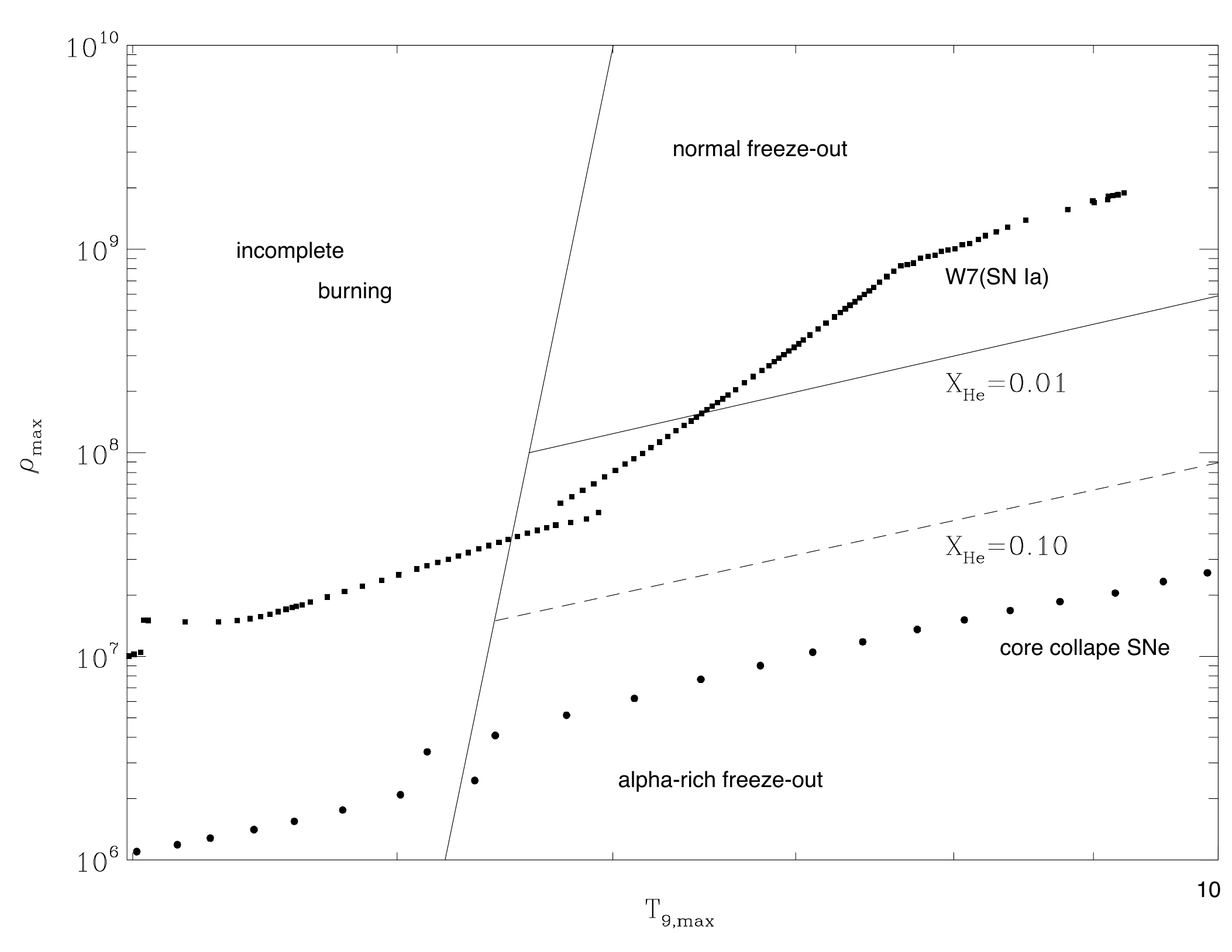}
\end{center}
\caption{Final results of explosive Si-burning as a function of maximum 
temperatures and densities attained in explosions before adiabatic expansion.
For temperatures in excess of 5$\times 10^9$K any fuel previously existing
is photodisintegrated into nucleons and $\alpha$~particles before re-assembling
in the expansion. For high densities this is described by a full NSE
with an Fe-group composition, favoring nuclei with maximum 
binding energies and proton/nucleon ratios equal to $Y_e$. For lower densities 
the NSE breaks into local equilibrium groups
(quasi-equilibrium, QSE) with group boundaries  
determined by reactions with an insufficiently fast reaction stream. 
Alpha-rich freeze-out (insufficient conversion of $\alpha$~particles 
into nuclei beyond carbon) is such a QSE-behavior. Lines with 1\% and 10\%
remaining $\alpha$-particle mass fraction are indicated as well as typical conditions for mass zones
in type~Ia and core-collapse supernovae.}
\label{4:explo2}
\end{figure}  

Many of the hydrostatic nuclear-burning processes \index{nucleosynthesis!explosive}
occur also under explosive conditions at higher temperatures and on
shorter timescales (see Fig.\ref{4:explo1}), when often the $\beta$-decay 
half-lives are longer than the explosive timescales, 
producing significant abundances of unstable isotopes as burning proceeds. 
This requires in general the additional knowledge of nuclear reactions
for and among unstable nuclei.
The fuels for explosive nucleosynthesis consist mainly of $N$=$Z$ nuclei like
$^{12}$C, $^{16}$O, $^{20}$Ne, $^{24}$Mg, or $^{28}$Si (the ashes of prior
hydrostatic burning),
resulting in heavier nuclei, again with
$N$$\approx$$Z$. At high densities also 
electron captures on nuclei $e^- + ^AZ \rightarrow ^AZ$-$1 + \nu$
can occur at substantial rates due to energetic, degenerate electrons
when Fermi energies are high, as already discussed for late hydrostatic \index{Fermi!energy}
burning stages. 

Explosive Si-burning \index{process!Si burning}
differs strongly from its hydrostatic counterpart
and can be divided into three different regimes:
(i) incomplete Si-burning and complete Si-burning with
either (ii) a normal (high density, low entropy) or (iii) an $\alpha$-rich
(low density, high entropy) freeze-out of charged-particle reactions
during cooling from NSE. \index{process!nuclear statistical equilibrium}
At initially-high temperatures or during a \emph{normal} freeze-out,
the abundances remain in a full NSE. The full NSE can break up in smaller
equilibrium clusters (quasi-equilibrium, QSE), for a detailed discussion
see \citet{1996ApJ...460..869H,1999ApJ...511..862H,2007ApJ...667..476H}. 
An example for such QSE-bevavior is an  $\alpha$-rich \index{process!$\alpha$-rich freeze out}
freeze-out, caused by the inability of the  $3\alpha$- reaction \index{process!3$\alpha$}
$^4$He(2$\alpha,\gamma)^{12}$C, and the $^4$He($\alpha$n,$\gamma)^9$Be reaction
to keep light nuclei like n, p, and $^4$He, and nuclei beyond $A$=12 in an NSE 
during declining temperatures, when densities are low. This causes
a large $\alpha$-particle abundance after freeze-out of nuclear reactions. 
This effect, most pronounced for core collapse supernovae,
depends on the entropy of the reaction environment, being proportional to $T^3/\rho$ in a radiation
dominated plasma (see Fig.~\ref{4:explo2}).

$r$-process nucleosynthesis (\emph{rapid} neutron capture)  \index{process!r process}
relates to environments of
explosive Si-burning, either with low or high entropies, where matter experiences
a normal or $\alpha$-rich freeze-out.
The requirement of a neutron to seed-nuclei ratio of 10 to 150 after freeze-out of
charged particle reactions\footnote{Such neutron/seed ratio is required in order to produce all, including the heaviest,
$r$-process nuclei via neutron capture from seed nuclei at their abundances before freeze-out.} translates into $Y_e$=0.12-0.3 for a 
normal freeze-out. For a moderate $Y_e$$>$0.40
an extremely $\alpha$-rich freeze-out is needed (see the disussion in Section
\ref{sec:4-4}). Under these conditions the large mass fraction in
$^4$He (with $N=Z$) permits ratios of remaining free neutrons to (small) 
abundances of heavier seed nuclei, which are sufficiently high to attain
$r$-process conditions. In many cases QSE-groups of neutron captures and
photodisintegrations are formed in the isotopic chains of heavy elements
during the operation of the $r$~process.

\section{Evolution of Massive Stars up to Core Collapse}
\label{sec:4-3}

In Section~\ref{sec:4-2} we have discussed nuclear burning processes in 
detail, including
also individual reactions which are of relevance during the evolution of
massive stars. This 
relates to the main focus of this book, the production of (radioactive) nuclei in 
astrophysical environments. In the present section we will discuss the physics of 
stellar evolution and major related observational features; but we refer to review articles or 
textbooks for technical descriptions of treatments of mass,
energy, and momentum conservations equations as well as energy transport 
(via radiation or convective motions) 
\citep{2009pfer.book.....M,2010RvMP...??..???B,2003ApJ...591..288H,2000ApJS..129..625L,2003ApJ...592..404L,2006ApJ...647..483L,2008AIPC..990..244O,2009SSRv..147....1E} (but see also the hydrostatic stellar struture / evolution equations in 
spherical symmetry, as presented in Ch.~3).
Stellar-evolution 
calculations as discussed here are based on the \emph{Geneva code} of A. Maeder and
G. Meynet and their students 
\citep{2009pfer.book.....M,2010RvMP...??..???B}.
This numerical implementation of stellar evolution includes (i) an adaptative reaction network for the advanced burning stages,
which is capable to follow the detailed evolution of $Y_e$ 
and a large set of
nuclei; (ii) a discretization of the stellar-structure equations, modified 
in order to damp instabilities occurring during the advanced stages of evolution; 
(iii) the treatment of dynamical shear in addition to the other mixing
processes (such as, e.g., horizontal turbulence, secular shear and meridional
circulation); and (iv) the treatment of convection as a diffusive process from
O-burning onwards. This allows to follow the evolution of
massive stars from their birth until the stage of Si-burning, including all nuclear
burning stages discussed in Sect.~\ref{sec:4-2}, for a wide range of
initial masses, metallicities and stellar rotation.
Here the treatment of rotation and mixing effects still utilizes methods
based on spherical symmetry. Full multi-dimensional calculations
of mixing processes during stellar evolution have recently been established
\citep{2007ApJ...667..448M,2009ApJ...690.1715A} and might open up a new
era for our understanding of the evolution of stars. 

\subsection{Stellar Evolution with Rotation}

The evolution of all stars (including massive stars discussed here) is \index{stars!rotation}
initiated by core H-burning, during which the star is found on the so-called
main sequence (MS) in the Hertzsprung-Russell (HR) diagram, 
\index{stars!Hertzsprung-Russell diagram} which relates the 
stellar luminosity to the stellar surface temperature (color).
The observational appearance of a star after the completion of core H-burning \index{process!H burning}
is affected by the fact that the H-burning region continues to move outward 
as a burning shell. The He-core contracts and ignites core He-burning
in the center, which produces mainly C and O. The star's trajectory in the HR diagram leaves the main sequence,
and its radius increases due to the increased radiation pressure. Depending
on the resulting surface temperature it becomes a blue or red supergiant 
(BSG or RSG). Radiation pressure can rise to such
extreme values that stars (more massive than 20-30 M$_\odot$) blow off their outer parts through strong \emph{stellar
winds} of velocities up to 2000~km~s$^{-1}$, exposing the more-interior parts of the star, 
the helium (or in some cases, the carbon) shell. \index{stars!Wolf Rayet}
Such a Wolf-Rayet (WR) star loses between $10^{-6}$ and 
a few times $10^{-5}$~M$_\odot$ 
per year,  in comparison to our Sun losing $10^{-14}$~of its M$_\odot$ per year through its solar wind.
For non-rotating stars, the
transition to the WR phase appears through the so-called Luminous
Blue Variable stars (LBVs). \index{stars!luminous blue variable}
LBVs are massive, intrinsically bright stars which display different scales of
light and color variability, ranging from rapid microvariations to rare
outbreaks of catastrophic mass loss. They represent a very short-lived
(perhaps as little as 40,000 years) strongly mass-losing phase in the evolution
of massive stars, during which they undergo deep erosion of the outer layers
before they enter the Wolf-Rayet phase.
For rotating stars, the WR phase may start before the star ends its main
sequence, since rotation enhances mass loss and rotation-induced mixing
further reduces the hydrogen content at the surface 
\citep{2003A&A...404..975M,2005A&A...429..581M}.
In the following we discuss how these evolutionary phases depend
on the initial properties of a star. Late burning phases progress
much more rapidly than the H burning of the main sequence state. This is because the carriers of the star's energy loss, 
which drives the evolution of a star, change from photons to neutrinos, which escape immediately at the densities discussed here, while photons
undergo a multitude of scattering processes until they finally escape at the photosphere\footnote{It takes a photon about 10$^5$ years to reach the surface, after it has been launched in the hot core of, e.g., our Sun.}.
The characteristics of late-burning stages are essentially identified by the size of a star's C+O-core
after core He-burning.

\begin{figure}[!tbp] 
\centering
\includegraphics[width=5.5cm]{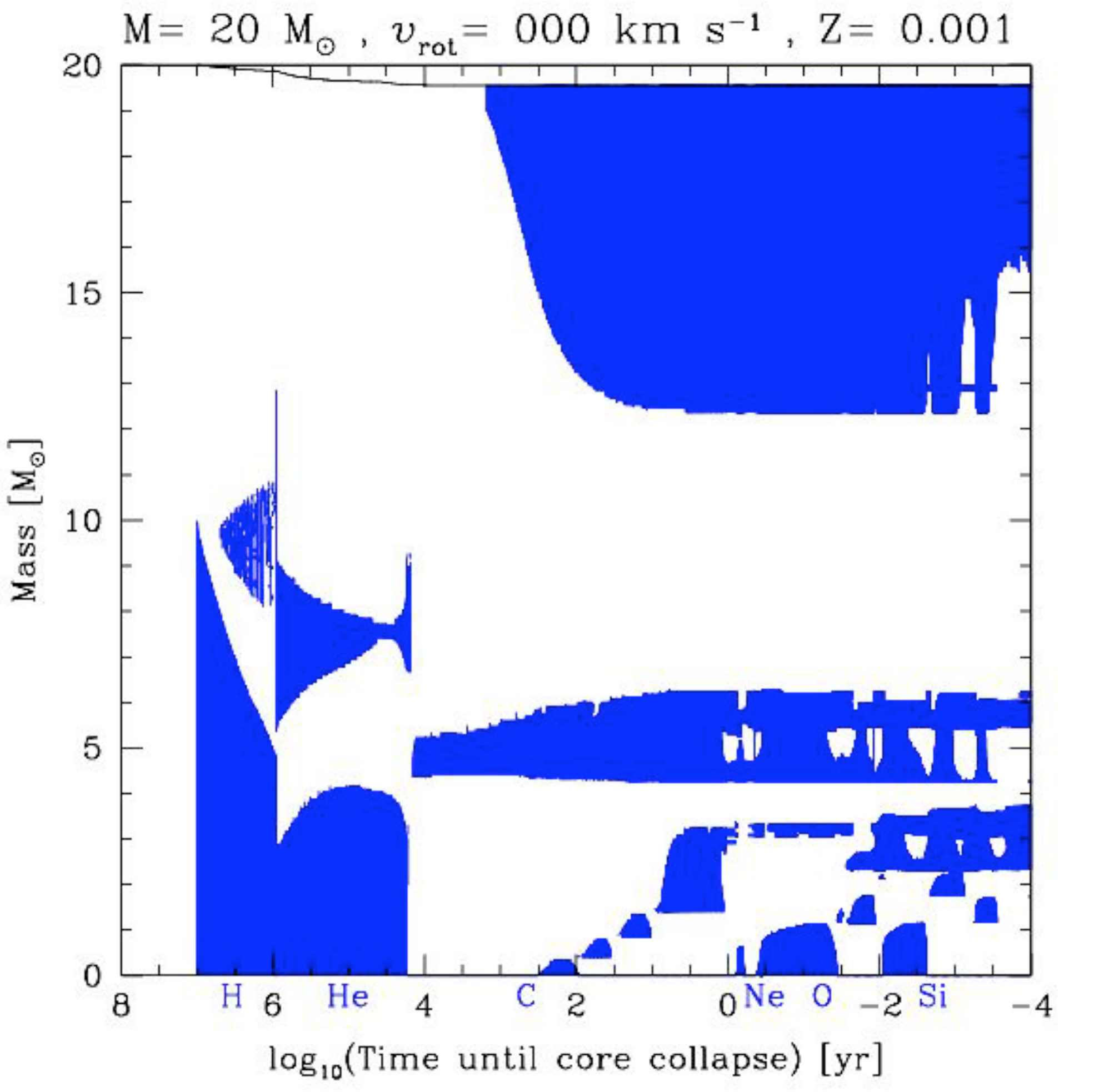}\includegraphics[width=5.5cm]{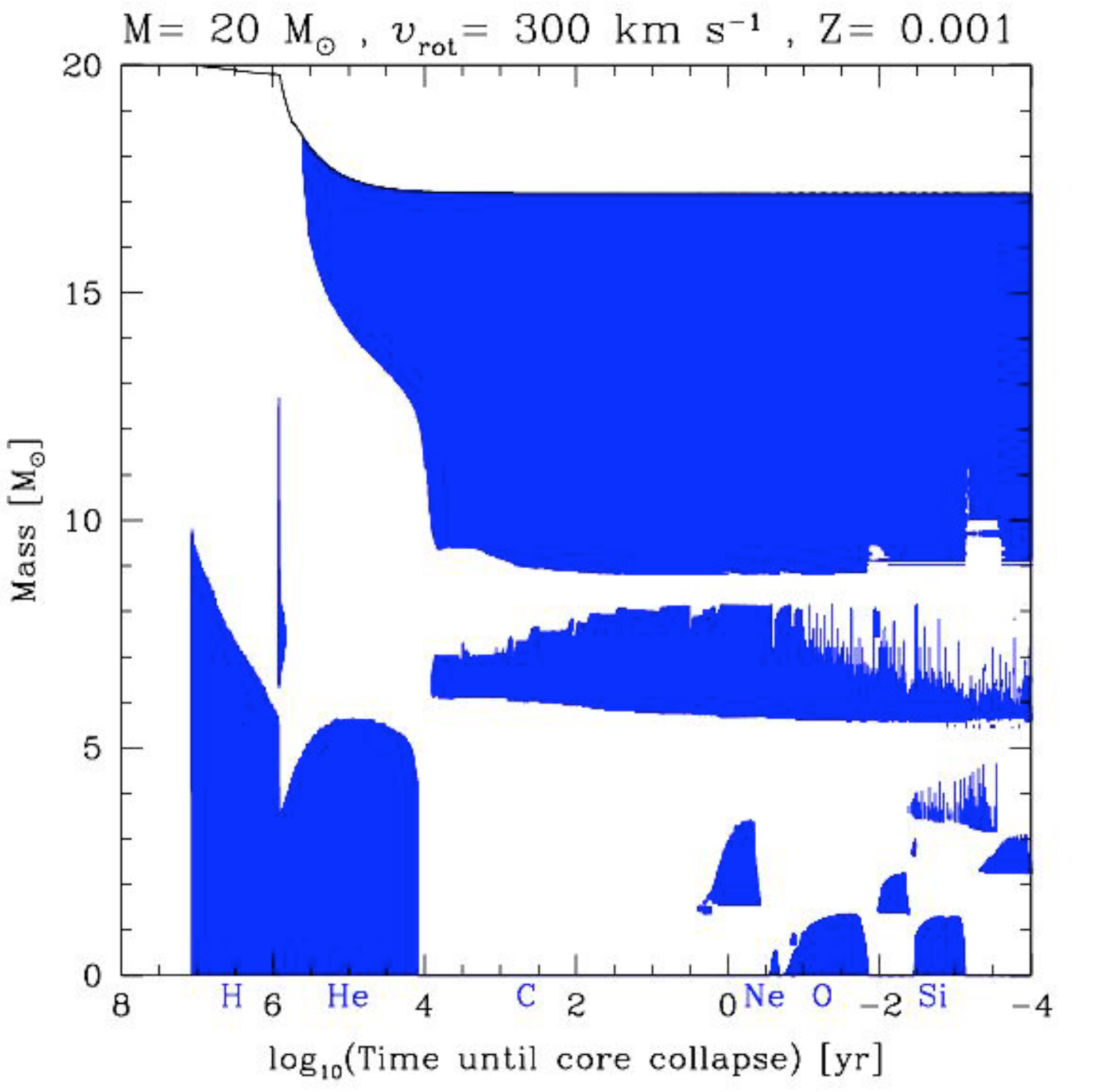}
\includegraphics[width=5.5cm]{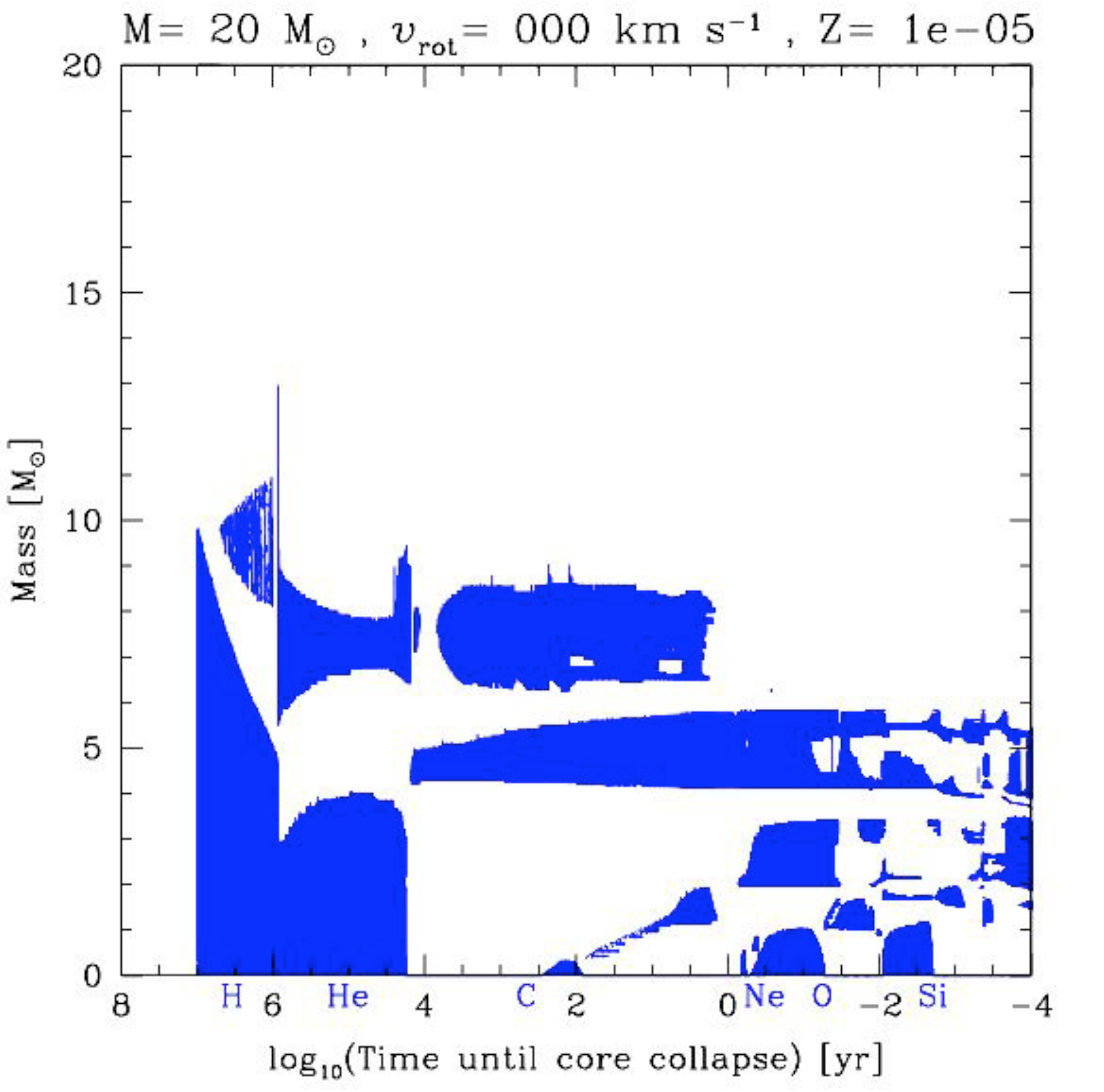}\includegraphics[width=5.5cm]{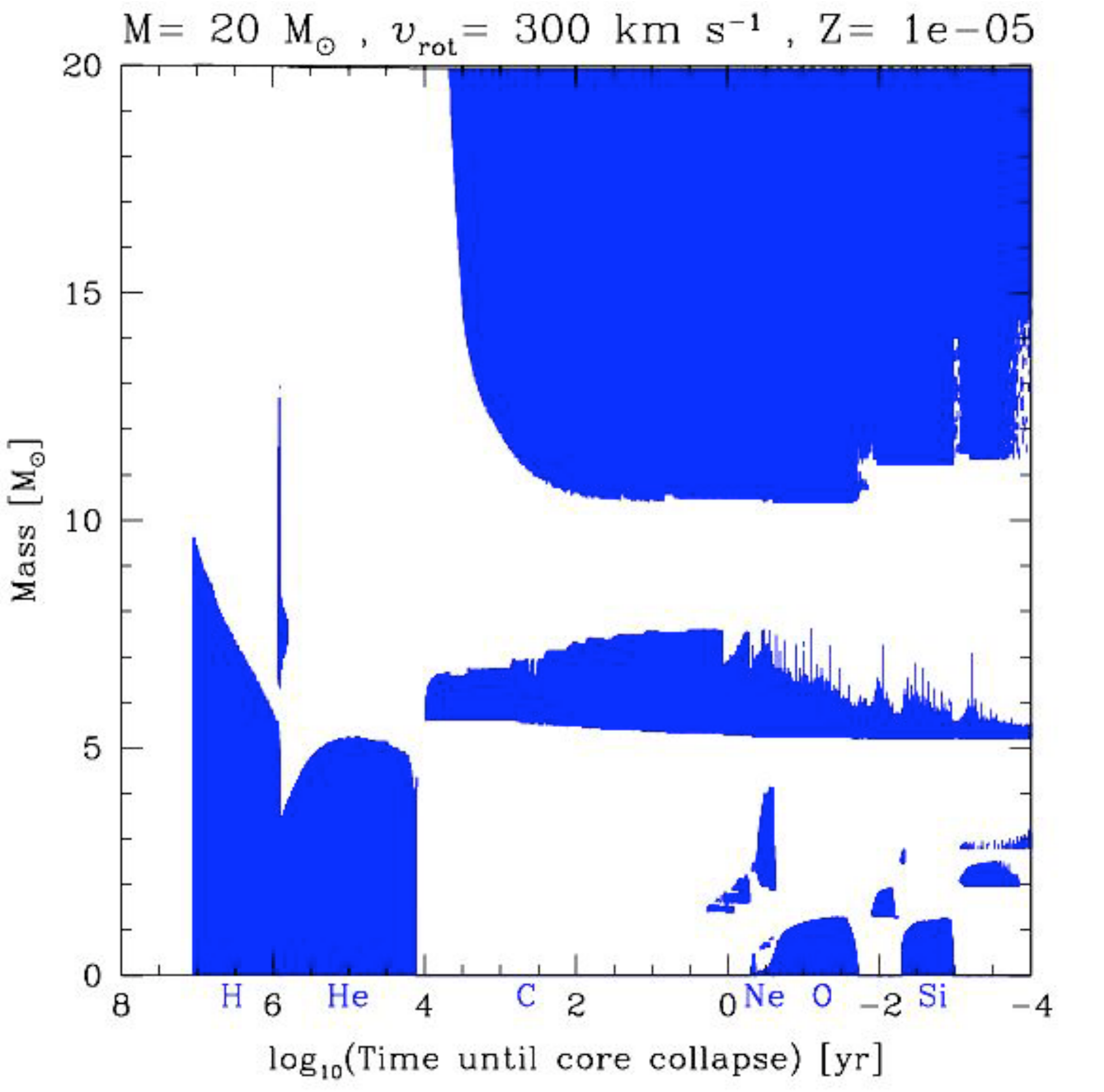}
\includegraphics[width=5.5cm]{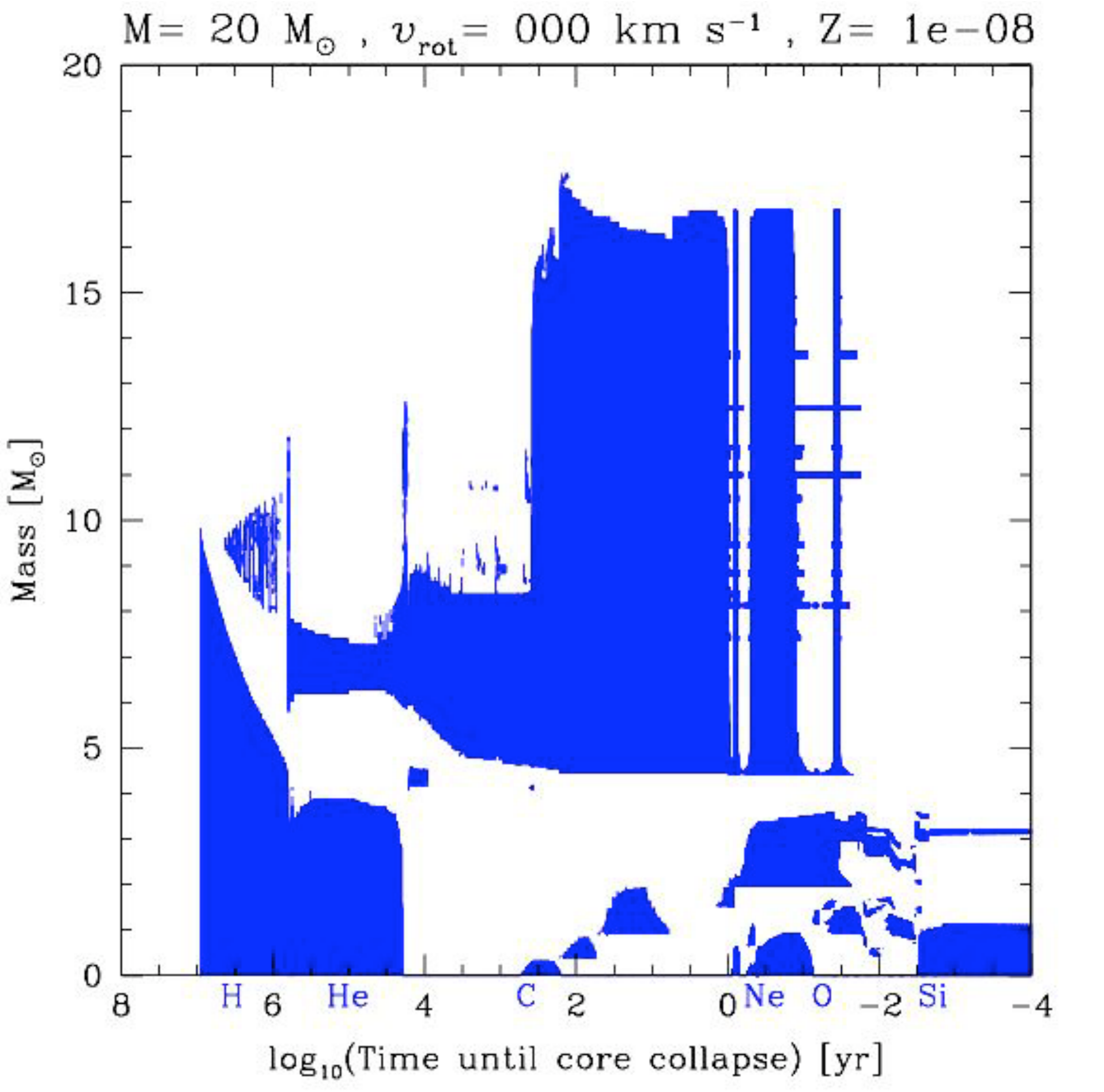}\includegraphics[width=5.5cm]{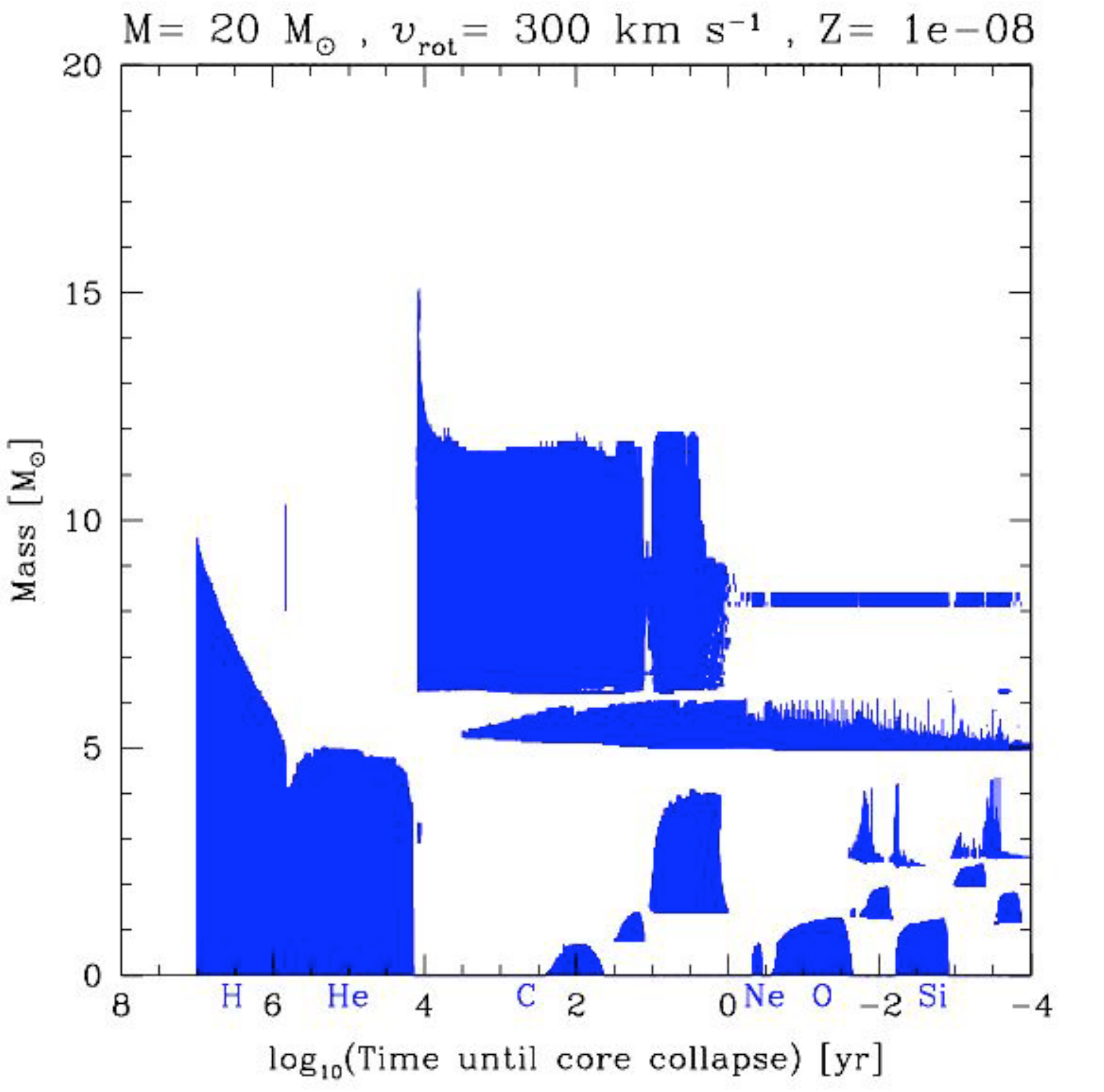}
\caption{Stellar structure (Kippenhahn) diagrams, which show the \index{stars!Kippenhahn diagram}
evolution of the structure as a function of the time left until the core
collapses after the completion of core Si-burning.
The initial parameters of the models are given on the top of each plot.
Coloring (shading) marks convective zones, and
the burning stages are denoted below the time axis. Non-rotating and moderately rotating 20~M$_\odot$ star models are shown, for different
metallicities $Z$. $v_{rot}$ indicates the rotation velocity at the surface of the star.}
\label{4:kip}
\end{figure}  

The evolution of stars is governed mainly by three initial parameters: (1) its mass $M$, (2) its metallicity ($Z$, i.e. the mass fraction of pre-existing 
elements heavier than He from earlier stellar generations), and (3) the rotation 
rate or surface rotation velocity $v_{rot}$. 
Solar metallicity corresponds to\footnote{The current value of solar metallicity is believed to be Z=0.014, see Ch.~1; the value of Z=0.02, which had been established before and was in common use till$\sim$2005, remains a reference for comparisons, though.} $Z=0.02$.
The evolution can also be influenced by interior magnetic fields, and by
a close binary companion. Rotation significantly affects
the pre-supernova state, through the impact it has on the
H and He-burning evolution. Two mass groups are distinguishable: Either
rotationally induced mixing dominates (for 
$M<$30~M$_{\odot}$), or rotationally
increased mass loss dominates (for 
$M>$30~M$_{\odot}$).
For massive stars around solar metallicity, mass loss plays a crucial role, in 
some cases removing more than half of the initial mass. Internal mixing, 
induced mainly by convection and rotation, also has a significant effect on the 
evolution of stars. 
An important result is the production of primary $^{14}$N (via the CNO-cycle) and $^{22}$Ne (via
$\alpha$-captures in He-burning), 
due to mixing of burning products (such as $^{12}$C) with hydrogen or $\alpha$'s, respectively (see the discusssion in Sect.~\ref{sec:4-2}).

The general impact of metallicity can be summarized in the following way:
Lower metallicity implies a (slightly) lower luminosity due to the
lack of CNO-cycling in hydrogen burning, which leads to slightly 
smaller convective cores. A lower metallicity also implies a lower opacity 
due to the lack of heavier elements with their many spectral lines, reducing
therefore also radiation pressure and hence mass loss (as long as the chemical 
composition has not been changed by burning or mixing in the part of the 
star under consideration). This results in lower metallicity stars being 
more compact and experiencing less mass loss. 
Prescriptions for mass loss as used in the Geneva
stellar evolution code are described in detail in 
\citet{2005A&A...429..581M}. Mass loss rates depend on metallicity as 
${ dM/dt} \propto (Z/Z_\odot)^{0.5...0.86}$, where $Z$ is the mass fraction
of heavy elements at the surface of the star.
The effects can be seen in Fig.{\ref{4:kip}} which shows the interior structure of stars through so-called \emph{Kippenhahn
diagrams} of 20 M${_\odot}$ models for different metallicities and rotation
velocities of the stars. These diagrams indicate regions
(in radial mass coordinates) where matter is unstable against convection; here the energy 
transport is dominated by transporting hot matter 
rather than through the propagation of photons.
The implications of such a behavior have already been described in Ch.~3, the
evolution of low and intermediate mass stars, and the physical
origin and treatment of these effects are addressed in Ch.~8.

With the exception of the outer convection
zone, convective regions  in most cases indicate \emph{burning zones},
such as \emph{core H-burning, core He-burning, core C-burning} etc.. They testify 
also the ignition of outward moving burning shells of the same nuclear 
burning stages. When comparing models for decreasing metallicities (without rotation, left column of 
Fig.{\ref{4:kip}}) one notices only minute reductions of the core sizes, but it 
is clearly seen that the outer (H-)burning shell moves \index{stars!structure}
further in towards smaller radial mass zones. In the third figure in this 
column we see merging of the H- and He-burning shells due to this effect,
which leads to a largely-increased energy generation and  
extension of these combined burning zones.

How does rotation change this picture, and how do rotation-induced processes 
vary with metallicity? At all metallicities, rotation usually increases the \index{stars!rotation}
core sizes via rotational mixing. The supply of more H-fuel leads to more
energy generation and therefore a higher luminosity. The higher luminosity
increases the radition pressure and stellar mass loss.  \index{stars!mass loss} \index{stars!core size}
The effect of increased core sizes (and smaller density gradients) can be
viewed in all models with $v_{rot}$=300 km s$^{-1}$ in the second column
of Fig.{\ref{4:kip}}. Clearly the convective core sizes are increased and
the shell burning zones have moved outward.  In the lowest
metallicity case,  the H/He-layers are separated again. In the intermediate metallicity case
$Z=10^{-5}$, the outer convection zone reaches the surface, and the star 
becomes a red supergiant.
For metallicities $Z=0.001$ (top row), the increased 
luminosity causes a sufficient increase in radiation pressure so that the mass
loss is substantially enhanced (see the decrease of the stellar mass indicated
by the top line).
Mass loss becomes gradually unimportant for decreasing metallicities. For the
rotating 20 M$_\odot$ models the stellar fraction lost  is more than 50\% for
solar metallicities, 13\% at $Z=0.001$, less than 3\% for $Z=10^{-5}$, and less
than 0.3\% for  $Z=10^{-8}$.

\begin{figure}[!tbp] 
\centering
\includegraphics[width=5.5cm]{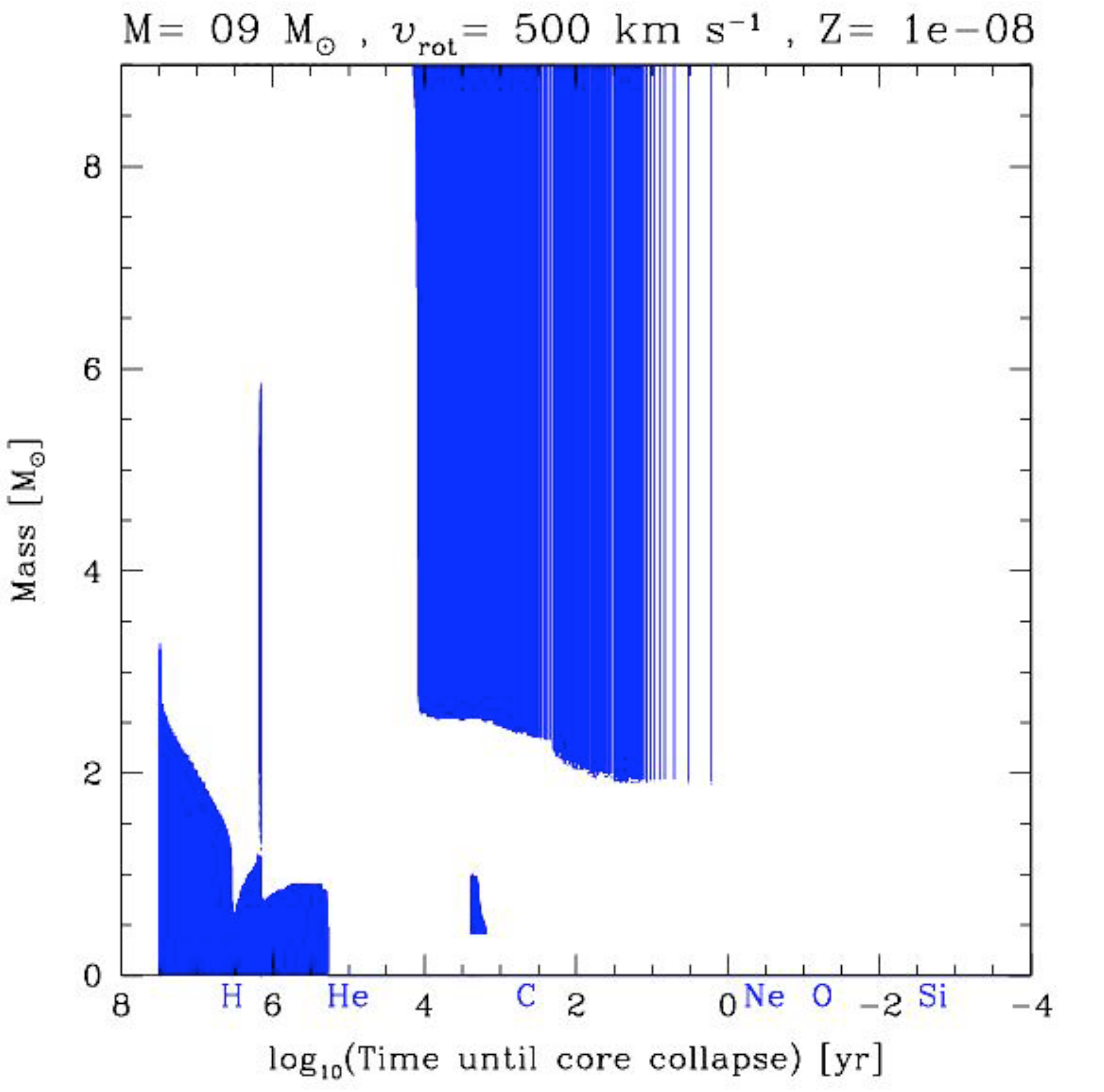}\includegraphics[width=5.5cm]{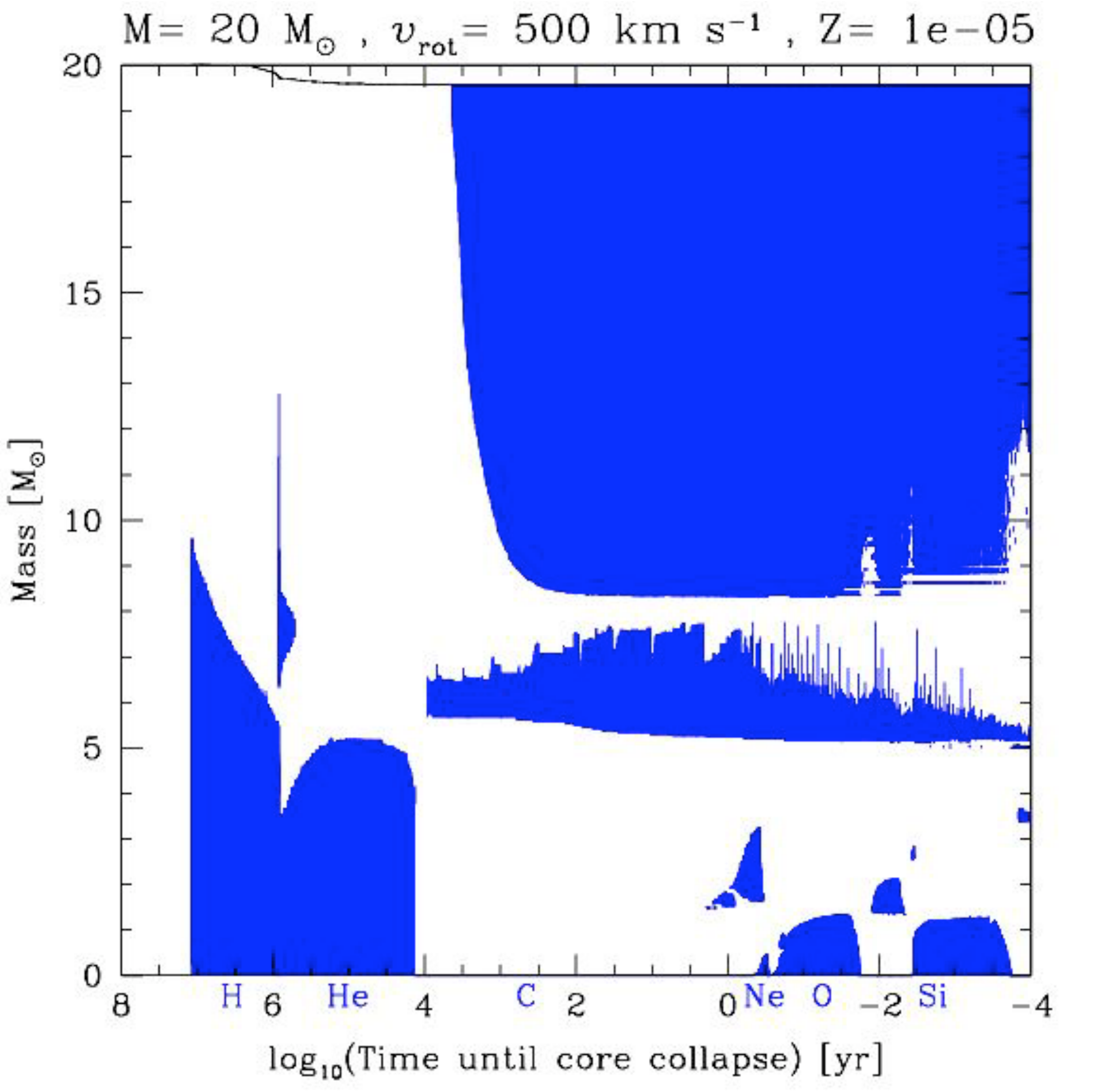}
\includegraphics[width=5.5cm]{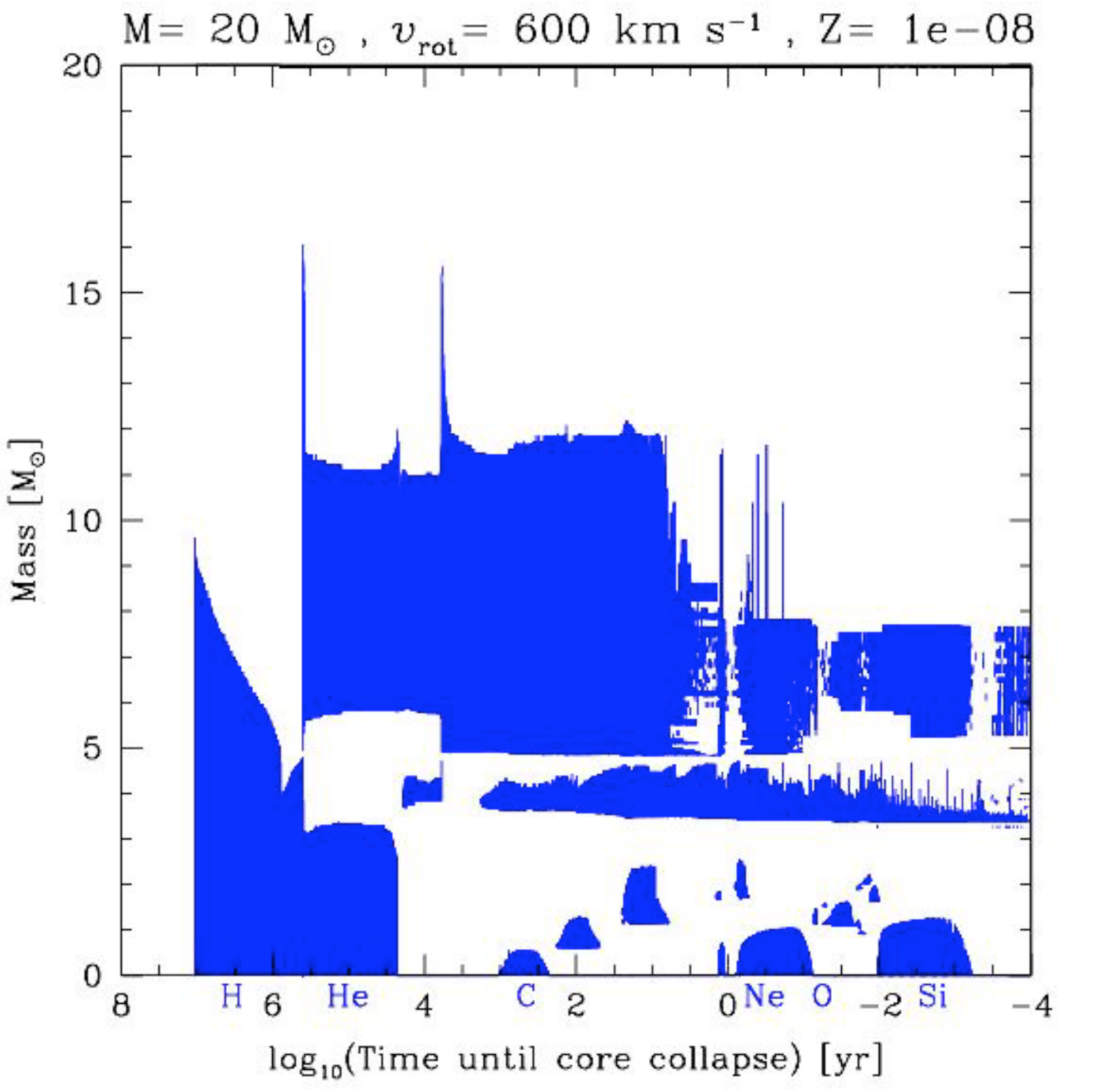}\includegraphics[width=5.5cm]{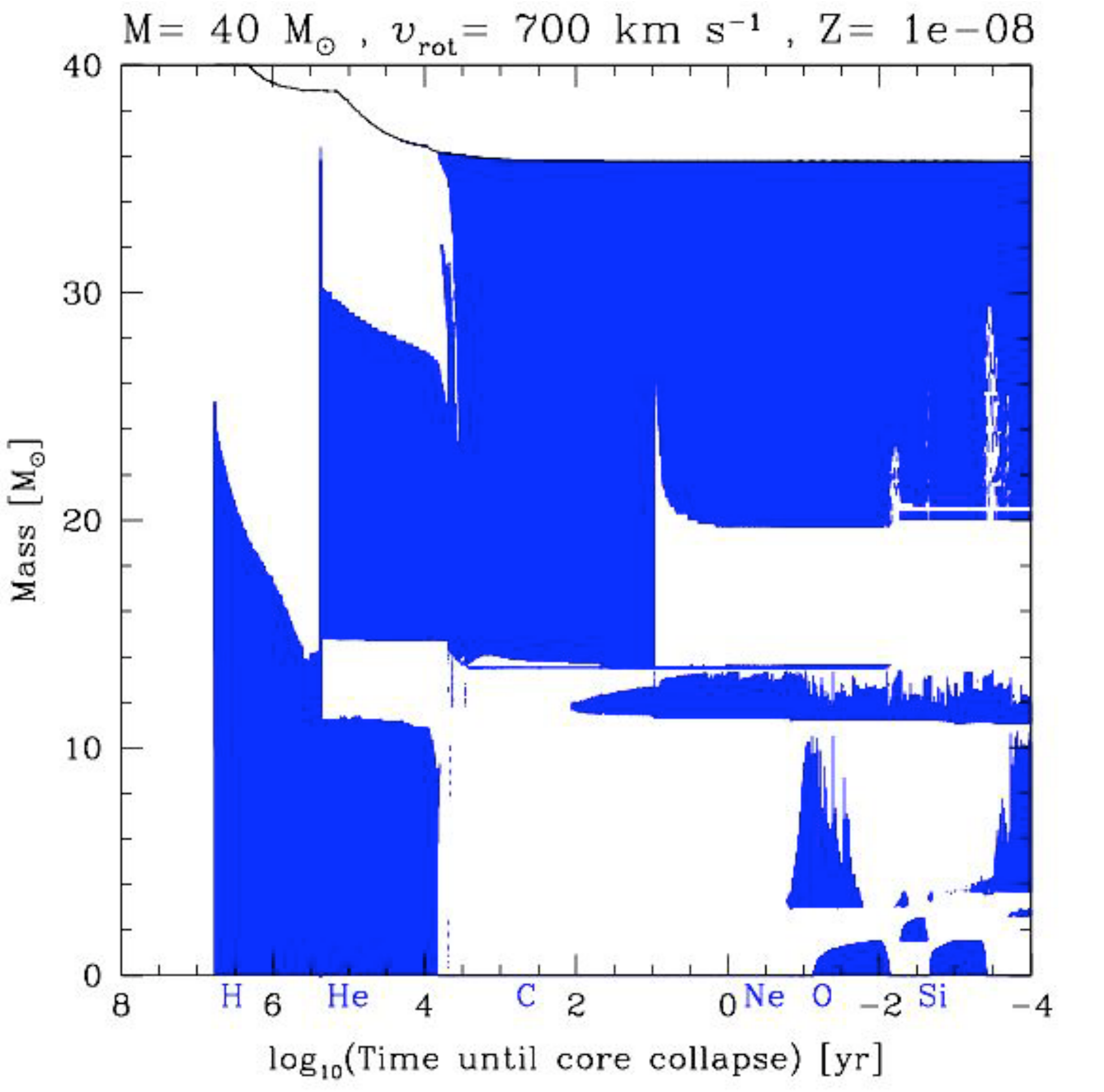}
\includegraphics[width=5.5cm]{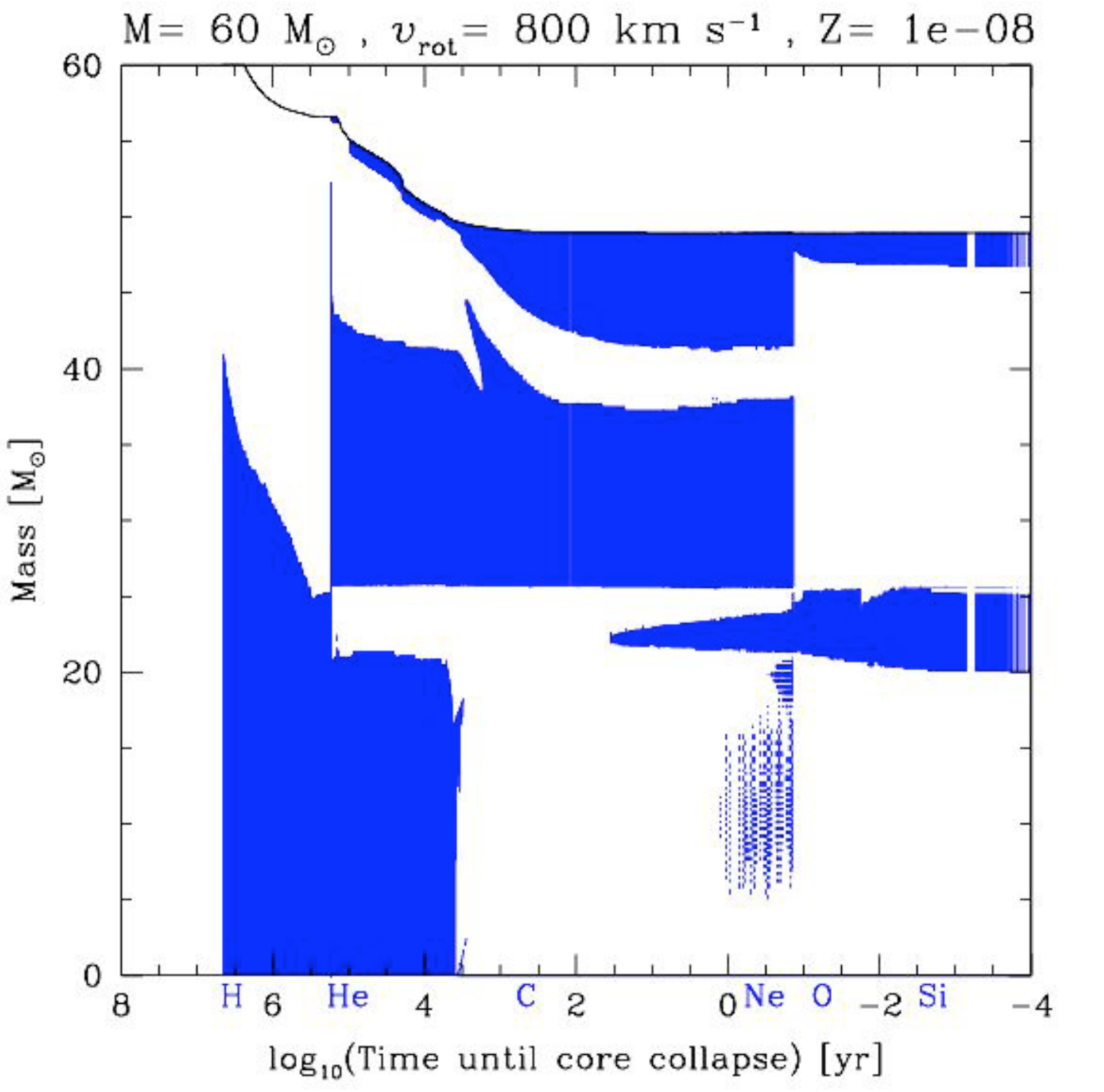}\includegraphics[width=5.5cm]{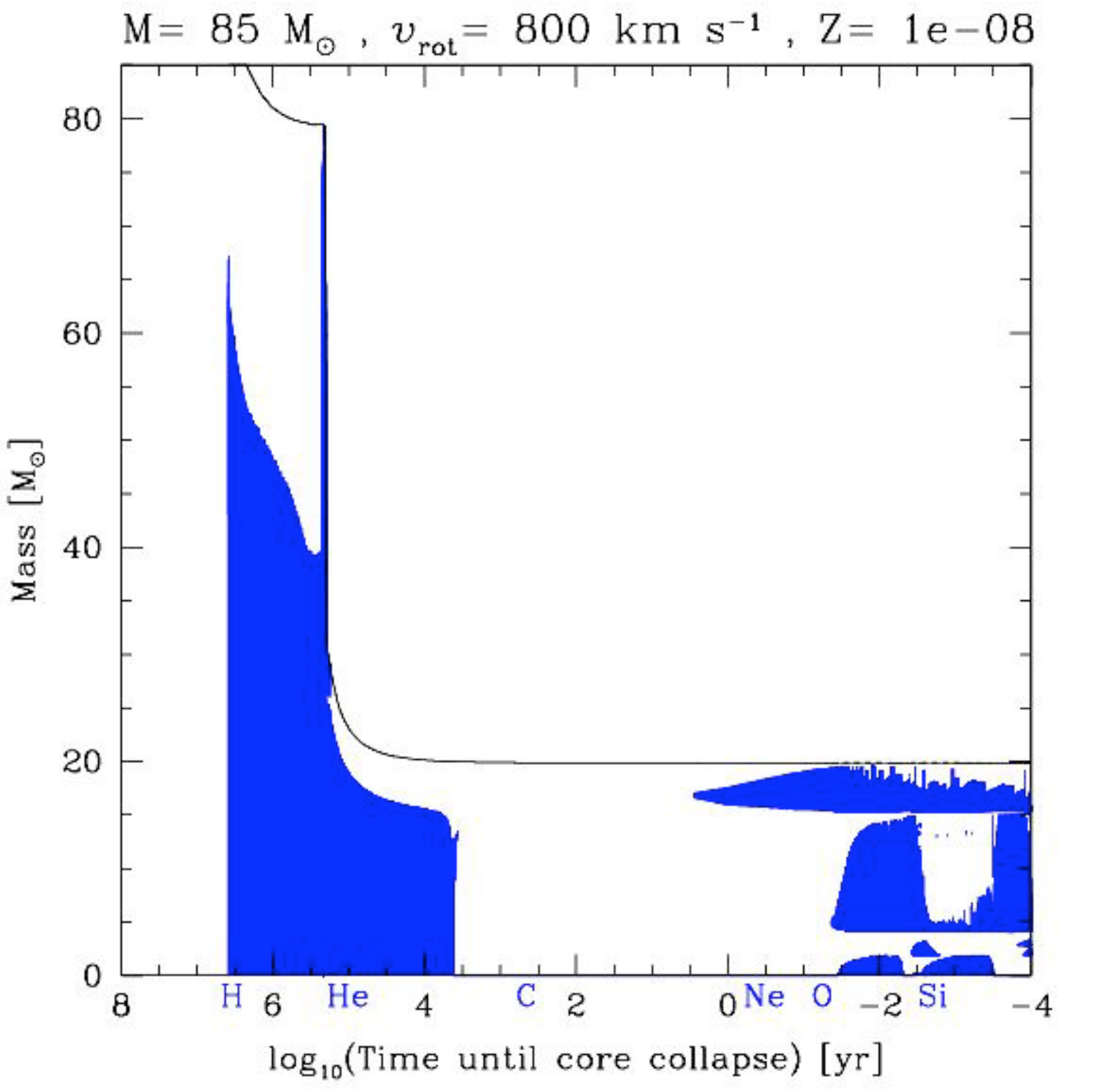}
\caption{The same as Fig. \ref{4:kip}. Stellar structure diagrams for rapidly
rotating stars of metallicity $Z=10^{-8}$, over a mass range from 9 to 
85~M$_\odot$. We see the drastically increasing amount of mass loss with 
increasing
mass (enhancing mixing of burning products to the surface, and increasing
opacities, i.e. acting like increased metallicities, plus some mass loss from
critical rotation for the most massive stars). The two metallicity cases shown for
the 20~M$_\odot$ star show again that stars are less compact and show more 
and enhanced mass loss for higher metallicities (becoming a RSG which leads to the appearance of a large convective envelope).}
\label{4:kip2}
\end{figure} 

This can be different for more massive stars 
\citep{2006A&A...447..623M}.
In Fig.{\ref{4:kip2}},
 we show results for low metallicity stars with 
$Z=10^{-8}$ and fast rotation (500-800~km~s$^{-1}$) from 9 to 85~M$_\odot$.
The surface layers of massive stars usually accelerate 
due to internal transport of angular momentum 
from the core to the envelope. Since at low $Z$, stellar winds are weak, this 
angular momentum dredged up by meridional circulation remains inside the star, 
and \index{stars!critical rotation}
the star reaches critical rotation more easily. At the critical limit, matter 
can be launched into a Keplerian disk which probably dissipates under 
the action of the strong radiation pressure of the star. 
Such an effect can be seen for the 85 M$_\odot$ star, which loses in total
more than 75\% of its initial mass, and initially about
10\% due to critical rotation.
The remaining mass loss occurs during the red
supergiant phase after rotation and convection have enriched the surface
in primary CNO elements. 
We can also see that this effect becomes vanishingly
small for stars with masses $M<30$M$_\odot$. The two 20 M$_\odot$ models with
varying metallicities and degrees of rotation again indicate the
influence of metallicity and rotation on the compactness and mass loss
of stars. In both cases the mass loss is negligible. 

We have not shown here the evolution of extremely low metallicity stars.
Below a metallicity of about $Z = 10^{-10}$,
the CNO cycle cannot operate when \index{stars!evolution at low metallicity}
H-burning starts after the star has been formed. The star therefore
contracts until He-burning ignites, because the energy generation rate of H burning through the pp-chains
cannot balance the effect of the gravitational force. Once
enough C and O is produced, the CNO cycle
can operate, and the star behaves like stars with $Z > 10^{-10}$
for the rest of the main sequence. Metal-free
stellar evolution models are presented in 
\citet{2004ApJ...608..405C,2002ApJ...567..532H,2005ApJ...619..427U,2008A&A...489..685E}.

Including the effects of both mass loss and 
rotation, massive star models reproduce many observables of stars with metallicities around solar $Z$. For example, 
models 
with rotation allow chemical surface enrichments already on the main sequence 
of core hydrogen burning (MS), whereas without the inclusion of rotation, 
self-enrichment is only possible during advanced burning evolution
such as the red supergiant RSG stage 
\citep{2000ApJ...544.1016H,2000A&A...361..101M}. 
Rotating star models also better reproduce the ratio of star types, for the ones which
retain their hydrogen surface layer (O stars), which lose the hydrogen layer
completely (WR stars),  and which even lose their helium layer. The latter affects
also the appearance of later core collapse supernova explosions of massive
stars. Indeed,
rotation changes the supernova type due to the mass loss of the hydrogen
envelope (turning such an event in optical observations from a type II
supernova with a strong plateau phase to a IIb event with a smaller
plateau, or even a Ib event for the case of complete loss of the hydrogen
envelope, and a Ic event with the additional loss of the He-envelope). 
This is discussed in more detail in Sect.~\ref{sec:4-4}.
Both aspects, the chemical surface enrichment in MS stars as well as the 
ratio of type Ib+Ic to type II supernovae, as a 
function of metallicity, are drastically changed 
compared to non-rotating models, which underestimate these ratios 
\citep{2009A&A...502..611G,2005A&A...429..581M}. 
The value of 300 km s$^{-1}$, used as the initial 
rotation velocity at solar metallicity, corresponds to an average velocity of 
about 220 km s$^{-1}$ on the main sequence (MS), which is close to the average 
observed value 
\citep{1982PASP...94..271F,2008IAUS..252..317M}.  
Observed ratios of stars of different types in the Magellanic clouds, as compared to our 
Galaxy 
\citep{1999A&A...346..459M,2007A&A...462..683M}, 
point to  stars rotating faster at lower metallicities.  
Fast initial rotation velocities in the range of
600 -- 800 km s$^{-1}$ \citep{2005A&A...443..581H} are supported by 
observations of very low-$Z$ stars \citep{2006A&A...449L..27C}.

\begin{figure}[!tbp] 
\centering
\includegraphics[width=12cm]{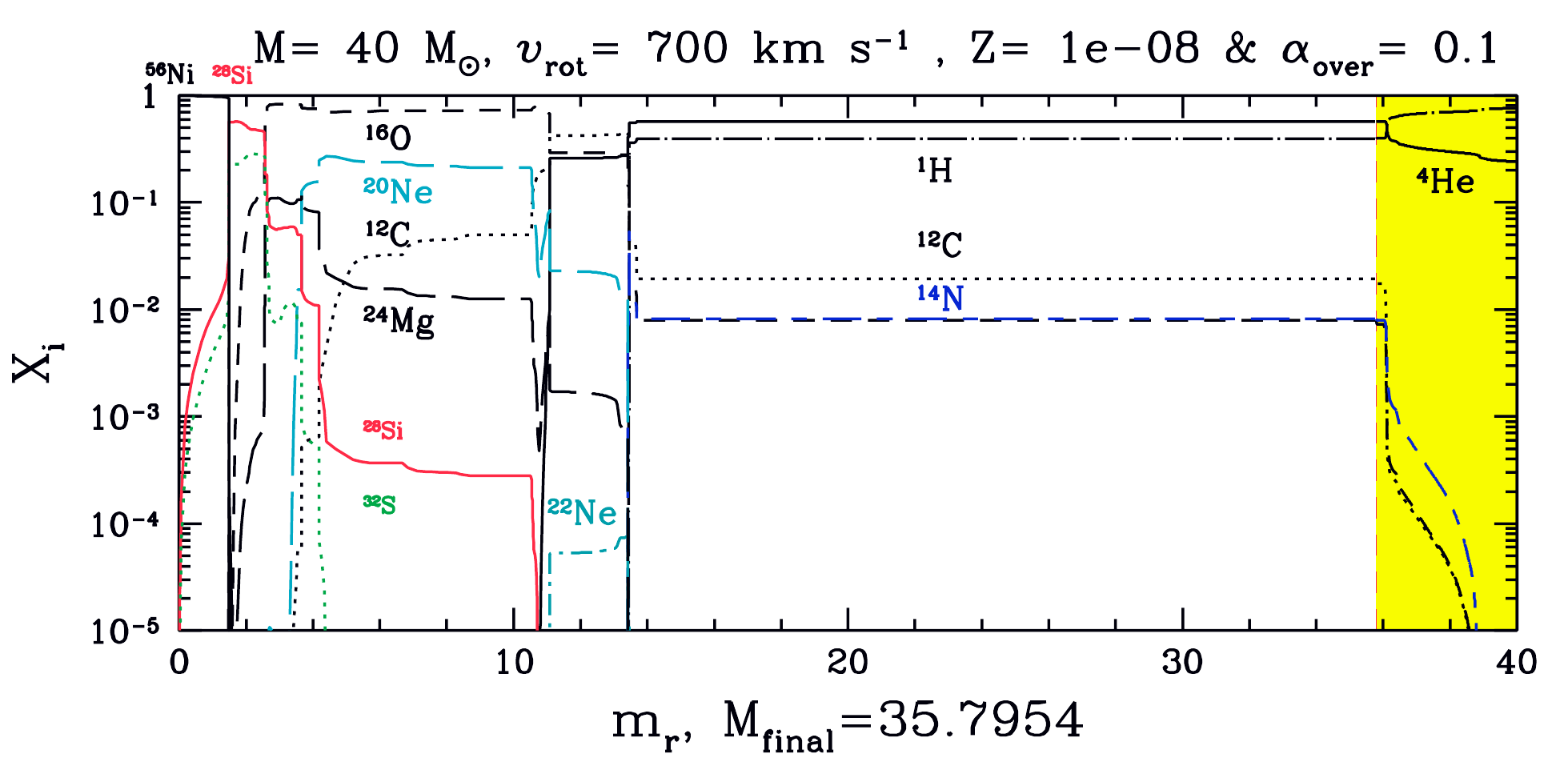}
\includegraphics[width=12cm]{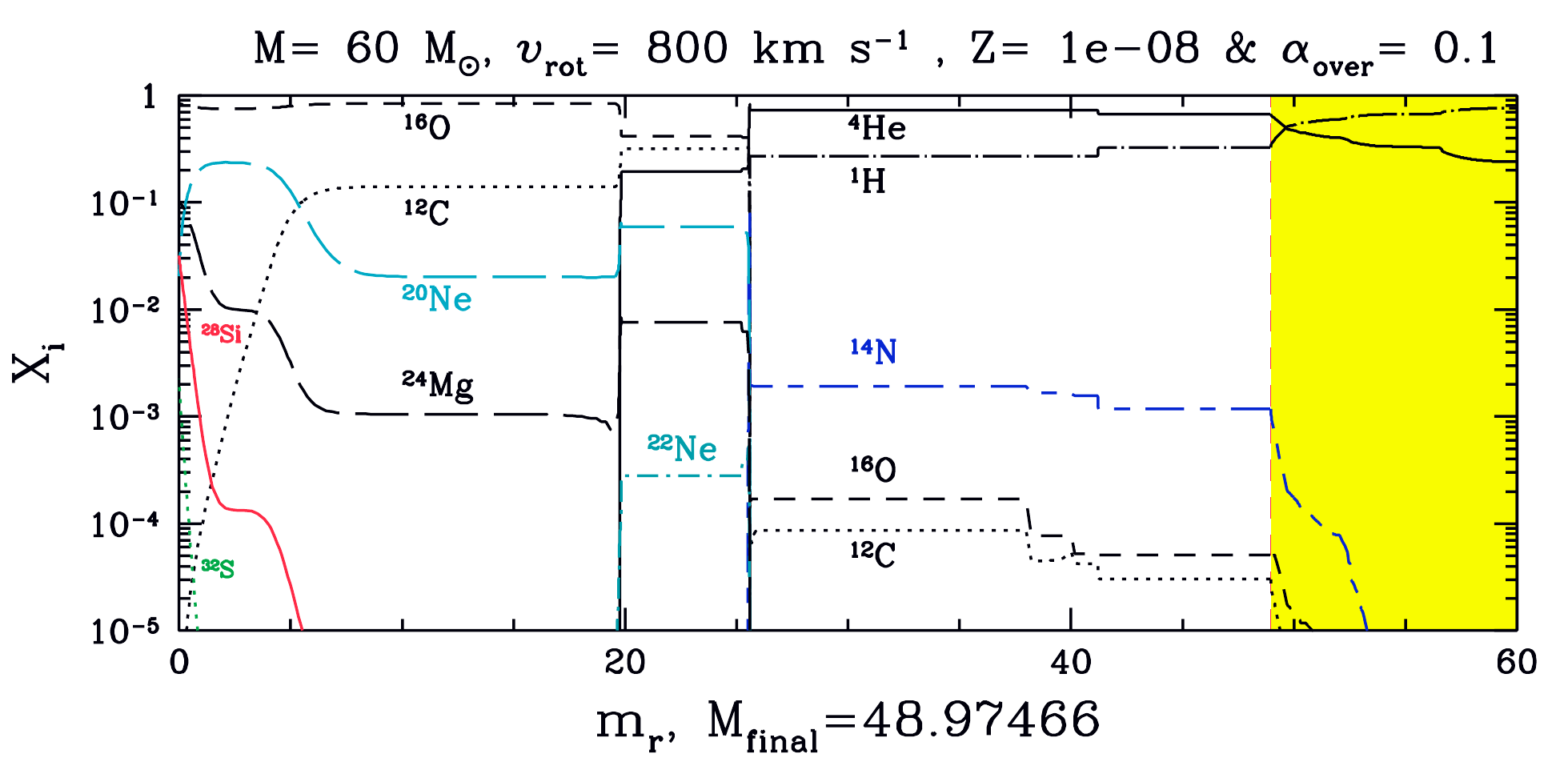}
\includegraphics[width=12cm]{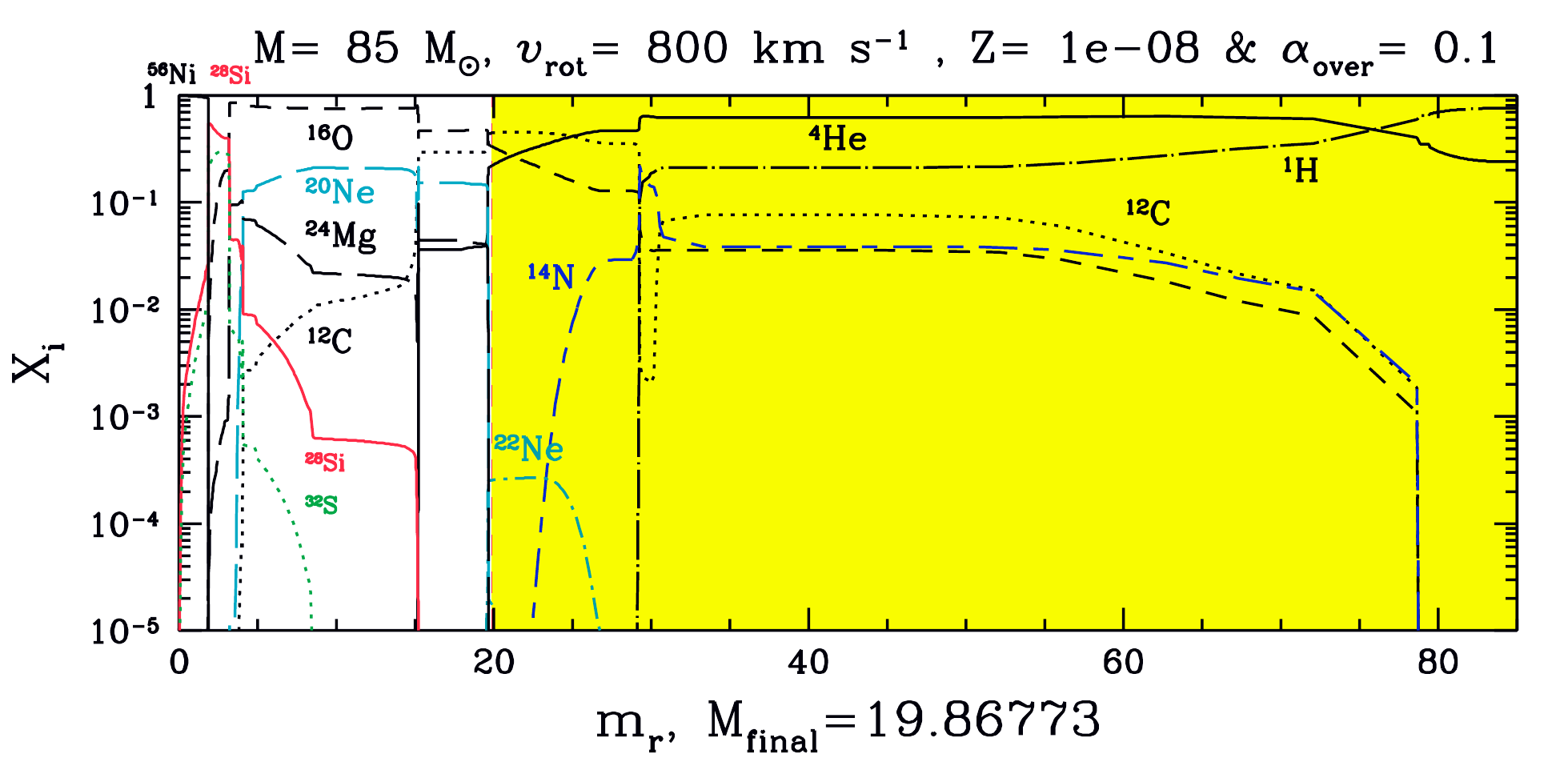}
\caption{Abundance profiles for the 40 ({\it top}), 60 ({\it middle}) and 85 ({\it
bottom}) $M_\odot$ models. The pre--SN and wind  (yellow shaded
area) chemical compositions are
separated by a vertical dashed line located at the pre--SN total mass
($M_{\rm final}$), given below each plot.}
\label{4:absw}
\end{figure} 

Rotation affects all burning stages and the resulting Fe-core \index{stars!core size}
(we will discuss this issue further in the next subsection, see also
Fig. \ref{4:presn}).
The size of the Fe-core in turn determines the final fate,
whether a supernova explosion with neutron star formation or the collapse to
a black hole occurs. The effects of rotation on pre-supernova models
are most spectacular for stars between 15 and 25 $M_{\odot}$. 
It changes the total size/radius of progenitors (leading
to blue instead of red supergiants) and the helium and CO core
(bigger by a factor of $\sim 1.5$ in rotating models). 
The history of convective zones (in particular the convective zones associated 
with shell H-burning and core He-burning) is strongly affected by rotation 
induced mixing \citep{2005A&A...443..581H}. 
The most important rotation induced mixing 
takes place at low $Z$ while He is burning inside a convective core. Primary C 
and O are mixed from the convective core into the H-burning shell. 
Once the enrichment is strong enough, the H-burning shell is boosted (the CNO 
cycle depends strongly on the C and O mixing at such low initial 
metallicities). The shell becomes convective and leads to an important \index{isotopes!14N}
primary $^{14}$N production while the convective core mass decreases,
leading to a less massive CO-core after He-burning than in non-rotating 
models. Convective and rotational mixing brings the primary CNO to the surface 
with interesting consequences for the stellar yields. The yield of $^{16}$O, 
being closely correlated with the mass of the CO-core, is reduced.
At the same time the C yield is slightly increased 
\citep{2005A&A...443..581H},
both due to the slightly lower temperatures in core He-burning. 
This is one possible 
explanation for the high [C/O] ratio observed in the most metal-poor 
halo stars (see Fig. 14 in \citet{2005A&A...430..655S} and 
\citet{2009A&A...500.1143F}) and in 
damped Lyman-$\alpha$ systems DLAs \citep{2008MNRAS.385.2011P}.

The fate of rotating stars at very low $Z$ is therefore probably the following:
$M < 30-40$ M$_\odot$ : Mass loss is insignificant and matter is only ejected into
the ISM during the SN explosion. 
30-40 M$_\odot < M < 60 $M$_\odot$ : Mass loss (at critical rotation and in the 
RSG stage) removes 10-20\% of the initial mass of the star. The star probably 
dies as a black hole without a SN explosion and therefore the feedback into the
ISM is only due to stellar winds.
$M > 60$ M$_\odot$: A strong mass loss removes a significant amount of mass and
the stars enter the WR phase. These stars therefore end as type Ib/c SNe and 
possibly as GRBs. This behavior is displayed in Fig. \ref{4:absw}.
At a metallicity $Z=10^{-8}$,
corresponding to an Fe/H ratio 
$log_{10}[(Fe/H)/(Fe/H)_\odot]=$[Fe/H]$\sim-6.6$,
C and O are shown in models to be mixed into the H-burning shell during He-burning. This raises the 
importance of the shell, and
leads to a reduction of the CO-core size. Later in the evolution,
the H-shell deepens and produces large amounts of primary nitrogen.
For the most massive stars ($M > 60$~$M_\odot$), significant mass
loss occurs during the red supergiant stage, caused by the surface enrichment in CNO elements from rotational and convective
mixing.

The properties of non-rotating low-$Z$ stars are 
presented in \citet{2003ApJ...591..288H,2008IAUS..255..297H}, and several 
groups have calculated their stellar yields 
\citep{2002ApJ...567..532H,2004ApJ...608..405C,2007ApJ...660..516T}.  
All results for the non-rotating stars (whether at solar metallicity
or for low-$Z$ models) are consistent
among these calculations, differences are understood from 
the treatments of convection and the rates used for
$^{12}$C$(\alpha,\gamma)^{16}$O.
The combined contributions to stellar yields by the wind and the
later supernova explosion (see Sect.~\ref{sec:4-4}) will be provided 
separately.
The results for stellar models with metallicities $Z$ close to solar
can be described as follows:
Rotating stars have larger yields in their stellar winds than
the non-rotating ones, because of the extra mass loss and mixing due to
rotation, for masses below $\sim 30\,M_{\odot}$.
The $^{12}$C and $^{16}$O yields are increased by a factor 1.5--2.5 by
rotation. At high
mass loss rates (above $\sim 30$~M$_{\odot}$), the rotating and
non-rotating models show similar yield values.
When the wind and explosive contributions are added, the 
total metal production of rotating stars
is larger by a factor 1.5-2.5 (see Sect.~\ref{sec:4-4}). For very
massive stars, the situation varies due to the
extreme mass loss, as shown in Fig.\ref{4:absw}.

\begin{table}[!tbp]
\caption{Stellar Properties (Limongi \& Chieffi 2006)}
\begin{center}
\begin{tabular}{c c c c}
\hline
\hline
$M_{ini}$/M$_\odot$ & $M_{fin}$/M$_\odot$ & $M_{He}$/M$_\odot$ & $M_{CO}$/M$_\odot$ \\
\hline
11 & 10.56 & \ 3.47 & \ 1.75 \\
15 & 13.49 & \ 5.29 & \ 2.72 \\
20 & 16.31 & \ 7.64 & \ 4.35 \\
30 & 12.91 & 12.68 & \ 8.01 \\
40 & 12.52 & 16.49 & \ 8.98 \\
60 & 17.08 & 25.17 & 12.62 \\
80 & 22.62 & 34.71 & 17.41 \\
\hline
\end{tabular}
\end{center}
\label{4:limongi}
\end{table}

\begin{table}[!tbp]
\caption{Stellar Properties (Hirschi et al. 2007)}
\begin{center}
\begin{tabular}{c c c c c c}
\hline
\hline
$M_{ini}$/M$_\odot$ & $Z$ & \ \ $v_{rot}$\ \  & $M_{fin}$/M$_\odot$ & $M_{He}$/M$_\odot$ & $M_{CO}$/M$_\odot$ \\
\hline
\ 9 & $1\times 10^{-8}$ & 500 & \ \ 9.00 & \ \ 1.90 & \ \ 1.34 \\
20 & $2\times 10^{-2}$ & 300 & \ \ 8.76 & \ \ 8.66 & \ \ 6.59 \\
20 & $1\times 10^{-3}$ & \ \ 0 & 19.56 & \ \ 6.58 & \ \ 4.39 \\
20 & $1\times 10^{-3}$ & 300 & 17.19 & \ \ 8.32 & \ \ 6.24 \\
20 & $1\times 10^{-5}$ & 300 & 19.93 & \ \ 7.90 & \ \ 5.68 \\
20 & $1\times 10^{-5}$ & 500 & 19.57 & \ \ 7.85 & \ \ 5.91 \\
20 & $1\times 10^{-8}$ & 300 & 20.00 & \ \ 6.17 & \ \ 5.18 \\
20 & $1\times 10^{-8}$ & 600 & 19.59 & \ \ 4.83 & \ \ 4.36 \\
40 & $1\times 10^{-8}$ & 700 & 35.80 & 13.50 & 12.80 \\
60 & $1\times 10^{-8}$ & 800 & 48.97 & 25.60 & 24.00 \\
85 & $1\times 10^{-8}$ & 800 & 19.87 & 19.90 & 18.80 \\
\hline
\end{tabular}
\end{center}
\label{4:hirschi}
\end{table}
In order to give a quantitative impression of the influence of initial mass,
metallicity and rotation on the evolution of stars, 
we present in Tables
\ref{4:limongi} and \ref{4:hirschi} results for (a) non-rotating solar
metallicity stars 
\citep{2006ApJ...647..483L} and (b) rotating stars for varying metallicities \citep{2007A&A...461..571H}.
Table \ref{4:hirschi} corresponds to the
models shown in Figs. \ref{4:kip}, \ref{4:kip2}, and \ref{4:absw}. Given are 
the initial and final mass
(in order to give an impression of the mass loss), as well as the core size
after central H-burning (the He-core) and after central He-burning (the
CO-core), and in table \ref{4:hirschi} also the metallicity $Z$ and 
initial rotational surface velocity in km~s$^{-1}$. As all burning stages after 
He-burning occur on significantly
shorter timescales than the earlier burning phases, the CO-core size is
the important quantity in order to determine the final outcome/fate of the
star.

After this general discussion of stellar evolution, as 
it varies with initial mass, metallicity and rotation, we now focus on two 
long-lived isotopes $^{26}$Al and $^{60}$Fe, which have important
contributions from the earlier burning stages in explosion ejecta. 

\subsubsection*{$^{26}$Al}

Long-lived $^{26}$Al is produced in core and shell H-burning via the
NaMgAl-cycle (see Ch.~3) in the $^{25}$Mg($p,\gamma)^{26}$Al
reaction and will be eventually ejected in the stellar wind during the WR-phase.
Gamma-ray observations of the 1.809 MeV decay line of $^{26}${Al} in systems
like the Wolf-Rayet binary system $\gamma$2~Vel, being the closest known
Wolf-Rayet (WR) star, serve as a constraint to nucleosynthesis in Wolf-Rayet
stars.  From  observations of the $\gamma$2~Vel binary system including a WR star, 
\citet{2000A&A...353..715O} 
claimed that \index{stars!$\gamma^2$ Velorum} \index{isotopes!26Al}
such WR stars must emit of the order $6\times 10^{-5}$M$_\odot$ of \Al by stellar
winds. The amount of $^{26}${Al} ejected into the interstellar medium is very
sensitive to metallicity, initial stellar mass, rotation and mass loss rate, 
related to one or more of the physical effects discussed above.
Results of detailed calculations can be found in 
\citet{1995Ap&SS.224..275L,1997A&A...320..460M,2005A&A...429..613P,2006ApJ...647..483L,2009arXiv0908.4283T}. 
\citet{2006ApJ...647..483L}  
provide an extended overview for the contribution
from 11 to 120~M$_\odot$ stars. The dominant source
for the $^{26}$Al production during stellar evolution is the 
$^{25}$Mg($p,\gamma)^{26}$Al reaction. Therefore the resulting abundance
depends (i) on this reaction rate converting $^{25}$Mg into $^{26}$Al, (ii) on 
the amount of $^{25}$Mg available, i.e. the total amount of matter in \index{process!Ne-Na cycle}
the NeNaMgAl-cycle (either in terms of the abundance/metallicity or in terms 
of the H-core size), and finally (iii) on the amount of $^{26}$Al distruction.
In the part of the He-core (after H-burning) which undergoes He-burning,
neutrons are produced via $(\alpha,n)$-reactions which destroy $^{26}$Al
via $^{26}$Al($n,p)^{26}$Mg and $^{26}$Al($n,\alpha)^{23}$Na. A further 
question is related to the amount
of matter being ejected in winds (i.e. mass loss) during
stellar evolution before $^{26}$Al can decay inside the star via 
$\beta^+$-decay with a half-life of $7.17\times 10^5$ y. 

He-burning, with its neutrons released, is destructive for 
$^{26}$Al, but shell C-burning is again a source of $^{26}$Al, also
via $^{25}$Mg($p,\gamma)^{26}$Al,  which is effective
 due to protons released in 
$^{12}$C($^{12}$C,$p)^{23}$Na (see table \ref{4:carbon} in 
Sect.~\ref{sec:4-2}). Convection in the C-burning shell brings in fresh
$^{12}$C fuel and $^{25}$Mg which has been also produced in prior He-burning
in the $^{22}$Ne($\alpha,n)^{25}$Mg reaction. \index{process!Ne burning}
$^{26}$Al production may be effective also in Ne-burning, based on 
$^{25}$Mg left over from C-burning and protons released via 
$^{23}$Na($\alpha,p)^{26}$Mg (see table \ref{4:neon}). This $^{26}$Al only 
survives if rapidly convected outwards to lower temperature environments (\Al may
decay rapidly in hot regions due to thermal population of its short-lived isomeric
state; see Fig. 1.3 in Ch.~1).

A fraction of the $^{26}$Al produced during stellar evolution will again be destroyed,
 when a shock front is released in a supernova explosion and propagates through the 
 stellar envelope; in particular,  material from C and Ne-burning, being close to the Fe-core, will be affected.
 But there are also source processes for explosive $^{26}$Al
production. The total yields, hydrostatic-evolution yields
combined with the destruction and contribution from explosive burning, are given in Sect.~\ref{sec:4-5}.

\subsubsection*{$^{60}$Fe}

$^{60}$Fe is produced by neutron captures on $^{59}$Fe, \index{isotopes!60Fe}
and destroyed again via  $^{60}$Fe($n,\gamma)^{61}$Fe, i.e. during the $s$~process. \index{process!s process}
Generally, slow capture of neutrons released from the 
$^{22}$Ne$(\alpha,n)^{25}$Mg reaction in core He-burning leads to the so-called 
\emph{weak} $s$~process, producing nuclei up to nuclear mass numbers of around A=90.
$^{59}$Fe is beta-unstable, thus in order for neutron capture to compete with 
this reaction branching (equating the neutron capture and beta-decay rates) 
requires a typcial neutron density of about $3\times 10^{10}$cm$^{-3}$.
These are relatively high neutron densities for an $s$~process, which also
ensure that the destruction of $^{60}$Fe via neutron captures dominates
over its decay with its half-life of $2.6\times 10^6$y (Fig.~7.22 in Ch.~7, \citet{2009PhRvL.103g2502R}). Core He-burning will
not provide sufficiently high-temperatures for the 
$^{22}$Ne$(\alpha,n)^{25}$Mg reaction to produce such high neutron densities.
It requires the conditions in shell He-burning to do so. 
Apparently conditions are most favorable during shell He-burning at late evolutionary
times when central O-burning has already active and a C-burning shell is
existent as well 
\citep[see][]{1995ApJS..101..181W,2002ApJ...576..323R,2006ApJ...647..483L,2009arXiv0908.4283T}. 
$^{60}$Fe yields are very sensitive to
uncertainties in He-destruction reactions (such as the 3$\alpha$-rate and
$^{12}$C($\alpha,\gamma)^{16}$O) which compete with the neutron source reaction 
$^{22}$Ne$(\alpha,n)^{25}$Mg and neutron(-capture) \emph{poisons} which compete
with the production and destruction rates of $^{60}$Fe via neutron captures
\citep{2002ApJ...576..323R,2009arXiv0908.4283T,2010AIPC.1213..201G,2009PhRvL.102o1101U}. 
Such uncertainties amount to
factors of up to 5 from present rate uncertainties.
Another possible effect which has not really been looked into, yet,
is the amount of $^{22}$Ne available in He-burning.
An important effect in low metallicity stars is the production of primary
$^{14}$N (not enherited from CNO of previous stellar generations,
but produced inside the star due to mixing of He-burning products with
H). This causes the production of $^{22}$Ne in
He-burning and can at low metallicities (with small seed abundances of Fe)
permit sizable $s$~processing, affecting again the abundance of $^{60}$Fe.

\subsection{Late Burning Stages and the Onset of Core Collapse}

Stars more massive than about 8~M$_\odot$ will, after finishing 
core and shell H- and He-burning, lead to CO-cores which exceed
the maximum stable mass of white dwarfs (the Chandrasekhar mass). For later
burning stages, when the partial or full degeneracy of the electron gas
is important, this critical limit $M_{Ch}(\rho Y_e,T)$ decides upon further 
contraction and the central ignition of subsequent burning stages, i.e.
C-, Ne-, O- and Si-burning. Dependent on the Fermi energy
\index{stars!Chandrasekhar mass} \index{Fermi!energy} 
of the degenerate electron gas, electron capture on the C-burning products
$^{20}$Ne and $^{24}$Mg can initiate a collapse, leading directly via nuclear 
statistical
equilibrium to a central Fe-core. This evolution path occurs for stars in the 
range 8-10 M$_\odot$ \citep{1987ApJ...322..206N}.
More massise stars will proceed through all burning stages
until Si-burning will finally produce an Fe-core.
All burning stages after core H- and He-burning proceed on timescales
which are shorter by orders of magnitude. The reason is that the energy carried
away by freely escaping neutrinos dominates over radiation losses by
photons which undergo a cascade of scattering processes before their final
escape. Most of these neutrinos are created 
when central densities and temperatures permit neutrino production via new
particle reactions, different from beta-decay or electron capture on nuclei.
Following neutrino production reactions are relevant: \index{neutrino!production}
(i) $e^-+e^+$-pair annihilation (\emph{pair neutrinos}), (ii) electron-photon 
scattering with neutrino-antineutrino pair creation (photo neutrinos),
and (iii) neutrino-antineutrino pair creation from plasma oscillations (\emph{plasmon 
neutrinos}). Neutrinos dominate the energy loss in
stellar evolution from this point on, and lead to increasingly shorter 
burning timescales, although the photon radiation 
luminosity of the star remains roughly constant.
The timescales for the individual burning stages are given in table 
\ref{4:burntime} in section \ref{sec:4-2}; these values refer to a 20~M$_\odot$ star with solar 
metallicity and no mass
loss \citep{1993PhR...227...65W}. 
Effects of mass loss, rotation and 
metallicity can change these timescales somewhat (up to 20\%).
Due to the large difference in evolution timescales, the dominant mass loss
by stellar winds occurs during H- and He-burning, and the final outcome
of stellar evolution is determined by the CO-core size after He-burning.
\index{stars!evolution}
Therefore, given all dependencies of stellar evolution via initial 
metallicities and rotation, the initial main sequence mass of a star is
less indicative for the final outcome than the size of its CO-core.

In the late phases of O- and Si-burning
(discussed in Sect.~\ref{sec:4-2}), 
electrons are moderately to strongly degenerate, dependent on the
initial stellar mass, and will be characterized by increasing
Fermi energies. This will allow for electron captures on burning products, and will
make matter more neutron-rich, i.e decrease $Y_e$, the electron or proton
to nucleon (neutrons plus protons) ratio. In high density O-burning 
\index{process!O burning} \index{process!electron capture} \index{process!Si burning}
($\rho>2\times 10^7$ g cm$^{-3}$) two electron capture reactions become important
and lead to a decrease in $Y_e$, $^{33}$S($e^-,\nu)^{33}$P and
$^{35}$Cl(($e^-,\nu)^{35}$S. Such effects become more extensive
at even higher densities in Si-burning and a large range of nuclei
has been identified to be of major importance
$^{55-68}$Co, $^{56-69}$Ni, $^{53-62}$Fe, $^{53-63}$Mn, $^{64-74}$Cu, 
$^{49-54}$Sc, $^{50-58}$V, $^{52-59}$Cr, $^{49-54}$Ti, $^{74-80}$Ga, 
$^{77-80}$Ge, $^{83}$Se, $^{80-83}$As, $^{50-58}$V, and $^{75}$Zn
\citep{1994ApJS...91..389A}. 
The amount of electron capture
and the resulting $Y_e$ has consequences for core sizes.
(The core sizes of the late burning stages are shown in Figs. {\ref{4:kip}} and {\ref{4:kip2}}).
The final size of the inner Fe-core represents the maximum mass
which can be supported by the pressure of the degenerate electron gas.
It is a function of $Y_e$, but also reflects temperature effects
if the electron gas is not completely degenerate 
\citep{1990RvMP...62..801B},
with $S_e$ being the entropy in electrons per baryon

\begin{equation}
M_{Ch}(Y_e,S_e)=1.44(2Y_e)^2 [1+({S_e\over \pi Y_e})^2] {\rm M}_\odot
. 
\label{4:Mchan}
\end{equation}

Stars with masses exceeding roughly 10 M$_\odot$ reach a point in their 
evolution where their Si-burning core (which will turn eventually into their Fe-core) exceeds this critical mass. 
At this point they collapse and bounce, if not too massive, to 
explode in spectacular core collapse events known as type II or Ib/c supernovae. \index{supernova!core collapse}
These explosions create a neutron star or black hole at the end 
of the life of a star. They play a preeminent role in the nucleosynthesis and 
chemical evolution of a galaxy.  

The collapse is initiated by the capture of degenerate electrons on nuclei,
which reduces the dominant contribution of the pressure (i.e. the one from the degenerate 
electron gas). Alternatively, for lower densities and higher temperatures
(in more massive stars), the pressure supporting the core is reduced by endoergic photodisintegrations of nuclei, reducing the 
thermal energy. 
The evolution in the core is determined by 
the competition of gravity (that causes the collapse of the core) and
weak interaction (that determines the rate at which electrons are captured and
the rate at which neutrinos are trapped during the collapse).

The early phases of this final stage of stellar evolution are known as \emph{presupernova evolution}. They follow the late-stage 
stellar evolution, and proceed until core densities of about $10^{10}$~g~cm$^{-3}$ and 
temperatures between 5 and 10$\times 10^9$K are reached.  
Until this point, modeling stellar evolution 
requires the consideration of extensive nuclear reaction networks, but 
is simplified by the fact that neutrinos need only be treated as a sink of 
energy and lepton number (due to their immediate escape). 
At later time and towards the collapse, this is no longer 
valid: As the weak interaction rates increase with the increasing density, 
the neutrino mean free paths shorten, so that the neutrinos eventually 
proceed from phases of free streaming, towards diffusion, and trapping. An
adequate handling of the transitions between these transport regimes 
necessitates a detailed time- and space-dependent bookkeeping of the neutrino 
distributions in the core (see Ch.~8). During collapse, electron
capture, accompanied by $\nu_e$ neutrino emission, dominates over 
electron antineutrino emission because the positron abundance 
is very low under electron-degenerate conditions. Later in the evolution the 
electron degeneracy is partially lifted, and in addition to the
electron flavor neutrinos, also heavy neutrinos, $\nu_\mu$ and $\nu_\tau$ and 
their antiparticles, are usually included
in numerical simulations of core collapse and postbounce evolution.

Advantageously, the temperature during the collapse and explosion are high 
enough that the matter composition is given by nuclear statistical equilibrium 
(NSE), i.e. without the need of reaction networks for the strong and electromagnetic 
interactions. The transition from a rather complex
global nuclear reaction network, involving many neutron, proton and $\alpha$ 
fusion reactions and their inverses, to a quasi-statistical equilibrium, in 
which reactions are fast enough to bring constrained regions
of the nuclear chart into equilibrium, to final and global nuclear statistical 
equilibrium is extensively discussed by 
\citet{1996ApJ...460..869H,1999ApJ...511..862H,2007ApJ...667..476H}. 
In the late stages of Si-burning and the early collapse phase,
weak interactions are dominated by
electron captures on protons and nuclei. These are important  equally in 
controling the neutronization of matter $Y_e$ and, in a large portion, 
also the 
stellar energy loss. Due to their strong energy dependence
$\propto E_e^5$, the electron capture rates increase rapidly during the 
collapse while the density  and the temperature increase 
(the electron Fermi energy $E_F$ scales with 
$\rho^{2/3}$, see \ref{sec:4-2}). 

The main weak interaction processes during the final evolution of a massive 
star are electron capture and $\beta$-decays. Their determination requires the 
calculation of Fermi and Gamow-Teller (GT) transitions. 
\index{process!Fermi transition} \index{process!Gamow Teller transition}
While the treatment of 
Fermi transitions (important only for $\beta$-decays) is
straightforward, a correct description of the GT transitions is a difficult 
problem in nuclear structure physics. In astrophysical environments, nuclei are 
fully ionized. Therefore, electron capture occurs from the continuum of the degenerate 
electron plasma, and energies of the electrons are high enough to induce 
transitions to the Gamow-Teller resonance. 
Shortly after the discovery of 
this collective excitation,
\citet{1979NuPhA.324..487B} 
recognized its importance for 
stellar electron capture. 
$\beta^-$-decay converts a neutron inside the nucleus into a proton and
emits an electron. In a degenerate electron gas, with fully 
populated levels up to the Fermi energy $E_F$, all decays which would 
produce electrons with smaller energies than $E_F$ are not possible (\emph{blocked}). 
Then, the decay rate of a given nuclear state 
is greatly reduced or even completely blocked at high densities. 
However, there is another pathway, as high temperatures populate a distribution of nuclear states: 
If an excited and thermally populated state of the decaying
nucleus is connected by 
large GT transition probabilities to low-lying states in the daughter nucleus, producing
electrons above the Fermi energy, such transition path can contribute significantly
to the stellar $\beta$-decay rates. The importance of these states in 
the parent nucleus for $\beta$-decay in astrophysical environments was first recognized by
\citet{1980ApJS...42..447F,1982ApJ...252..715F,1985ApJ...293....1F}.

Recent experimental data on GT distributions in iron group nuclei, 
measured in charge exchange reactions, show that the GT strength is 
strongly \emph{quenched} (reduced), compared to the \emph{independent-particle-model} 
value, and fragmented over many states in the daughter nucleus. An accurate 
understanding of these effects is essential for a reliable evaluation of the
stellar weak-interaction rates, particularly for the stellar 
electron-capture rates 
\citep{1980ApJS...42..447F,2000NuPhA.673..481L}. 
The nuclear \emph{shell-model} is the
only known tool to reliably describe GT distributions in nuclei. 
When comparing the shell-model based rates (by Langanke and Martinez-Pinedo) 
with the those from Fuller et al., 
one finds that the shell-model based rates are almost always smaller  
at the relevant temperatures and densities, caused by the above mentioned
quenching of the Gamow-Teller strength,  and by a systematic misplacement of the
energy of the Gamow-Teller resonance. \index{Gamow Teller resonance}

\begin{figure}[htbp]  
  \includegraphics[width=0.3\linewidth]{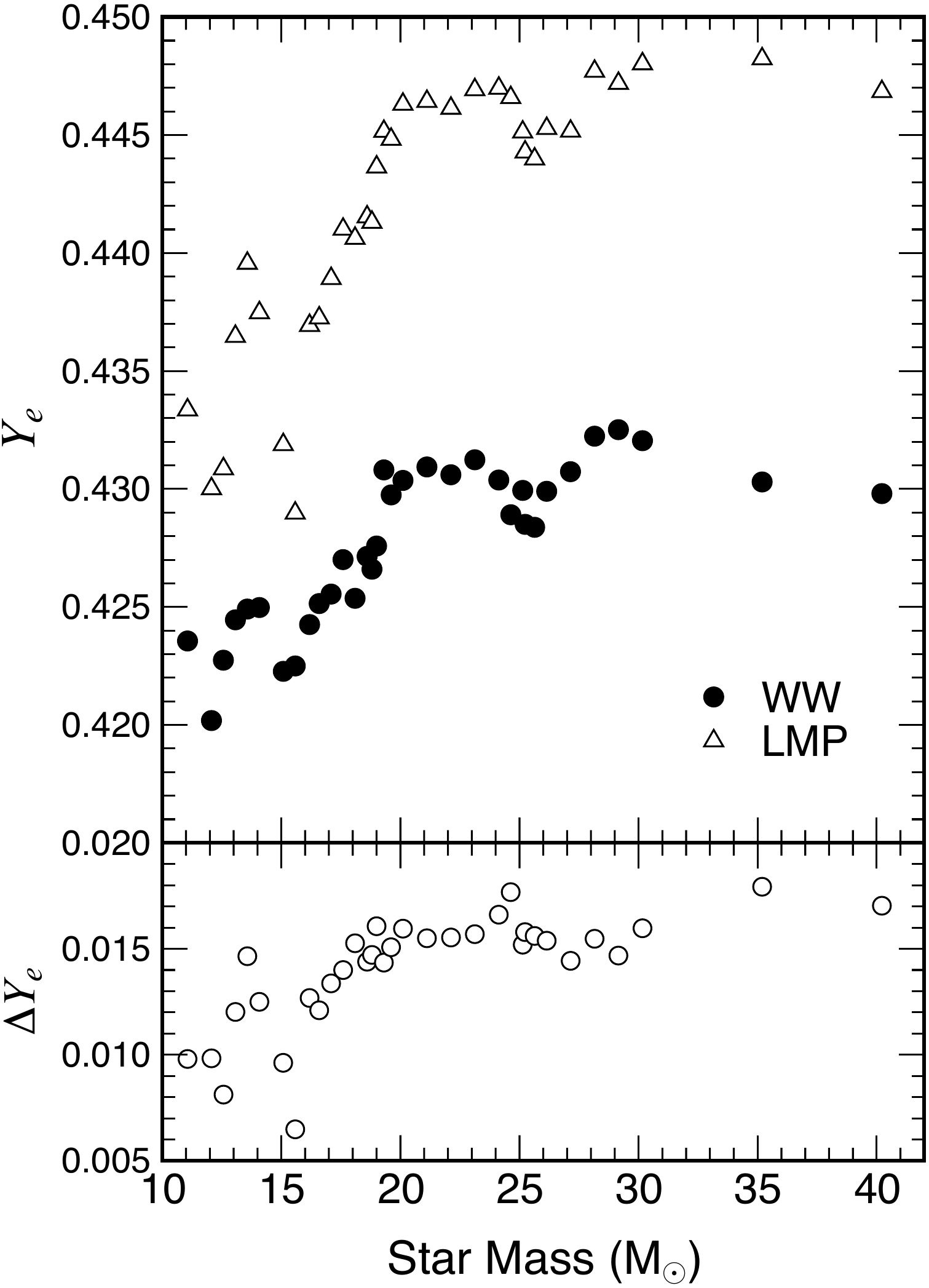}%
  \hspace{0.025\linewidth}%
  \includegraphics[width=0.3\linewidth]{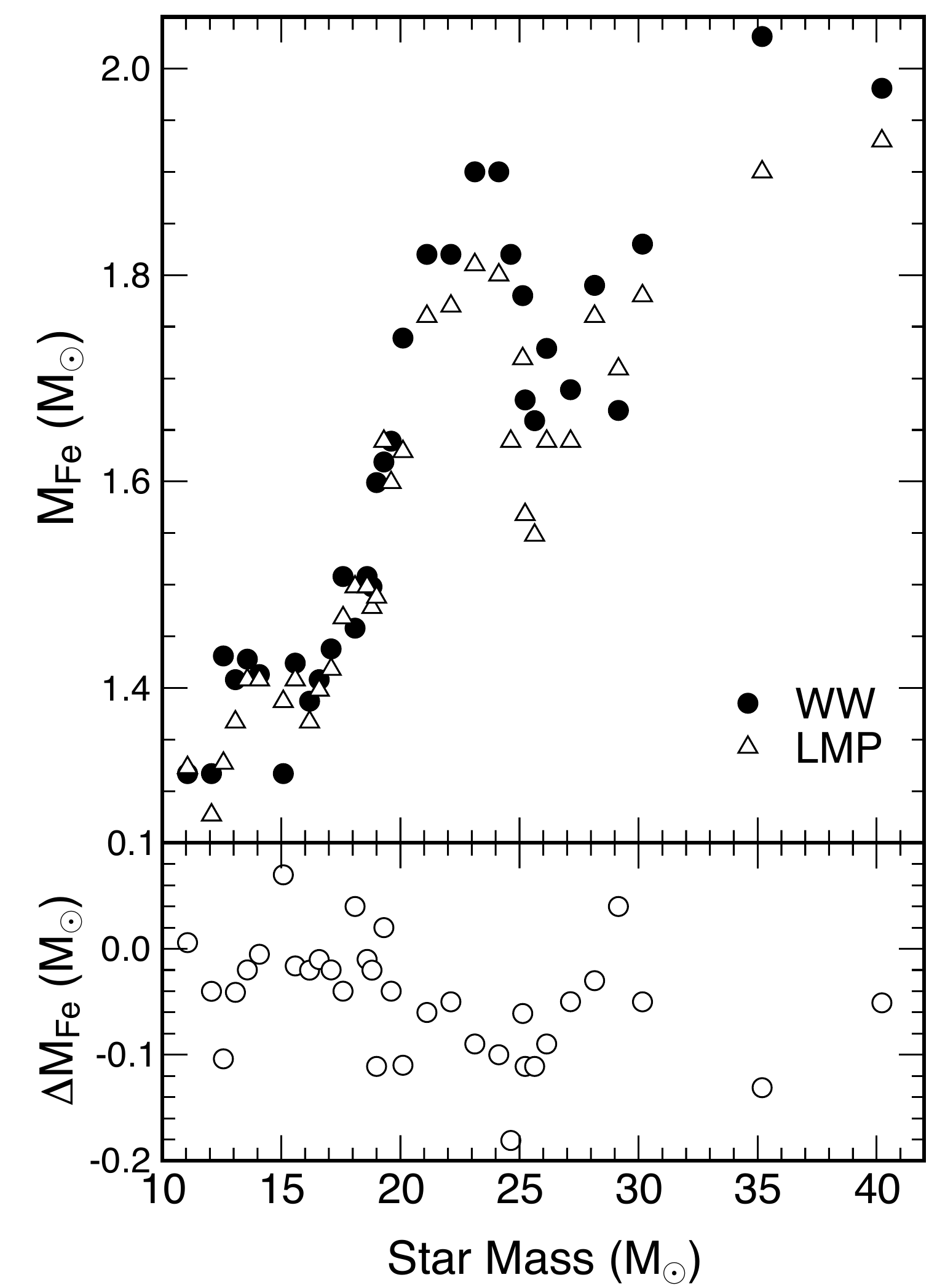}%
  \hspace{0.025\linewidth}%
  \includegraphics[width=0.3\linewidth]{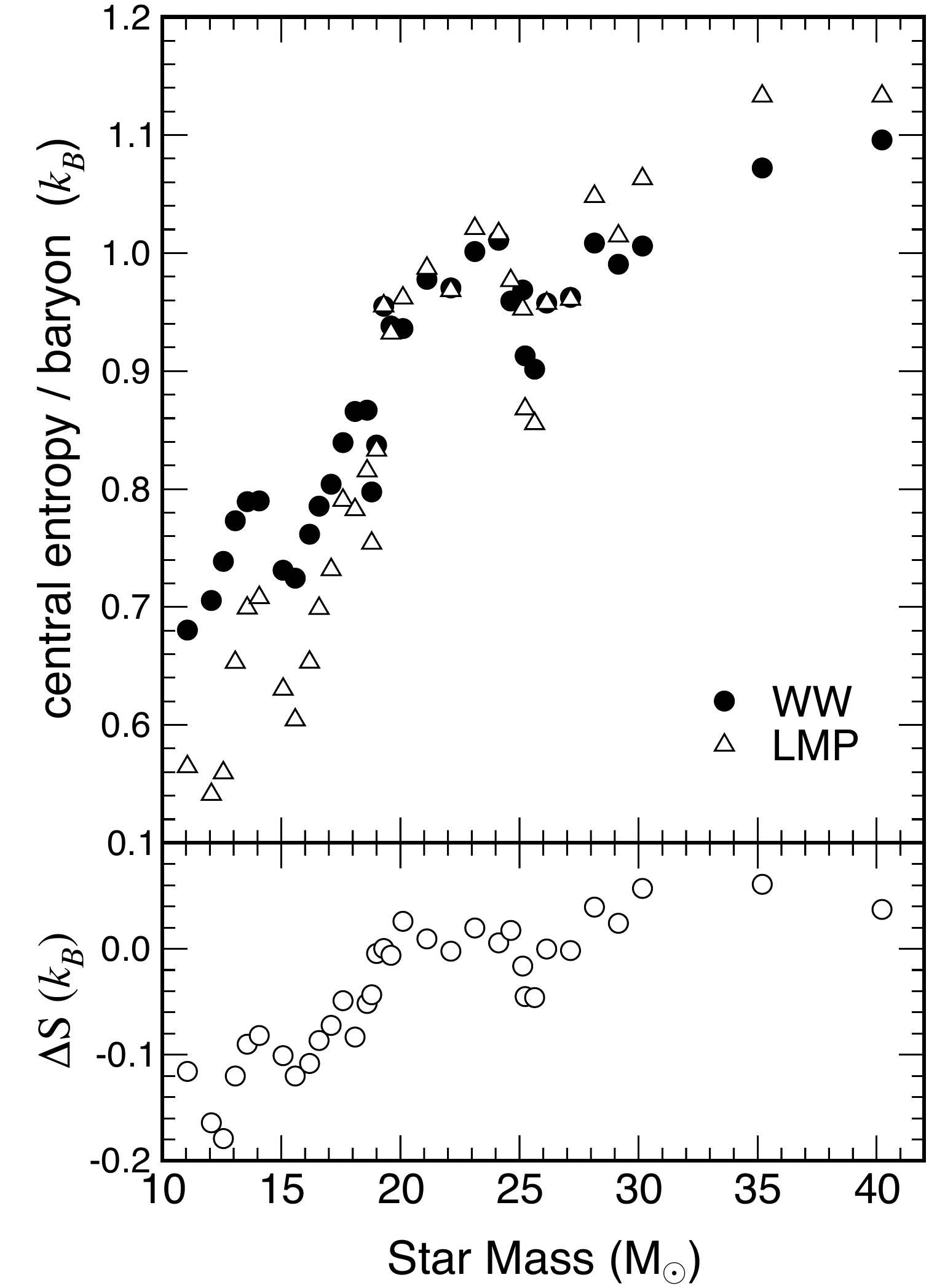}
  \caption{Comparison of the center values of $Y_e$ (left), the iron
    core sizes (middle) and the central entropy (right) for
    11--40 M$_\odot$ stars between the WW models and the ones
    using the shell model weak interaction rates (LMP)
    \citep{2001PhRvL..86.1678H}.  
    The lower parts define the changes
    in the 3 quantities between the LMP and WW models.
    \label{4:presn}}
\end{figure} 

The influence of these shell-model rates on the late-stage evolution of massive 
stars has been investigated by 
\citet{2001PhRvL..86.1678H,2001ApJ...560..307H} ,
and compared to earlier calculations 
\citep{1995ApJS..101..181W}. 
Fig. \ref{4:presn} illustrates the consequences of the shell model weak 
interaction rates for presupernova models in terms of the three decisive 
quantities: the central electron or proton to nucleon ratio $Y_e$, the entropy, and the 
iron core mass. The central values of $Y_e$ at the onset of core collapse 
increased by 0.01-0.015 for the new rates. This is a significant effect. For 
example, a change from $Y_e$ = 0.43 in the Woosley \& Weaver model for a 
20 M$_\odot$ star to 
$Y_e$ = 0.445 in the new models increases the respective Chandrasekhar
mass by about 0.075 M$_\odot$ (see Equ.\ref{4:Mchan}). The new models also 
result in lower core entropies for stars with $M< 20$ M$_\odot$, while for 
$M> 20$ M$_\odot$, the new models actually have a slightly larger
entropy. The Fe-core masses are generally smaller, where 
the effect is larger for more massive stars ($M> 20$ M$_\odot$), while for the 
most common supernovae ($M< 20$ M$_\odot$) the
reduction is by about 0.05 M$_\odot$ (the Fe-core is here defined as the mass 
interior to the point where the composition is dominated by more than 50\% 
of Fe-group elements with $A\ge 48$). This reduction seems opposite to the
expected effect due to
slower electron capture rates in the new models. It is, however, related 
to changes in the entropy profile during shell Si-burning which reduces 
the growth of the iron core just prior to collapse.

\begin{figure}[htbp] 
  \includegraphics[width=\linewidth]{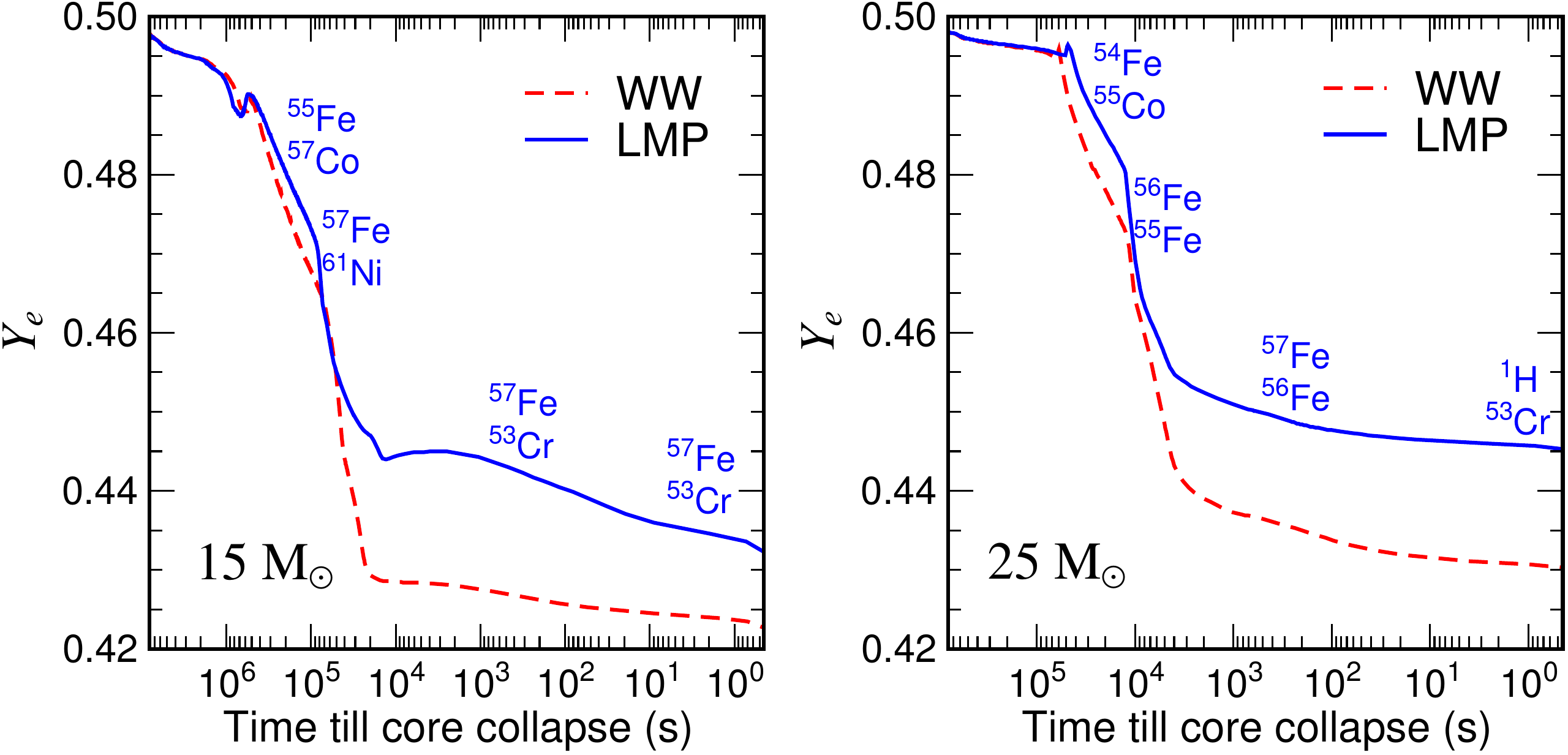}
  \caption{Evolution of the $Y_e$ value in the center of a
    15~$M_\odot$ star (left panel) and a 25~$M_\odot$ star (right
    panel) as a function of time until bounce.  The dashed line shows
    the evolution in the Woosley and Weaver models
    (WW)~\citep{1995ApJS..101..181W}, while the solid line shows the
    results using the shell-model based weak-interaction rates of
    Langanke and Mart{\'\i}nez-Pinedo (LMP). The two most important
    nuclei in the determination of the total electron-capture rate, for the
    calculations adopting the shell model rates, are displayed as a function
    of stellar evolution time.\label{4:Ye}}
\end{figure} 

The evolution of $Y_e$ during the presupernova phase is plotted
in Fig. {\ref{4:Ye}}. Weak processes become particularly important in 
reducing $Y_e$
below 0.5 after oxygen depletion ($\approx 10^7$ s and $10^6$ s before core collapse for 
the 15 M and 25 M stars, respectively)
and $Y_e$ begins a decline, which becomes precipitous during Si-burning. 
Initially electron captures occur much more rapidly than beta-decays. As the 
shell model rates are generally smaller, the initial reduction of $Y_e$ is 
smaller in the new models. The temperature in these models is correspondingly 
larger as less energy is radiated away by neutrino emission.
An important feature of the new models is shown in the left panel of 
Fig. {\ref{4:Ye}}.
For times between $10^4$ and $10^3$ s before core collapse, $Y_e$ increases due to 
the fact that $\beta$-decay becomes competitive with electron capture 
after Si-depletion in the core and during shell Si-burning. 
The presence of an important $\beta$-decay contribution has two effects
\citep{1994ApJS...91..389A}. 
Obviously it counteracts the reduction of $Y_e$
in the core, but also acts as an additional neutrino source,
causing a stronger cooling of the core and a reduction in entropy. This 
cooling can be quite efficient, as often the average neutrino energy from
the $\beta$-decays involved is larger than for the competing electron captures. 
As a consequence the new models have significantly lower core temperatures.
At later stages of the collapse 
$\beta$-decay becomes unimportant again as an increased electron Fermi energy
blocks/reduces its role. The shell model 
weak interaction rates predict the presupernova evolution to
proceed along a temperature-density-$Y_e$ trajectory where the weak processes 
involve nuclei rather close to stability which will permit to test these 
effects in the next-generation 
radioactive ion-beam facilities.

Fig. {\ref{4:Ye}} identifies 
the two most important nuclei (the ones with the largest value for the product 
of abundance times rate) for the electron capture during various stages of the 
final evolution of 15 M$_\odot$ and 25 M$_\odot$ stars. An
exhaustive list of the most important nuclei for both electron capture and 
beta-decay during the final stages of stellar evolution for stars of different 
masses is given in \citet{2001ApJ...560..307H}.
In total, the weak interaction processes shift the matter composition to 
smaller $Y_e$ values 
and hence more neutron-rich nuclei, subsequently affecting the 
nucleosynthesis. Its importance for the elemental abundance distribution, 
however, strongly depends on the location of
the mass cut in the supernova explosion. It is currently assumed that the 
remnant will have a larger baryonic mass than the Fe-core, but smaller than 
the mass enclosed by the O-shell 
\citep{2002RvMP...74.1015W}. 
As the reduction of $Y_e$ occurs mainly during Si-burning, it is essential 
to determine how much of this material will be ejected.

\section{Supernovae from Massive Stars and the Role of Radioactivity}
\label{sec:4-4}

\subsection{The Explosion Mechanism}

Supernova explosions are an application of numerical astrophysical
modelling that has a long tradition. Continued
improvements of the models are motivated by the following points:
(i) open questions regarding the explosion mechanism;
(ii) availability of observations for inidividual supernova explosions;
(iii) interesting input physics that tests matter
under conditions that are not accessible on earth;
(iv) visibility in light and other photon wavelengths,
cosmic rays, neutrino emission, decay gamma-rays of radioactive products,
perhaps gravitational wave emission;
(v) visibility on cosmological distances with improving statistical
information on the events and (vi) their
impact on the interstellar matter (e.g. abundances of metal-poor stars)
and Galactic evolution.

\begin{figure}[!tbp]
\centerline{\includegraphics[width=10cm]{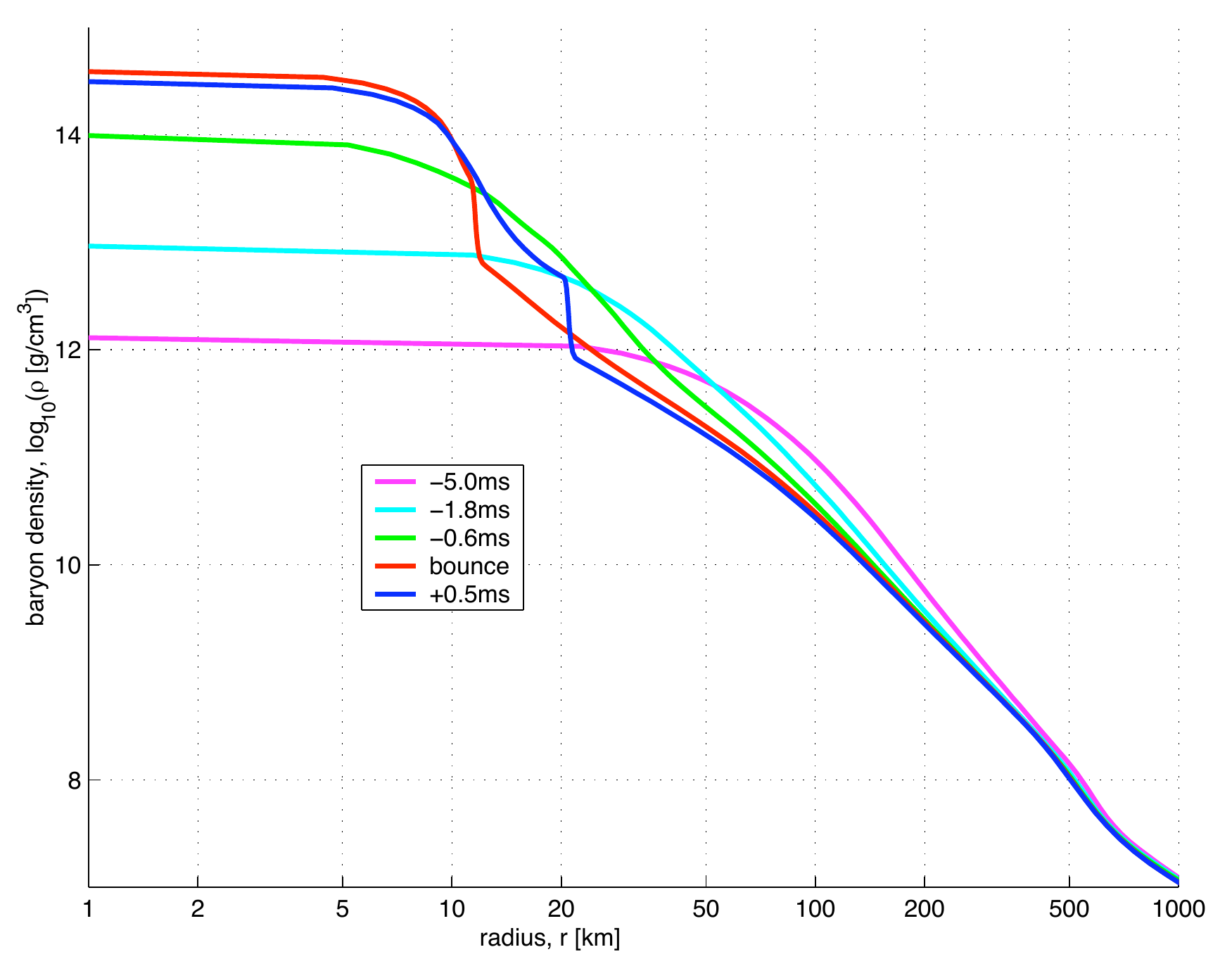}}
\caption{A sequence of density profiles of a 13~M$_\odot$ star before and
after core bounce. For such a relatively low mass supernova with a small
Fe-core the bounce occurs at a maximum density of less than twice nuclear
matter density. At the bounce one recognizes the size of the homologous core
(with roughly constant density). Thereafter the emergence of an outward
moving density (shock) wave can be witnessed.}
\label{4:shock}
\end{figure}

As discussed in the previous sections, 
the death of massive stars \( \approx 8-40 \) M\( _{\odot } \) proceeds
through several evolutionary and dynamical phases. At first, the 
modeling of a star must include the evolution through all nuclear burning stages 
until the resulting
inner iron core grows beyond the maximum mass which can be supported
by the dominant pressure of the degenerate electron gas. At this point,
the inner stellar core enters a dynamical phase of gravitational collapse,
during which it compactifies by \( \sim 5 \)
orders of magnitude. The nuclear saturation density \index{nuclear density} (i.e. the density
of stable nuclei $\approx 2\times 10^{14}$g cm$^{-3}$) is exceeded at the
center of the collapse and a protoneutron star (PNS) is formed. \index{stars!proto-neutron star} The dynamical
time scale reduces from a few hundreds of milliseconds at the onset
of collapse to a few milliseconds after the core has bounced back
at nuclear densities (see Fig. \ref{4:shock} from 
\citet{2003NuPhA.719..144L}.

The ensuing accretion phase onto the protoneutron star with fluid 
instabilities and radiative transfer phenomena, like the transport of neutrinos, 
is not well understood. It may last \( 0.5-10 \)
seconds and can therefore be interpreted as a second evolutionary
stage (much longer than the dynamical or transport time scale). Eventually
it will lead to the observed vigorous supernova explosion, a dynamic
phase where heavy elements are produced by explosive nucleosynthesis
in an outward propagating shock wave. The processed matter is mixed by fluid
instabilities and ejected into the interstellar medium, where it contributes
to Galactic evolution. The remaining PNS
at the center enters another evolutionary phase during which
it cools by neutrino emission and contracts or even collapses to
a black hole in a last dynamical phase.

While initially such calculations were performed in spherical symmetry and
therefore lacked the consistent treatment of turbulent motion, presently
performed research is done with multidimensional supernova models
\citep{2003PhRvL..91t1102H,2005ApJ...620..840L,2005A&A...443..201M,2006ApJ...640..878B,2007ApJ...667..382S,2008PhRvL.100a1101L,2009ApJ...694..664M}. 
The main ingredients are radiation (neutrino) transport, (relativistic)  \index{process!hydrodynamics}
hydrodynamics, and the nuclear equation of state at such high densities.
Recent progress has been made in the exploration of multidimensional
hydrodynamics with idealized input physics. A refreshing
view on the supernova mechanism has recently been suggested by
pointing out that in certain axisymmetric simulations vibrational
(so called PNS g-)modes \index{supernova!sound wave modes}
are excited so that sound waves are emitted into the heating region.
These sound waves are postulated to revive the stalled shock by
dissipation of sound energy 
\citep{2006ApJ...640..878B}.
Other efforts explore the role of magnetic fields
and rotation in two-dimensional simulations with simplified input
physics.
One kind of proof-of-principle models is carried out in spherically
symmetric approaches. The assumption of spherical symmetry is for
many supernovae not compatible with observational constraints.
However, one important advantage of spherically symmetric models
is that sophisticated treatments of the neutrino-matter interactions
can be included and that
the neutrino spectra and transport are correctly treated in general
relativistic
space-time.  Models of
this kind try to address the question of how many neutrinos are
emerging from the compactification of an inner stellar core, how
is their emission distributed as a function of time and how do
these neutrino fluxes generically affect the cooling, heating, or
nucleosynthesis in the outer layers of the star without the
complication of 3D dynamical fluid instabilities
\citep{2003NuPhA.719..144L,2004ApJS..150..263L,2009A&A...499....1F,2009arXiv0908.1871F}.
The attempt to combine all these aspects with forefront methods is ongoing
in order to achieve the final goal of understanding the multi-D explosion
mechanism with up to date microphysics from the equation of state to all
neutrino and nuclear interactions\footnote{For a review of the corresponding tools see Ch.~8.)}. 

The phase of stellar core collapse has intensively been studied in
spherically symmetric simulations with neutrino transport.
The crucial weak processes during the collapse and postbounce evolution are 
$ \nu + (A, Z) \leftrightarrow \nu + (A, Z)$,  
$ \nu + e^\pm \leftrightarrow  \nu +e^\pm$,    
$ p + e^- \leftrightarrow n + \nu_e$,
$(A, Z) + e^- \leftrightarrow  (A, Z-1) + \nu_e$ ,  
$\nu + N     \leftrightarrow \nu + N$,
$ n+ e^+ \leftrightarrow   p + \bar{\nu_e}$,
$ (A, Z) + e^+ \leftrightarrow   (A, Z + 1) + \bar{\nu_e}$,
$ \nu + (A, Z)  \leftrightarrow \nu + (A, Z)^*$,
$ (A, Z)^* \leftrightarrow   (A, Z) + \nu + \bar{\nu}$,
$   N +N \leftrightarrow     N + N + \nu + \bar{\nu}$,
$   \nu_e + \bar{\nu_e} \leftrightarrow \nu_{\mu,\tau} + \bar{\nu_{\mu,\tau}}$,
$  e^+ + e^- \leftrightarrow   \nu + \bar{\nu}$.
Here, a nucleus is symbolized by its mass number $A$ and charge $Z$, $N$ 
denotes 
either a neutron or a proton and $\nu$ represents any neutrino or antineutrino.
We note that, according to the generally accepted collapse picture 
(Bethe 1990; \citet{1979NuPhA.324..487B}), elastic
scattering of neutrinos on nuclei is mainly responsible for the trapping, as it
determines the diffusion time scale of the outwards streaming neutrinos. 
Shortly after trapping, the neutrinos are thermalized by energy downscattering,
experienced mainly in inelastic scattering off electrons. The relevant cross 
sections for these processes are discussed in 
\citet{2006NuPhA.777..395M}. 
The basic neutrino opacity in core \index{neutrino!scattering}
collapse is provided by neutrino scattering off nucleons. Depending
on the distribution of the nucleons in space and the wavelength of
the neutrinos, various important coherence effects can occur: Most
important during collapse is the binding of nucleons into nuclei
with a density contrast of several orders of magnitude to the
surrounding nucleon gas. Coherent scattering off nuclei dominates
the scattering opacity of neutrinos (and scales with $A^2$). Moreover,
these neutrino opacities should be corrected by an ion-ion correlation
function, this occurs if the neutrino wavelength is comparable to the
distances of scattering nuclei and quantum mechanical intererence effects
appear \citep{2005PhLB..630....1S,2006NuPhA.777..356B}.
Even if current core collapse models include
a full ensemble of nuclei in place of the traditional apprach with one
representative heavy nucleus, it remains non-trivial to adequately
determine correlation effects in the ion mixture.
Depending on the Q-value of an electron-capturing nucleus,
neutrinos are emitted with a high energy of the order of the electron chemical
potential/Fermi energy. As the neutrino opacities scale with the squared
neutrino energy, the initially trapped neutrinos will downscatter
to lower energies until the diffusion time scale becomes comparable
to the thermalization time scale. The thermalization in current
collapse models occurs through neutrino-electron scattering because
the energy transfer per collision with the light electron is more
efficient than with the heavier nucleons. The contribution of inelastic
scattering of neutrinos off heavy nuclei depends on the individual
nuclei and affects only the high-energy tail of the neutrino spectrum.

\citet{1980ApJ...238..991G} 
have shown that only the inner $M_{Ch}(Y_e)$
(see the definition in Eq. \ref{4:Mchan}) undergo a homologous collapse
($v_{collapse}(r)\propto r$), while at the edge of this core the velocity
becomes supersonic and a fraction of the free-fall velocity. The inner
core, falling at subsonic velocities where matter can communicate with
sound speed, cannot communicate with the free-falling envelope.
After the neutrinos are trapped, electron captures and neutrino captures are
in equilibrium ($e^- + p \leftrightarrow n + \nu_e$) and the total lepton
fraction $Y_L=Y_e+Y_\nu$ stays constant. $Y_e$ stops to decrease and
$M_{Ch}$ stops shrinking. Typical values (with the most recent
electron capture rates 
 \citep{2003PhRvL..90x1102L}
of $Y_L\approx0.3$ are found in 
numerical collapse calculations 
\citep{2003PhRvL..91t1102H,2005A&A...443..201M} 
which correspond to $M_{Ch}\approx 0.5$M$_\odot$.
As soon as nuclear densities are reached at the center of the
collapsing core, repulsive nuclear forces dominate the pressure in
the equation of state. The collapse comes to a halt and matter
bounces back to launch an outgoing pressure wave through the core.
It travels through the subsonic inner core and steepens to a shock
wave as soon as it faces supersonic infall velocities. Hence the
matter in the PNS remains at low entropy \( \sim 1.4 \) k$_B$ per
baryon while the supersonically accreting layers become shock-heated
and dissociated at entropies larger than \( \sim 6 \) k$_B$ per baryon.
Numerical simulations based on standard input physics and accurate
neutrino transport exclude the possibility that the kinetic energy
of the hydrodynamical bounce at nuclear densities drives a prompt
supernova explosion because of dissociation and neutrino losses.

This can be seen in Fig. \ref{4:shock} presenting spherically symmetric 
calculations of a 13~M$_\odot$ star. The inner core contains about 
0.6~M$_\odot$ of the initial Fe-core. The transition to free nucleons occurred
only in this inner, homologous core and the outward moving shock runs through
material consiting of Fe-group nuclei. The dissociation takes 8.7 MeV/nucleon
or $8\times 10^{18}$erg g$^{-1}$. Based on initial shock energies of
$(4-8)\times 10^{51}$erg, this is sufficient for passing through 0.25-0.5~M$_\odot$ 
and leads
in essentially all cases to a stalling of the prompt shock. Only recently
a possible exception was found \citep{2009PhRvL.102h1101S}. If a hadron-quark phase 
transition occurs in the collapsed core at the appropriate time, releasing
additional gravitational binding energy in the form of neutrinos from this
second collapse, prompt explosions can be attained.

While core collapse determines the state of \index{stars!proto-neutron star}
the \emph{cold} nuclear matter inside the PNS, the mass of the hot mantle
around the PNS grows by continued accretion. The infalling matter
is heated and dissociated by the impact at the
accretion front and continues to drift inward. At first, it can
still increase its entropy by the absorption of a small fraction
of outstreaming neutrinos (heating region). Further in, where the
matter settles on the surface of the PNS, neutrino emission dominates
absorption and the electron fraction and entropy decrease significantly
(cooling region).
The tight non-local feedback between the accretion rate and the luminosity
is well captured in computer simulations
in spherical symmetry that accurately solve the Boltzmann neutrino
transport equation for the three neutrino flavors.
All progenitor stars
between main sequence masses of \( 13 \) and \( 40 \) M\( _{\odot
} \) showed no explosions in simulations of the postbounce evolution
phase \citep{2003NuPhA.719..144L}.  
This indicates that the neutrino flux 
from the PNS does not have
the fundamental strength to blow off the surrounding layers for a
vigorous explosion.

Improved electron capture rates on heavy nuclei
overcame the idealized blocking of Gamow-Teller transitions in the
traditionally applied single-particle model. In the single-particle
picture of nuclei the so-called pf-shell is filled for $Z=40$ or $N=40$ 
for protons or neutrons respectively. Neutron numbers beyond $N=40$ require
a filling of the gd-orbits. If during core collapse nuclei ($Y_e$) become
so neutron-rich that nuclei with $Z<40$ and $N>40$ dominate the NSE
composition, electron capture would require the conversion of an $fp$ proton
to a $gd$ neutron as all $pf$ neutron orbits are filled. This Pauli-blocked
transition would lead to the dominance of electron capture on free protons rather
than nuclei and under such conditions. The recent finding, that configuration
mixing and finite temperature effects result in unfilled $pf$ neutron orbits, 
removes this Pauli-blocking and results in the fact that under these
condition electron capture rates on nuclei dominate those on free protons
\citep{2003PhRvL..90x1102L}. 
Thus, there are two effects due to
the new set of electron capture rates: 1. at low densities for less
neutron-rich nuclei the total amount of electron capture is reduced with an 
improved description of Gamow-Teller transitions (see the discussion
of the early collapse phase in Sect.~\ref{sec:4-3}), 2. at high densities
in the late collapse phase the total amount of electron capture is enhanced,
leading to smaller $Y_e$ and $Y_L$ values than before. Such changes
caused a reduction of homologous core sizes down to $M_{Ch}=0.5$~M$_\odot$
(see discussion above and \citet{2003PhRvL..91t1102H}).
This faster deleptonization in the collapse phase in comparison to 
captures on free protons alone thus resulted in a \( 20\% \) smaller inner 
core at bounce.

\begin{figure}[!tbp] 
\centering
\includegraphics[width=12cm]{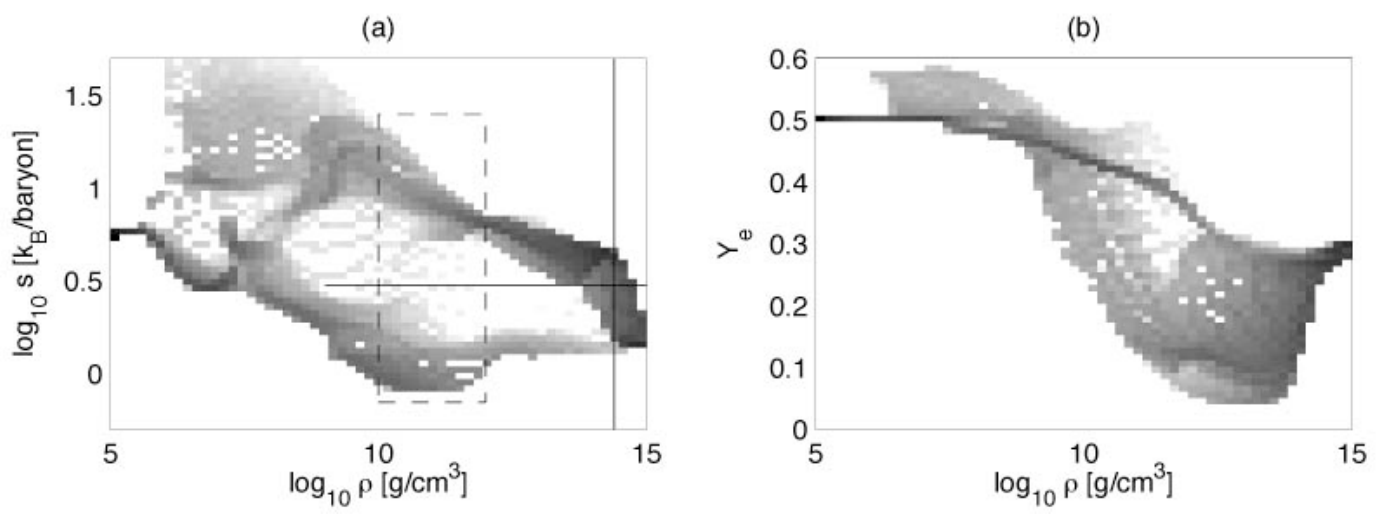}
\caption{Overview of the conditions attained in a simulation of the collapse, 
bounce, 
and explosion (artificially induced) of a 20~M$_\odot$ star. Shown are two 
histograms of the occurrence of
conditions as a function of density $\rho$, specific entropy $s$ and electron 
fraction $Y_e$ . The shading
of a given bin corresponds to $log_{10} (\int dm dt)$ in arbitrary units, where
the integral over mass is
performed over the mass $dm$ of matter whose thermodynamic state at a given 
time falls into the bin.
The integral over time extends over the duration of a simulation. Hence, 
regions of dark shading
correspond to states that are experienced by considerable mass for an extended 
time, while light or
absent shading corresponds to conditions that are rarely assumed in the 
supernova simulation. The
vertical black line indicates the nuclear density. The horizontal black line 
indicates an entropy of
3 k$_B$ /baryon beyond which ions are dissociated. It clearly separates the 
conditions
of cold infalling matter on the lower branch from the conditions of hot 
shocked matter on the upper
branch.}
\label{4:phasespace}
\end{figure}

Taking all this improved physics into acount leads in the entire simulations
(i.e. all mass zones invoved) to conditions in densities $\rho$, electron
abundance $Y_e$ and entropy $s$ per baryon, where properties like the
equation of state or other microscopic physics is needed in
current supernova simulations.
Fig. \ref{4:phasespace} provides this information for a simulation of a
20~M$_\odot$ star \citep{2009ApJ...698.1174L}.

Moreover, a comparison of the effects of the only two publicly available 
equations of state
by \citet{1991NuPhA.535..331L} and \citet{1998PThPh.100.1013S,1998NuPhA.637..435S}
is required.
\index{equation of state}
In simulations of massive progenitors that do not
explode and exceed the maximum stable mass of the accreting neutron star in
the postbounce phase, it was demonstrated that the neutrino signal changes
dramatically when the PNS collapses to a black hole
\citep{2009A&A...499....1F}.
Depending on the stiffness of the equation of state or the accretion
rate from the external layers of the progenitor star, this can
happen at very different time after bounce. Hence, the neutrino signal
carries a clear imprint of the stiffness of the equation of state
and the accretion rate to the observer of neutrinos.

The detailed treatment of the neutrino transport and interactions
is of great importance for the nucleosynthesis. \index{neutrino!transport}
This has been shown in several recent studies 
\citep{2006ApJ...637..415F,2006PhRvL..96n2502F,2005ApJ...623..325P,2006ApJ...644.1028P,2006ApJ...647.1323W}. 
This also opens an opportunity to investigate
neutrino flavor oscillations among electron, muon and tau neutrinos. On the one
hand side the long term explosion runs achieve (low) density structures
that allow for MSW (Mikheyev-Smirnov-Wolfenstein effect) neutrino flavor 
oscillations in the outer layers 
\citep{1978PhRvD..17.2369W,1985YaFiz..42.1441M}. 
These may give additional hints on the expansion velocity and density
distribution in case that the neutrinos can be observed from a
near-by supernova. On the other hand, collective flavor transitions
have recently been postulated in regions where the neutrino density
exceeds the electron density 
\citep{2006PhRvD..74l3004D,2007PhRvL..99x1802D,2007JCAP...12..010F}.
This condition will be achieved
in the evacuated zone that surrounds the PNS after the onset of an
explosion. The impact of these collective neutrino flavor oscillations
on the neutrino heating during the shock expansion, the neutrino
wind, and the nucleosynthesis are important open points that require
a detailed investigation under consideration of accurate neutrino
transport and spectra.

\begin{figure}[!tbp]  
\centering
\includegraphics[width=12cm]{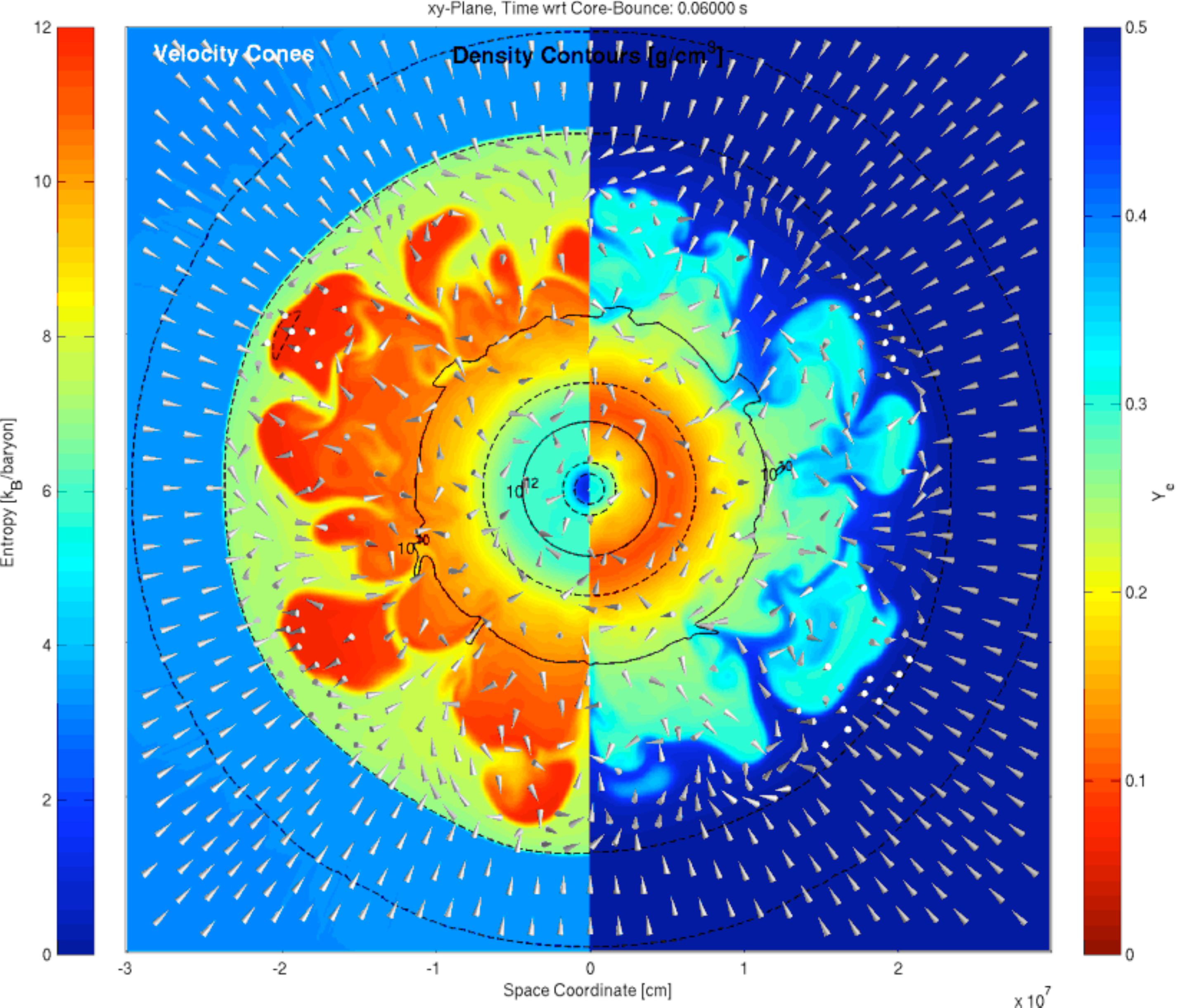}
\caption{Illustration of the early accretion phase in a three-dimensional 
simulation with a resolution of $600^3$ zones and the isototropic diffusion
source approximation for 3D neutrino transport 
\citep{2009ApJ...698.1174L}.
Shown are density contours as black lines for a 15~M$_\odot$ star from 
\citet{1995ApJS..101..181W}. 
\emph{Left:} The color
indicates the specific entropy and the cones the direction 
of the velocity. \emph{Right:} The color refers to the magnetic field strength and 
the cones to its direction. The
cool high-density interior of the PNS and the hot low-density accreted matter 
behind the standing accretion front are clearly distinguishable.}
\label{4:3D}
\end{figure} 

The difficulty to reproduce explosions in spherically symmetric
models of core-collapse and postbounce evolution stimulated the consideration 
of numerous modifications and alternatives to this basic scenario, mostly 
relying on multi-dimensional
effects that could not be treated in spherical symmetry.  
It was discussed whether convection inside the PNS could accelerate the 
deleptonization and increase the neutrino luminosity 
\citep{1993PhR...227...97W}. 
The convective overturn 
between the PNS and shock 
front was shown to increase the efficiency of neutrino energy deposition 
\citep{1994ApJ...435..339H}. 
Asymmetric instabilities of the standing accretion shock 
\citep{2003ApJ...584..971B,2009ApJ...694..820F} 
may help to 
push the shock to larger radii and g-mode oscillations of the PNS may 
contribute to neutrino heating by the dissipation of sound waves
between the PNS and the shock 
\citep{2006ApJ...640..878B}. 
Moreover, it has been suggested that magnetic fields
have an impact on the explosion mechanism 
\citep{2006RPPh...69..971K}. 
Most of the 
above-mentioned modifications of the explosion mechanism are essentially of a 
three-dimensional nature. In
order to illustrate the complexity of the crucial accretion phase we show in 
Fig. \ref{4:3D} a slice
through a three-dimensional simulation of core-collapse and postbounce 
evolution of a recent run 
\citep{2008JPhG...35a4056L}. 
Its input physics uses the Lattimer-Swesty equation of state
\citep{1991NuPhA.535..331L} 
and a parameterization of the neutrino physics for the collapse phase 
\citep{2005ApJ...620..840L}. 
The treatment of neutrino cooling and heating in the postbounce phase is  
based on multi-group diffusion (the isotropic diffion source approximation
of \citet{2009ApJ...698.1174L}).

Initially, spherically symmetric supernova models were the most realistic 
among all feasible computer representations of the event. With increasing 
observational evidence for the complexity of the explosions 
\citep{2003ApJ...582..905H} 
their primary purpose shifted 
from a realistic representation to the identification and understanding of 
the basic principles of the explosion
mechanism. After the emergence of axisymmetric simulations with 
sophisticated and computationally intensive spectral neutrino transport 
\citep{2003PhRvL..90x1101B,2005ApJ...626..317W} 
spherically symmetric models still have several assets. 
In the following subsection we will first describe purely phenomenological
calculations based on artificially induced explosions via a ``piston'' or
energy deposition in terms of a ``thermal bomb'', purely in order to
discuss nucleosynthesis effects. We will then also discuss still artificial
explosions in spherical symmetry, however resulting from a ``self-consistent''
treatment including neutrino transport, which permits to analyse the
effect of neutrinos on the nucleosynthesis of the innermost ejecta.

\subsection{Nucleosynthesis in Explosions}

\subsubsection*{Major Explosive Burning Processes}

Despite considerable improvements of stellar models and
numerical simulations in recent years, some fundamental problems
remain in nucleosynthesis predictions. It has become evident that
certain evolution aspects can only be followed in models going
beyond one-dimensional simulations, such as convection, rotation,
and the explosion mechanism. However, it is still not feasible
to directly couple full reaction networks, containing several
thousand nuclei, to multi-dimensional hydrodynamic calculations
due to the lack of required computing power, even in modern
computers. 
\index{nucleosynthesis!explosive}
Thus, postprocessing after explosion models with parameterized 
networks still
remains an important approach. One-dimensional models can directly
accommodate increasingly larger networks but they cannot capture
all of the necessary physics. As outlined in the previous subsection, it has 
become apparent that a
self-consistent treatment of core collapse supernovae in 1D
does not lead to successful explosions when using
presently known input physics while 2D models show some promise.
There are strong indications
that the delayed neutrino mechanism works combined with a multi-D convection
treatment for unstable layers (possibly with the aid of rotation, magnetic
fields and/or still existent uncertainties in neutrino opacities).
Therefore, hybrid approaches using certain parameterizations or
approximations have been and are still necessary when
predicting the nucleosynthetic yields required for the
application described above.

Supernova nucleosynthesis predictions have a long tradition.
All of these predictions relied on an artificially introduced
explosion, either via a piston or a thermal bomb
introduced into the progenitor star model.
The mass cut between the ejecta and the remnant
does not emerge from this kind of
simulations but has to be determined
from additional conditions.
While the usage of artificially introduced explosions is justifiable
for the outer stellar layers, provided we
know the correct explosion energy to be dumped into the shock front (on
the order of 10$^{51}$ erg seen in observations), it clearly is
incorrect for the innermost ejected layers which should be directly
related to the physical processes causing the explosion.
This affects the Fe-group composition, which has been 
recognized as a clear problem by many groups 
\citep{1995ApJS..101..181W,1990ApJ...349..222T,1996ApJ...460..408T,1999ApJ...517..193N,2001ApJ...555..880N,2006NuPhA.777..424N}.
The problem is also linked to the
so-called neutrino wind, emitted seconds after the supernova
explosion, and considered as a possible source of the $r$~process to
produce the heaviest elements via neutron captures 
\citep{1996ApJ...471..331Q}, 
as will be discussed below.

\begin{figure}[tbp] 
\centering
\includegraphics[width=12cm]{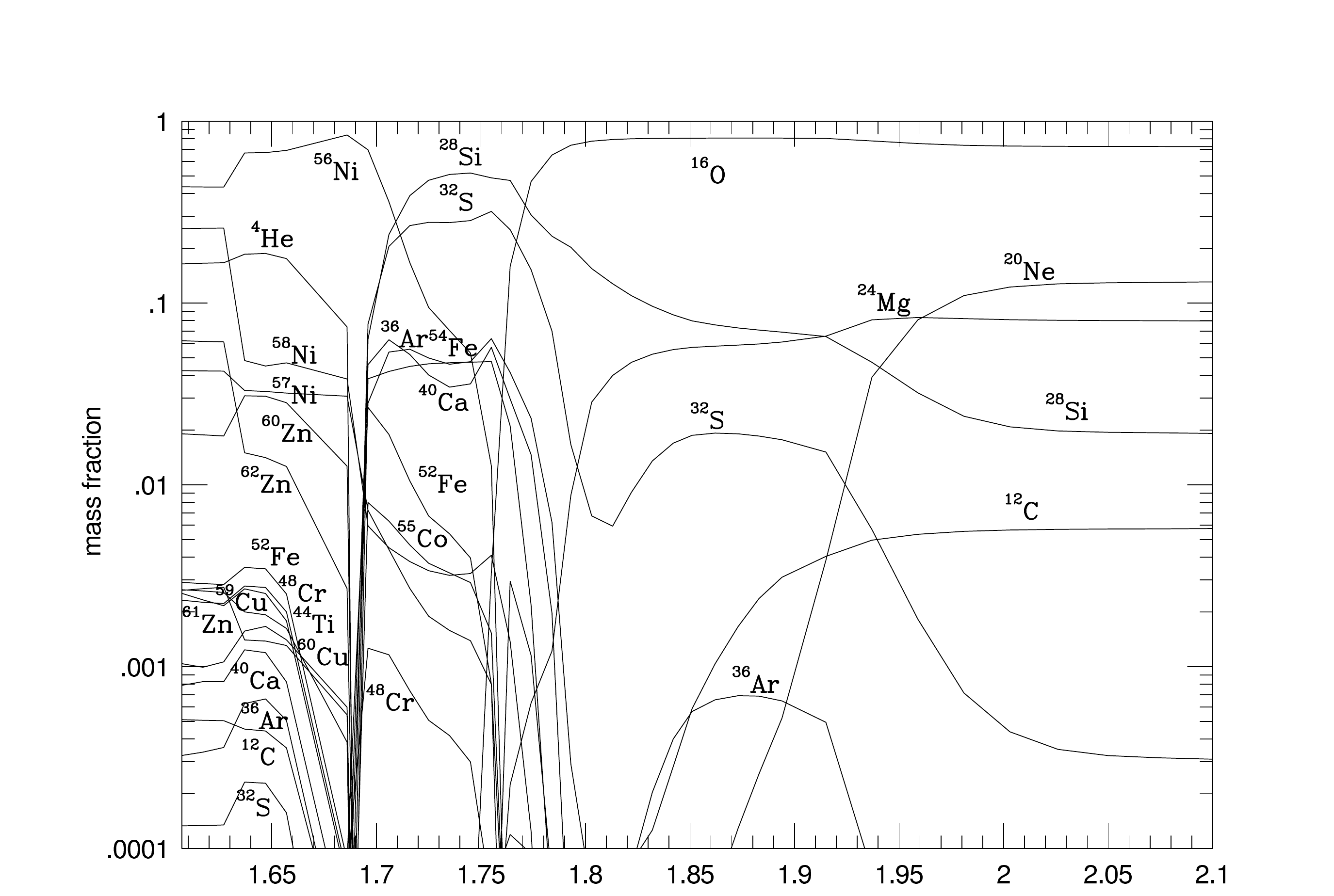}
\vspace{0.5cm}
\caption{Mass fractions of a few major nuclei after passage of the
supernova shockfront through a star with an intial mass of 20~M$_\odot$. 
Matter outside 2M$_\odot$ is essentially unaltered.
Mass zones further in experience explosive Si, O, Ne, and C-burning.
For ejecting 0.07M$_\odot$ of $^{56}$Ni the mass cut between
neutron star and ejecta is required to be located at 1.6M$_\odot$.}
\label{4:global20}
\end{figure} 

Given the above detailed discussion of the physics, problems and options
regarding core collapse supernovae, we will adopt the following approach in 
order to predict the most reliable nucleosynthesis predictions for the ejecta
in a 1D spherically symmetric treatment.
The multiplication of neutrino capture cross sections on nucleons with
a free parameter in 1D spherically symmetric calculations can mimic the
enhanced energy deposition which multi-D models show. The free parameter is
tuned to give correct explosion energies and $^{56}$Ni yields for a number
of well known supernovae. This approach provides clear predictions for the
mass cut between the remaining neutron star and the ejecta. It also includes
the effect neutrinos can have on the correct $Y_e$ in the ejecta and the
related nucleosynthesis. In the outer explosively burning layers, essentially
only the energy in the shock front matters. The behavior of these zones can
be easily understood from the maximum temperatures attained in the radiation
bubble and for a first discussion we will just focus on these features,
which can also be obtained with an artifically induced thermal bomb treatment.

For a given/known $Y_e$ and density $\rho$, the most significant parameter in 
explosive nucleosynthesis
is the temperature, and a good prediction for the composition can already be
made by only knowing $T_{max}$, without having to perform complex
nucleosynthesis calculations. 
\citet{1980NYASA.336..335W}
already recognized, that matter behind the shock front is strongly radiation
dominated. Assuming an almost homogeneous density and temperature distribution
behind the shock (which is approximately correct,
one can equate the supernove energy with
the radiation energy inside the radius $r$ of the shock front

\begin{equation}
E_{SN} = {{4\pi } \over 3} r^3 a T^4(r).
\label{4:totale}
\end{equation}

\begin{figure}[tbp] 
\centering
\includegraphics[width=8cm]{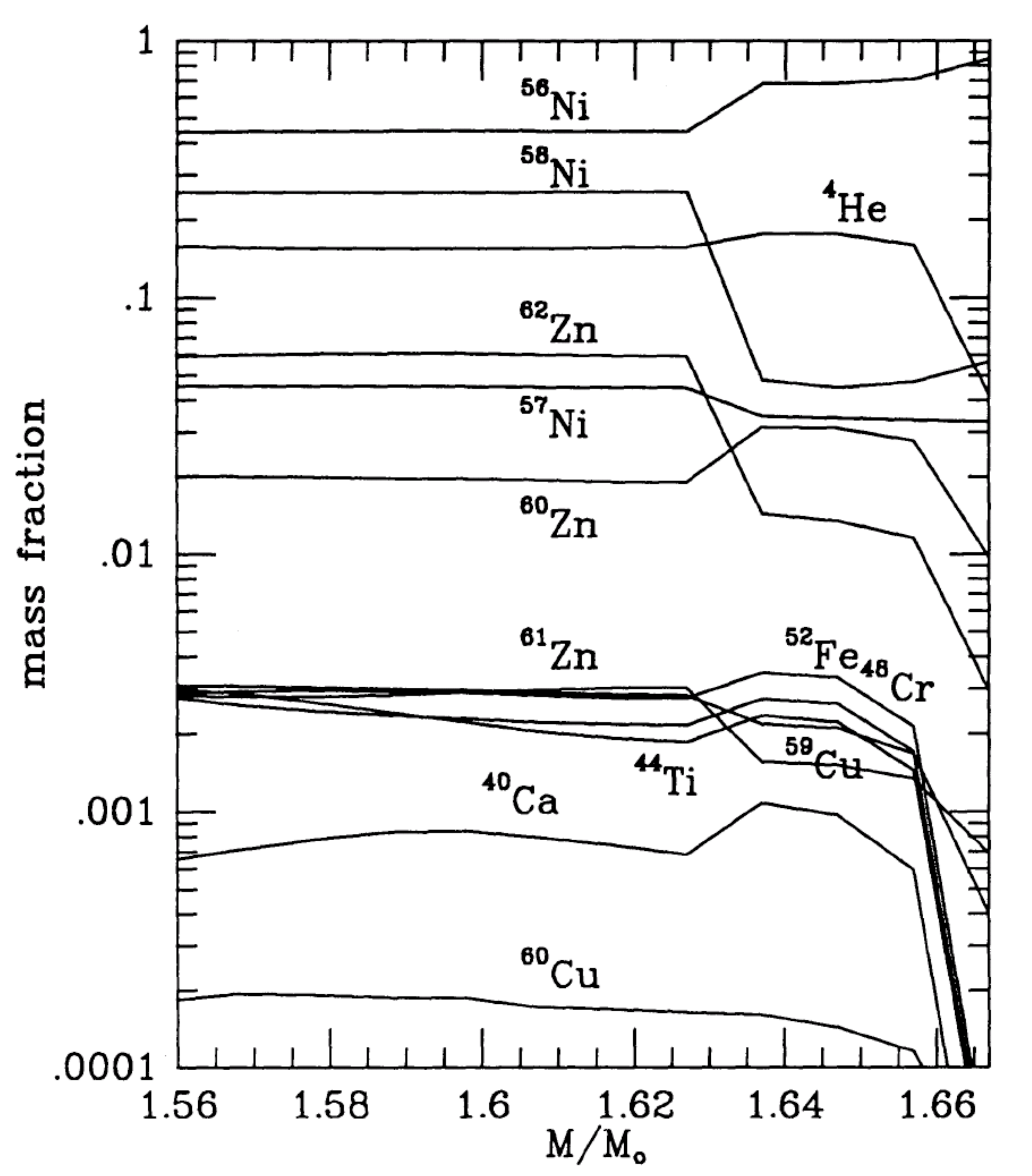}
\caption{Mass fractions of the dominant nuclei in zones which experience
$\alpha$-rich freeze-out. Notice the relatively large amounts of Zn and
Cu nuclei, which originate from $\alpha$-captures on Ni and Co. One can
recognize their strong decrease beyond 1.66M$_\odot$, which goes parallel
with the
decrease of the $^4$He-abundance and other $\alpha$-nuclei such as $^{40}$Ca,
$^{44}$Ti, $^{48}$Cr, and $^{52}$Fe. Nuclei which would dominate in a
nuclear statistical equilibrium like $^{56,57,58}$Ni stay constant or
increase even slightly. The increase of all nuclei with $N=Z$ at
1.63M$_\odot$
and the decrease of nuclei with N$>$Z is due to the change in Y$_e$ in the
original stellar model before collapse (see also Fig.\ref{4:global20})}
\label{4:alphar}
\end{figure} 

This equation can be solved for $r$. With $T=5\times 10^9$K, the
lower bound for explosive Si-burning with complete Si-exhaustion, and
an induced thermal bomb energy of $E_{SN}=10^{51}$erg, the result is 
$r\approx 3700$km.
For the evolutionary model by 
\citet{1988PhR...163...13N} 
of a
20M$_\odot$ star this radius corresponds to
1.7M$_{\odot}$, in excellent agreement with the exact hydrodynamic
calculation.
Temperatures which characterize the edge of the other explosive
burning zones correspond to the following radii: incomplete Si-burning
($T_9$=4, $r$=4980km), explosive O-burning (3.3, 6430), and explosive
Ne/C-burning (2.1, 11750). This relates to masses of 1.75, 1.81, and
2.05M$_\odot$ in case of the 20M$_\odot$ star.
The radii mentioned are model independent and vary only with the
supernova energy. In the following we present a number of plots which
show the different mass fractions $X_i=A_iY_i$ as a function of radial mass
$M(r)/$M$_\odot$, passing outwards through a 20M$_\odot$ star through all
explosive burning regions.

\begin{figure}[tbp] 
\centering
\includegraphics[width=10cm]{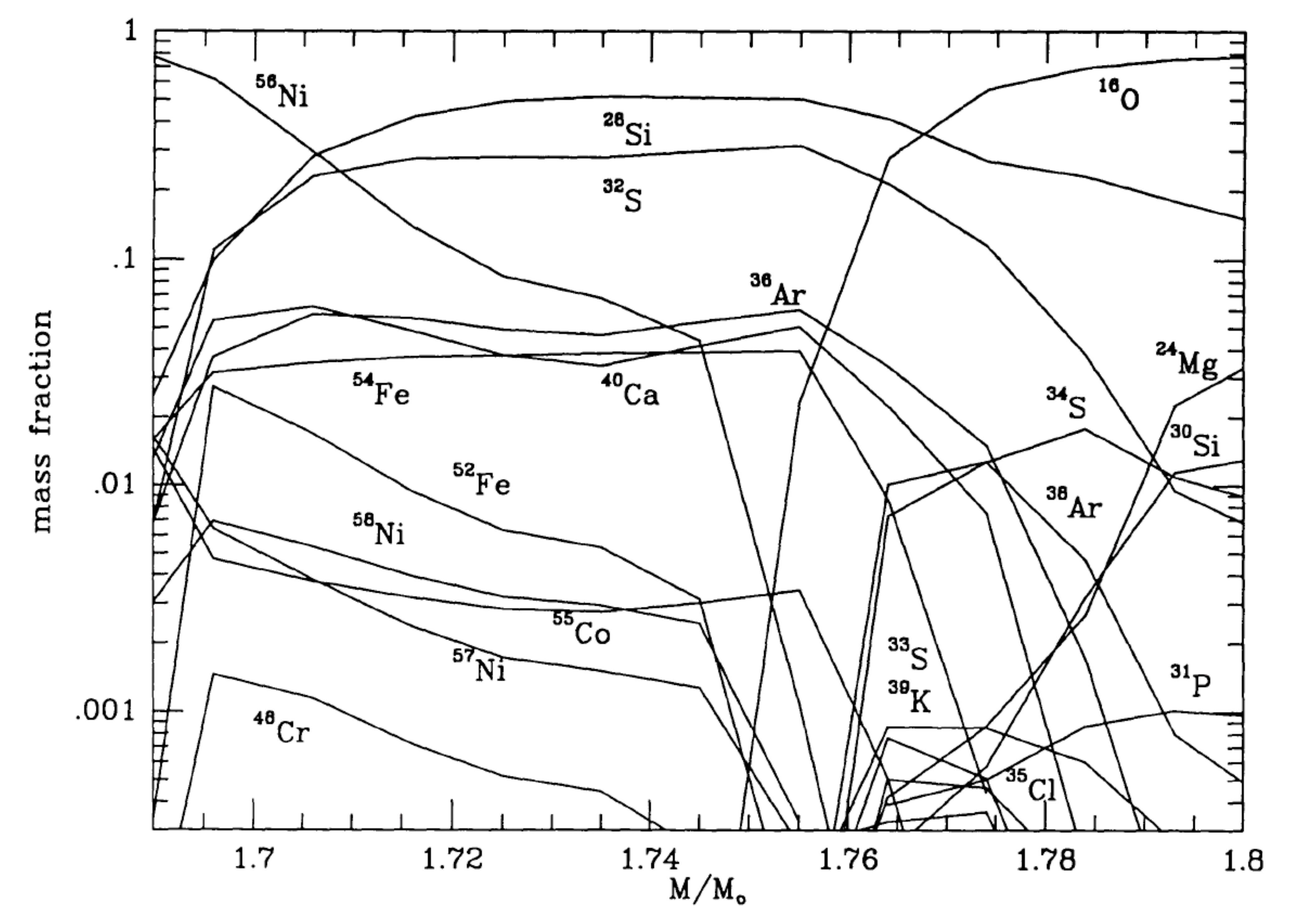}
\caption{Mass fractions of nuclei in the zones of incomplete Si-burning
M$<$1.74M$_\odot$ and explosive O-burning M$<$1.8M$_\odot$. The
Si-burning
zones are are characterized by important quantities of Fe-group
nuclei besides $^{28}$Si, $^{32}$S, $^{36}$Ar, and $^{40}$Ca. Explosive
O-burning
produces mostly the latter, together with more
neutron-rich nuclei like $^{30}$Si, $^{34}$S, $^{38}$Ar etc.}
\label{4:incompl}
\end{figure}  
\begin{figure}[tbp]
\centering
\includegraphics[width=10cm]{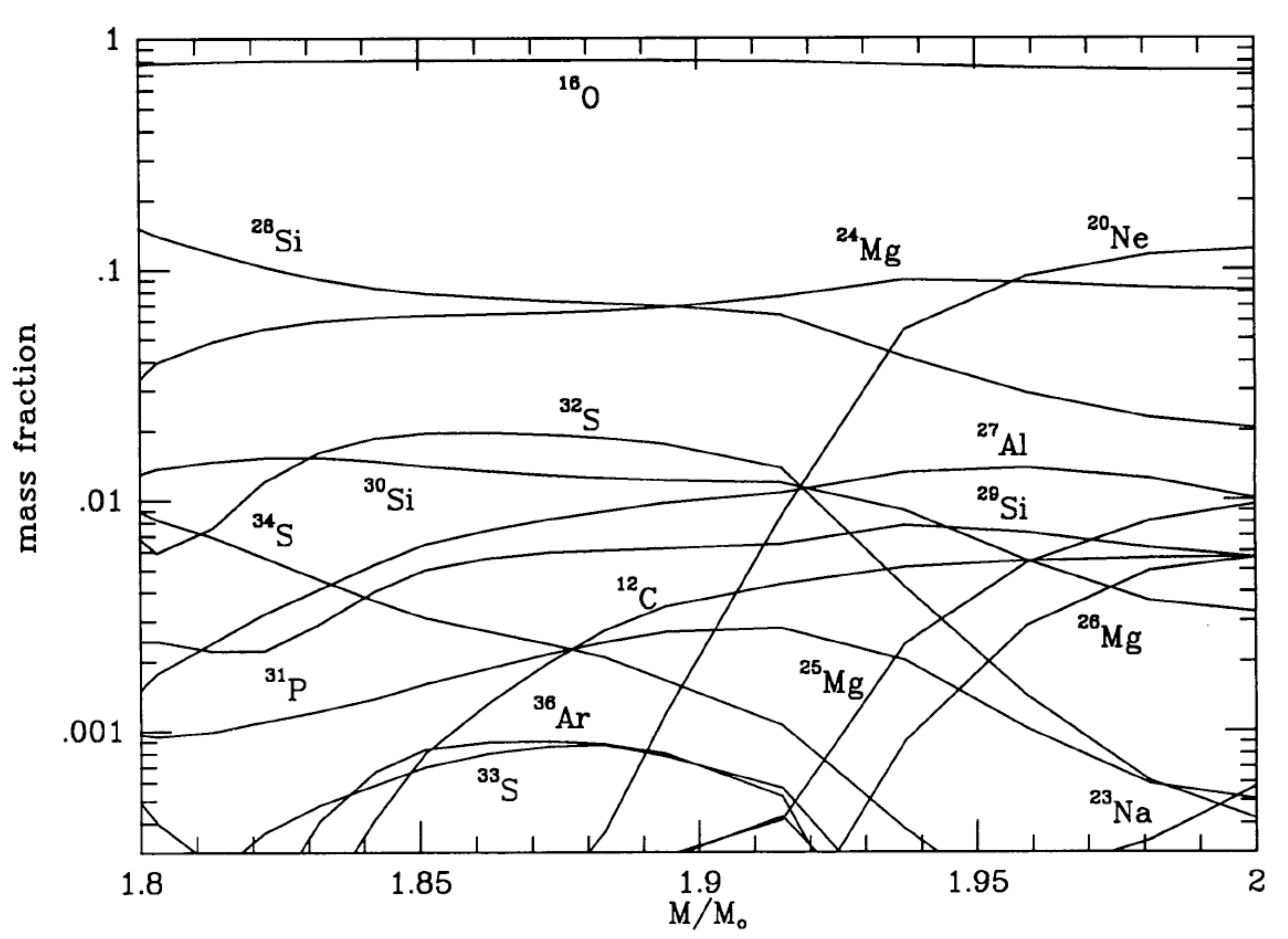}
\caption{Composition in mass zones of explosive Ne and C-burning. The
dominant products are $^{16}$O, $^{24}$Mg, and $^{28}$Si.
Besides the major abundances, mentioned above, explosive Ne-burning supplies 
also substantial amounts
of $^{27}$Al, $^{29}$Si, $^{32}$S, $^{30}$Si, and $^{31}$P.
Explosive C-burning contributes in addition the nuclei $^{20}$Ne,
$^{23}$Na, $^{24}$Mg, $^{25}$Mg, and $^{26}$Mg.}
\label{4:explone}
\end{figure} 

Matter between the mass cut \index{supernova!mass cut}
$M(r)$=$M_{cut}$ and the mass enclosed in the radius corresponding to
explosive Si-burning with complete Si-exhaustion is indicated with
$M$(ex Si-c). Then follows the zone of incomplete Si-burning until
$M$(ex Si-i), explosive O-burning until $M$(ex O), explosive Ne/C-burning
until $M$(ex Ne), and unprocessed matter from the C/Ne-core is ejected until
$M$(C-core).
The zones beyond explosive Ne/C-burning
($T_{max}<2.1\times 10^9$K) are essentially unaltered and the composition
is almost identical to the pre-explosive one.
When performing such calculations for a variety of progenitors over a
range of initial stellar masses, one can analyze the dependence of
the mass involved in these different burning regimes as a function initial
stellar mass (see Sect.~\ref{sec:4-5}).

Results for a 20M$_\odot$ star \citep{1988PhR...163...13N}
are given as examples for the abundance behavior in a series
of Figs \ref{4:global20}, \ref{4:alphar}, \ref{4:incompl}, \ref{4:explone}. 
It should be mentioned here that this still uses a simplified thermal
bomb treatment for the pre-collapse model rather than the results from
a 1D spherically symmetric simulation with enhanced neutrino capture rates,
which insures an explosion also in 1D.
The explosion energy used corresponds to a supernova
energy of $10^{51}$erg. As mentioned before, this treatment cannot predict
a self-consistent explosion and the position of the mass cut between neutron 
star and ejecta. Only the
observation of $0.07\pm 0.01$M$_\odot$ of $^{56}$Ni in SN1987A (a 20M$_\odot$
star) gives an important constraint, because $^{56}$Ni is produced in the
innermost ejected zones.
The explosive nucleosynthesis due to burning in the shock front
is shown in Fig. \ref{4:global20} for a few major nuclei. Inside 1.7M$_\odot$ 
all Fe-group
nuclei are produced in {\it explosive} Si-burning during the SN II event.
At 1.63M$_\odot$ Y$_e$ changes from 0.494 to 0.499
and leads to a smaller $^{56}$Ni abundance further inside, where more
neutron-rich Ni-isotopes share the abundance with $^{56}$Ni.
This is an artifact of the $Y_e$ gradient in the pre-collapse model which
can be changed in a consistent explosion treatment via neutrino interactions
with this matter.

In explosive Si-burning only\index{process!$\alpha$-rich freeze out} \index{process!Si burning}
 $\alpha$-rich freeze-out and incomplete Si-burning are encountered. Contrary
to SNe Ia,
densities in excess of $10^8$gcm$^{-3}$, which would result in a normal
freeze-out, are not attained in the ejecta of this 20M$_\odot$ star
(see also Fig.\ref{4:explo2}).
The most abundant nucleus in the  $\alpha$-rich freeze-out is
$^{56}$Ni.
For the less abundant nuclei the final  $\alpha$-capture plays a dominant
role transforming nuclei like $^{56}$Ni, $^{57}$Ni, and $^{58}$Ni
into $^{60}$Zn, $^{61}$Zn, and $^{62}$Zn (see Fig.\ref{4:alphar}).

The region which experiences
imcomplete Si-burning starts at 1.69M$_{\odot}$
and extends out to 1.74M$_{\odot}$.
In the innermost zones with temperatures close to $4\times 10^9$K there
exists still a contamination
by the Fe-group nuclei $^{54}$Fe, $^{56}$Ni, $^{52}$Fe, $^{58}$Ni,
$^{55}$Co, and $^{57}$Ni. Explosive O-burning occurs in the mass
zones up to 1.8M$_{\odot}$ (see Fig.\ref{4:incompl}).
The main burning products are $^{28}$Si,
\index{isotopes!28Si} \index{isotopes!58Ni} \index{isotopes!55Co} \index{isotopes!54Fe} \index{isotopes!52Fe} 
\index{isotopes!32S} \index{isotopes!36Ar} \index{isotopes!38Ar} \index{isotopes!34S} \index{isotopes!16O}   
$^{32}$S, $^{36}$Ar, $^{40}$Ca, $^{38}$Ar, and $^{34}$S.
With mass fractions less than $10^{-2}$ also $^{33}$S, $^{39}$K,
$^{35}$Cl, $^{42}$Ca, and $^{37}$Ar are produced.
Explosive Ne-burning leads to an $^{16}$O-enhancement over its hydrostatic
value in the mass zones up to 2M$_\odot$ (see Fig.\ref{4:explone}).

\subsubsection*{Explosive Burning off Stability}

{\it {\bf The p-Process}}

\index{process!p process}
Up to now we discussed the production of heavy nuclei beyond the Fe-group
only via slow neutron captures (the $s$~process) in hydrostatic stellar
evolution.
   A number of proton-rich (p-)isotopes of naturally occurring
stable heavy nuclei cannot be produced by neutron captures along
the line of stability. The currently
most favored production mechanism for those 35 p-isotopes
between Se and Hg is photodisintegration ($\gamma$~process) of intermediate and
heavy elements at high temperatures in late (explosive) evolution stages of
massive stars 
\citep{1978ApJS...36..285W,1990A&A...227..271R}. 
However, not all p-nuclides can be produced satisfactorily, yet. A well-known
deficiency in the model is the underproduction of the Mo-Ru
region, but the region 151$<$A$<$167 is also underproduced,
even in recent calculations 
\citep{2002ApJ...576..323R,2003PhR...384....1A,2006ApJ...653..474R,2008JPhG...35a4029D}.
There exist deficiencies in astrophysical modeling and the
employed nuclear physics. Recent investigations have shown
that there are still considerable uncertainties in the description
of nuclear properties governing the relevant photodisintegration rates. This 
has triggered a number of experimental
efforts to directly or indirectly determine reaction rates 
and nuclear properties for the p/$\gamma$~process 
\citep{2006PhRvC..73a5804R}.
Here it is important to investigate the  
sensitivity of the location of the $\gamma$-process path 
with respect to reaction rate uncertainties. 

Concerning the astrophysical modeling, only a range of temperatures has to 
be considered which are related to the explosive Ne/O-burning zones of a 
supernova explosion (see Figs.\ref{4:incompl} and \ref{4:explone}), where
partial (but not complete) photodisintegration of pre-existing nuclei occurs
(from prior hydrostatic evolution or inherited metallicity), i.e.
at $\approx 2-3\times$10$^9$K.
The $\gamma$~process starts with the photodisintegration of stable
seed nuclei that are present in the stellar plasma. 
During the
photodisintegration period, neutron, proton, and  $\alpha$-emission
channels compete with eachother and with beta-decays further away from
stability. In general, the process, acting like ``spallation'' of pre-existing
nucei, commences with a sequence of $(\gamma,n)$-reactions, moves the 
abundances to
the proton-rich side. At some point in a chain of isotopes,
$(\gamma,p)$ and/or $(\gamma,\alpha)$-reactions become faster than 
neutron emissions, and the flow branches and feeds other
isotopic chains. At late times photodisintegrations become less effective,
when decreasing temperatures shift the branching points and make beta-decays
more important. Finally the
remaining unstable nuclei decay back to stability. 
The branchings established by the dominance of proton and/or
 $\alpha$-emission over neutron emission are crucial in determining
the radioactive progenitors of the stable p-nuclei
and depend on the ratios of the involved reaction
rates. Numerous experimental and theoretical efforts have been undertaken
to improve the reaction input, especially with respect to open questions in
optical potentials for $\alpha$~particles and protons 
\citep{2006PhRvC..74b5805G,2007PhRvC..76e5807K,2008PhRvL.101s1101K,2009PhRvC..79f5801Y}.

\begin{figure}[tbh] 
\begin{center}
\includegraphics[width=0.9\textwidth]{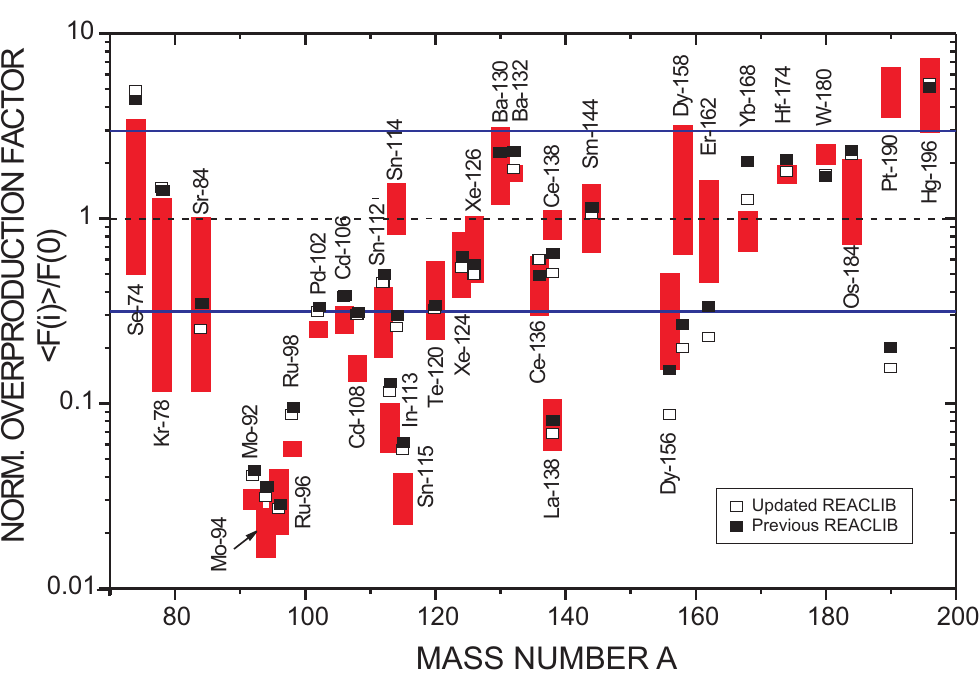}
\end{center}
\caption{Normalized overproduction factors of $p$-process nuclei derived with the 
\citet{2006ApJ...653..474R} 
(open squares) and 
\citet{2008JPhG...35a4029D} 
(full squares)
reaction library. In addition, the results from a range of stellar
models (10-25M$_\odot$) from 
\citet{1995A&A...298..517R} 
are given for comparison. A
value equal to unity corresponds to relative solar abundances.}
\label{4:pprocess}
\end{figure} 

Applications of $p$-process network calculations to the temperature profiles of
initiated explosions have been performed 
by \citet{1995A&A...298..517R,2006ApJ...653..474R,2008JPhG...35a4029D}. 
Here, in Fig.~\ref{4:pprocess} we present the results of a 25M$_\odot$ mass model
\citep{2008JPhG...35a4029D} 
with two reaction rate libraries without and with inclusion 
of all experimental improvements, existing at that point. It is noticed that 
the nuclear uncertainties cannot change the underproduction of especially
the light p-nuclei. Another process seems to be required to supply these
missing abundances.
\\

\noindent{\it {\bf The $\nu$p-Process}}

\begin{figure}[tbp] 
\includegraphics[width=\textwidth]{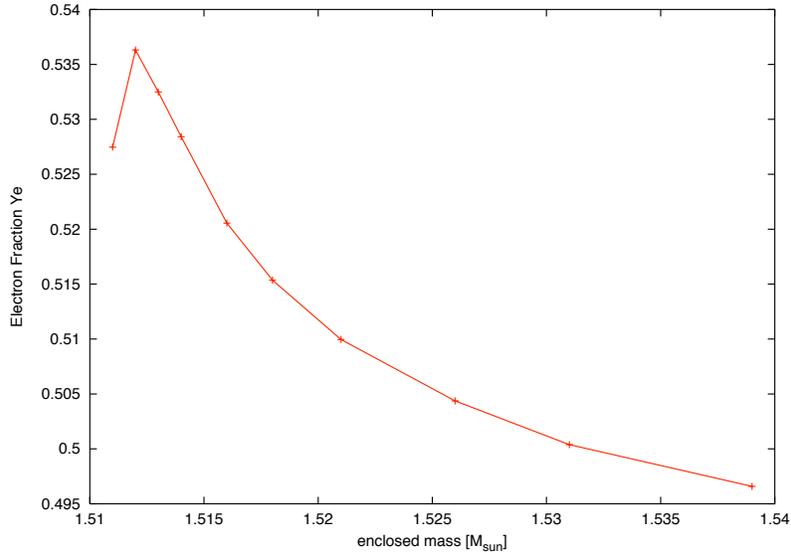}
\caption{$Y_e$ of the innermost ejecta due to neutrino interactions with matter.
At high temperatures electrons are not degenerate, thus the reduction of $Y_e$
due to electron captures is ineffective. For similar neutrino and antineutrino
spectra the neutron-proton mass difference favors
$\nu_e + n \leftrightarrow p + e^-$ over
$\bar\nu_e + p \leftrightarrow n + e^+$.}
\label{4:yeveut}
\end{figure}

\index{process!$\nu$p process} \index{process!rp process}
Neutron-deficient nuclei can also be produced by two other astrophysical 
nucleosynthesis 
processes: the  $rp$~process in X-ray bursts (which, however, does not eject
matter into the interstellar medium 
\citep{1981ApJS...45..389W,1998PhR...294..167S,2008ApJS..174..261F}
and the recently discovered 
$\nu p$~process in core collapse supernovae 
\citep{2006ApJ...637..415F,2006PhRvL..96n2502F,2006ApJ...644.1028P,2006ApJ...647.1323W}.
The $\nu p$~process occurs in explosive environments when
proton-rich matter is ejected under the influence of strong
neutrino fluxes. This includes the innermost ejecta of core-collapse supernova 
\citep{2006A&A...457..281B,2005ApJ...620..861T,2008JPhG...35a4056L} 
and possible ejecta from
black hole accretion disks in the collapsar model of
gamma-ray bursts \citep{2006ApJ...643.1057S}. 
The discussion of these innermost
ejected mass zones has been skipped above, when discussing the results
for explosive nucleosynthesis in a 20~M$_\odot$ star, utilizing a thermal
bomb but the pre-collapse stellar conditions with the corresponding $Y_e$.
Here, as discussed in the beginning of this subsection, we have boosted the 
energy deposition efficiencies by enhancing the neutrino
and anti-neutrino captures on neutrons and protons in a 1D simulation.
While this is not a fully self-consistant treatment,
no external (artificial) energy is required to produce a successful explosion
with a consistently emerging mass cut between neutron star and ejecta.
Moreover, this treatment guarantees provides a $Y_e$ that is consistently 
determined by all weak interactions processes.
The result is that explosions are obtained
and the neutrino interaction with matter leads to a
$Y_e$ enhanced beyond 0.5 (see Fig. \ref{4:yeveut}) which
overcomes nucleosynthesis problems for the Fe-group encountered previously.

\begin{figure}[htbp] 
\centerline{\includegraphics[angle=0,width=11cm]{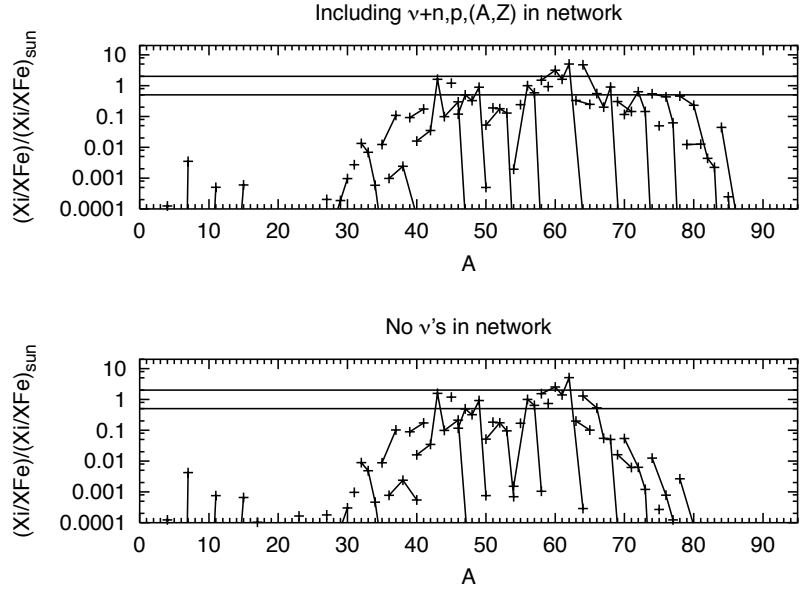}}
\caption{Final abundances in mass zones in the innermost ejecta
which experienced neutrino irradiation, leading to proton-rich conditions
($Y_e>0.5$). The upper part of the figure shows the neucleosynthesis results
in the innermost ejecta of explosive, 
after  $\alpha$-rich and proton-rich freeze-out from
Si-burning, normalized to solar after decay.
The bottom part of the figure also includes the interaction of 
anti-electron neutrinos with protons ($\bar \nu_e +p \rightarrow n + e^+$)
which produces neutons, permitting the late change of $^{64}$Ge via
$^{64}$Ge($n,p)^{64}$Ga. This feature permits further proton captures to 
produce havier nuclei (the so-called $\nu p$~process.
Here matter up to $A=85$ is produced.}
\label{4:nupi1}
\end{figure}

The matter in these ejecta is
heated to temperatures well above 10$^{10}$ and becomes
fully dissociated into protons and neutrons. The ratio
of protons to neutrons is mainly determined by neutrino
and antineutrino absorptions on neutrons and protons,
respectively. Similar neutrino and antineutrino energy spectra and fluxes 
produce proton-dominated matter in the reactions  \index{process!beta decay}
$\nu_e + n \leftrightarrow p + e^-$ and 
$\bar\nu_e + p \leftrightarrow n + e^+$,
due to the n-p mass difference.
When the matter expands and cools, the free
neutrons and protons combine into $\alpha$-particles. Later,
at temperatures around 5$\times$10$^9$K, $\alpha$-particles assemble 
into heavier nuclei via unstable intermediate nuclei, e.g. the 
triple-$\alpha$ reaction via unstable $^8$Be, 
but - depending on the entropy and the expansion of matter -
only a fraction of those form iron-group nuclei ( $\alpha$-rich freeze-out). In
case of a proton-rich environment, there are also still free protons
available at the time of the alpha freeze-out. Once 
the temperature drops to about 2$\times$10$^9$K, the composition of
the ejecta consists mostly of $^4$He, protons, and iron group nuclei with
N$\approx$Z (mainly $^{56}$Ni) in order of decreasing abundance.
Without neutrinos, synthesis of nuclei beyond the iron peak becomes very 
inefficient due to bottleneck (mainly even-even
$N=Z$) nuclei with long beta-decay
half-lives and small proton-capture cross sections. 
\index{isotopes!64Ge}  
Such a nucleus is $^{64}$Ge. Thus, with the $Y_e$ determined by neutrino
interactions with free neutrons and protons in the early very hot phase
of dissociated nuclei, the nucleosynthesis leads to an  $\alpha$- and proton-rich
freeze-out which does not stop at $^{56}$Ni but continues up to $^{64}$Ge
(which later decays to $^{64}$Zn. This part of the story enables core
collapse yeads which produce Fe-group nuclei up to essentially $^{64}$Zn.
The effect is seen in the upper portion Fig. \ref{4:nupi1}.

However, the matter is subject to a large
neutrino/antineutrino flux from the proto-neutron star.
While neutrons are bound in neutron-deficient $N=Z$
nuclei and neutrino captures on these nuclei are negligible due to energetics, 
antineutrinos are readily captured both on free protons and on heavy nuclei on 
a timescale of a few seconds. As protons are more abundant than
heavy nuclei, antineutrino captures occur predominantly
on protons, leading to residual neutron densities of 
$10^{14}--10^{15}$ cm$^{-3}$ for several seconds. These neutrons are 
easily captured by heavy neutron-deficient nuclei, for
example $^{64}$Ge, inducing $(n,p)$ reactions with time scales
much shorter than the beta-decay half-life. This permits further proton 
captures and allows the nucleosynthesis flow to continue to 
heavier nuclei (see lower part of Fig.~\ref{4:nupi1}). The $\nu p$~process 
\citep{2006PhRvL..96n2502F} 
is this sequence of 
$(p,\gamma)$-reactions, followed
by $(n,p)$-reactions or beta-decays, where the neutrons are supplied
by antineutrino captures on free protons.

\begin{figure}[htbp] 
\begin{centering}
\includegraphics[width=10cm]{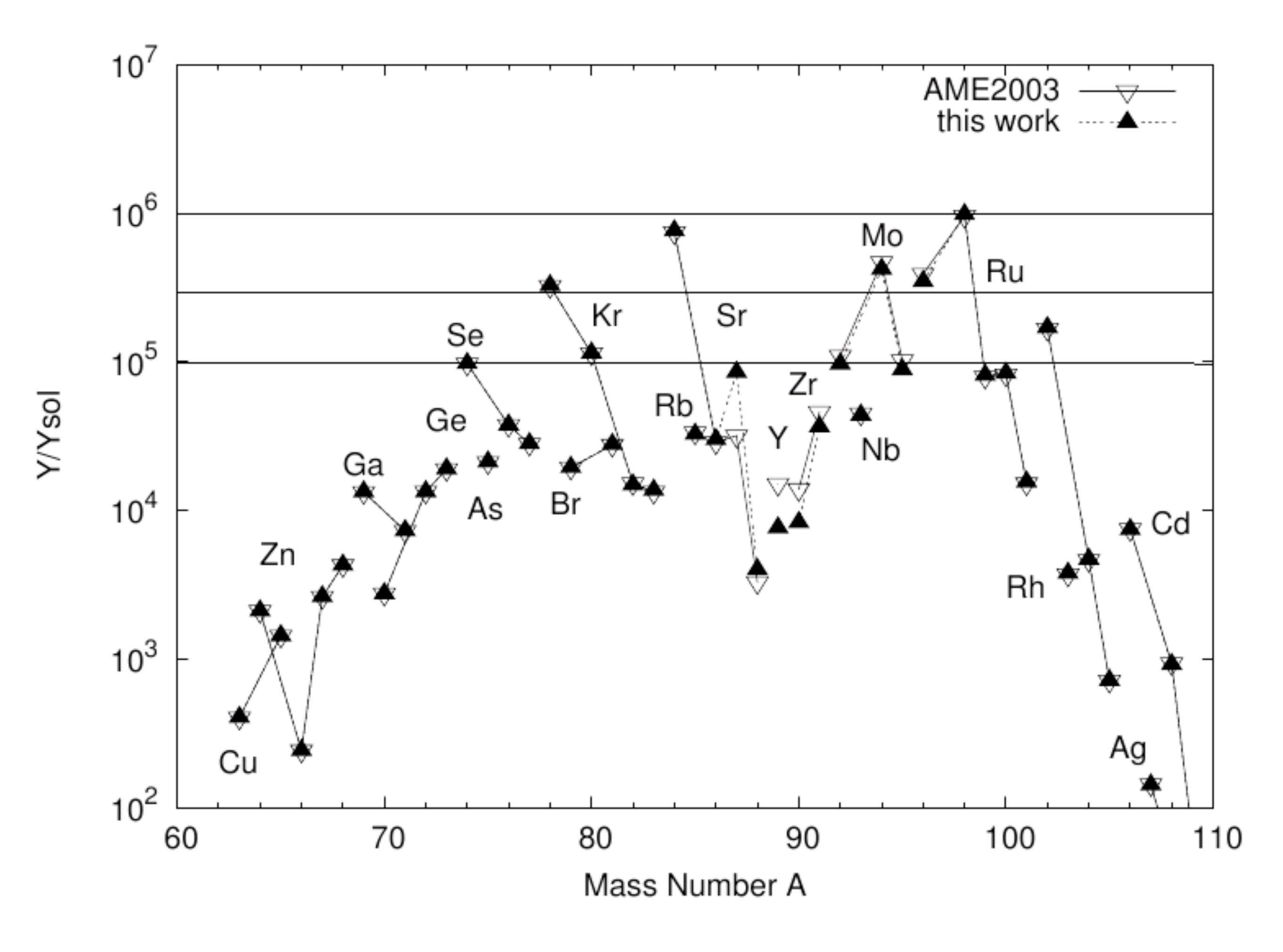}
\end{centering}
\caption{Final abundances in mass zones experiencing the
$\nu p$~process, i.e. the innermost ejecta of explosive,  $\alpha$-rich freeze-out
Si-burning, normalized to solar after decay 
for two sets of thermonuclear reaction rates/masses.
Matter up to $A=100$ can be produced easily.}
\label{4:nupabundance}
\end{figure} 

In Fig.\ref{4:nupabundance} we also show $\nu p$~process nucleosynthesis 
results from the innermost early neutrino wind ejecta produced in
the explosion of a 
15M$_\odot$  star \citep{2003NuPhA.718..269J}, 
also utilized in 
\citep{2006ApJ...644.1028P,2009ApJ...690L.135F}, 
which synthesizes efficiently nuclei even 
for $A>90$.
Two sets of astrophysical reaction rates were used in
the reaction network, both based on theoretical rates from the NON-SMOKER 
code \citep{2000ADNDT..75....1R,2004ADNDT..88....1R}, 
but once with the latest excited state 
information and masses from the AME2003 compilation 
\citep{2003NuPhA.729..337A} 
and another set also with the latest mass measurements 
\citep{2008PhRvC..78e4310W}.
Fig.\ref{4:nupabundance} shows the final abundances normalized to
solar abundances after decay to stability for these two sets
of thermonuclear reaction rates. Only nuclei produced
in the p-rich ejecta are shown. As is clearly seen,
there is no difference in the yields for the two
different sets of rates except for a few nuclei in the mass
\index{isotopes!87Sr} \index{isotopes!88Sr} \index{isotopes!89Y} \index{isotopes!90Zr} \index{isotopes!91Zr} \index{isotopes!88Tc}    
range $85<A<95$, namely $^{87,88}$Sr, $^{89}$Y, and $^{90,91}$Zr.
This can be directly traced back to the large change in
the mass of $^{88}$Tc ($\Delta M$= -1031 keV). This change in mass leads to 
an increase in the reaction rate for $^{88}$Tc$(\gamma, p)^{87}$Mo at the 
relevant temperatures and therefore a relative suppression of the
opposite capture rate.
These results show that the $\nu p$~process can easily produce the light 
p-nuclei of Mo and Ru, which are deficient in $p$-process calculations. Further
processing depends on the expansion (speed) of matter and the
overlying mass of ejecta.
\\
 
\noindent{\it {\bf The r-Process}}

\index{process!r process}
A rapid neutron-capture process ($r$~process) in an explosive environment is 
traditionally believed to be responsible for the nucleosynthesis of about half 
of the heavy elements above Fe. While in recent years the 
high entropy (neutrino) wind (HEW) of core-collapse supernovae has been 
considered to be \index{supernova!high-entropy wind}
one of the most promising sites, hydrodynamical simulations still encounter 
difficulties to reproduce the astrophysical conditions under which this 
process occurs. The classical \emph{waiting-point} approximation, with the basic 
assumptions of an Fe-group seed, an $(n,\gamma)-(\gamma,n)$-equilibrium for 
constant neutron densities $n_n$ at a chosen temperature $T$, over a process 
duration $\tau$, and an instantaneous freezeout, has helped to gain improved 
insight into the systematics 
of an $r$~process in terms of its dependence on nuclear-physics input and 
astrophysical conditions 
\citep{1991PhR...208..267C,1993ApJ...403..216K,2007ApJ...662...39K}.
This corresponds to a set of quasi-equilibria with each QSE group being
represented by an isotopic chain. \index{process!quasiequilibrium}
Taking a specific seed nucleus, the solar $r$-process pattern peaks can be
reproduced by a variation/superposition of neutron number densities $n_n$ and
durations $\tau$. Whether the solar $r$-process abundances are
fully reproduced in each astrophysical event, i.e., whether each such
event encounters the full superposition of conditions required, is a matter of 
debate
\citep{1996ApJ...466L.109W,2001NuPhA.693..282P,2003ApJ...591..936S,2006ApJ...643.1180H,2007PhR...442..237Q,2009ApJ...694L..49F,2010ApJ...712.1359F}.
In realistic astrophysical environments with time variations in $n_n$ and
$T$, it has to be investigated whether at all and for which time
duration $\tau$ the supposed $(n,\gamma)-(\gamma,n)$-equilibrium of the \index{process!n$\gamma$ equilibrium}
classical approach will hold and how freeze-out effects change this behavior. 
In general, late neutron captures may alter the final abundance distribution. 
In this case neutron capture reactions will be important. Also $\beta$-delayed 
neutrons can play a role in forming and displacing the 
peaks after freeze-out.

For many years since  \citet{1994ApJ...433..229W,1994A&A...286..857T,1996ApJ...471..331Q}
the high entropy wind has been considered as the most promising (realistic?)
environment, expelled from newly formed (hot) neutron stars in
core-collapse supernovae, which continue to release neutrinos after the 
supernova shock wave is launched. These neutrinos interact with matter 
of the outermost proto-neutron star layers which are heated and ejected in a 
continuous wind. The late neutrino flux also leads to moderately neutron-rich 
matter~\citep{1996ApJ...471..331Q} 
via interactions with neutrons and protons and 
causes matter ejection with high entropies. (However, there are recent studies
\citep{2009arXiv0908.1871F} from collapse calculations
which predict a proton-rich wind composition for more than the first 10s
after collapse.)
Problems were encountered to attain entropies sufficiently high in order to 
obtain high neutron/seed ratios which can produce the heaviest $r$~process nuclei 
\citep{2001ApJ...562..887T,2001ApJ...554..578W,2002ApJ...578L.137T}. 
Recent hydrodynamic simulations for core-collapse supernovae support the idea 
that these entropy constraints can be fulfilled in the late phase (after the 
initial explosion) when a reverse shock is forming
\citep{2006ApJ...646L.131F,2007A&A...467.1227A,2007PhR...442...23B,2007PhR...442...38J,2009A&A...494..829P}.

The question is whether such high entropies occur at times with sufficiently 
high temperatures when an $r$~process is still underway
\citep{2008ApJ...672.1068K}.
Exploratory calculations to obtain the necessary conditions for an $r$~process
in expanding high-entropy matter have been undertaken by a number of groups
\citep{1997ApJ...482..951H,1997ApJS..112..199M,2000ApJ...533..424O,2001ApJ...554..578W,2002ApJ...578L.137T,2004ApJ...606.1057W,2004ApJ...600..204Y,2007ApJ...666L..77W,2008ApJ...672.1068K}.
Recent investigations 
\citep{2009ApJ...694L..49F,2010ApJ...712.1359F}
focussed (a) on the effects of varying nuclear 
physics input [mass models FRDM (Finite Range Droplet Model, 
\citet{1995ADNDT..59..185M}), 
ETFSI-1 (Extended Thomas-Fermi with Strutinsky Integral)~\citep{1995ADNDT..61..127A}, 
ETFS-Q with quenching of shell closures far from stability
\index{nuclei!mass model} \index{nuclear mass model!FRDM} \index{nuclear mass model!ETFS} \index{nuclear mass model!Duflo-Zuker} 
\citep{1996PhLB..387..455P}, 
the mass formula by Duflo \& Zuker (DUFLO-ZUKER, 
\citet{1995PhRvC..52...23D}) 
and HFB-17 (a rencent Hartree-Fock-Bogoliubov approach)
\citep{2009PhRvL.102x2501G}] 
and (b) the
detailed understanding of the nuclear flow through the chart of nuclides,
testing equilibria, freeze-out and delayed neutron capture.
To investigate these effects we have applied a full network containing up to 
6500 nuclei and the corresponding nuclear masses, cross sections and 
$\beta$-decay properties. 

\begin{figure}[!tbh] 
\begin{centering}
\includegraphics[width=10cm]{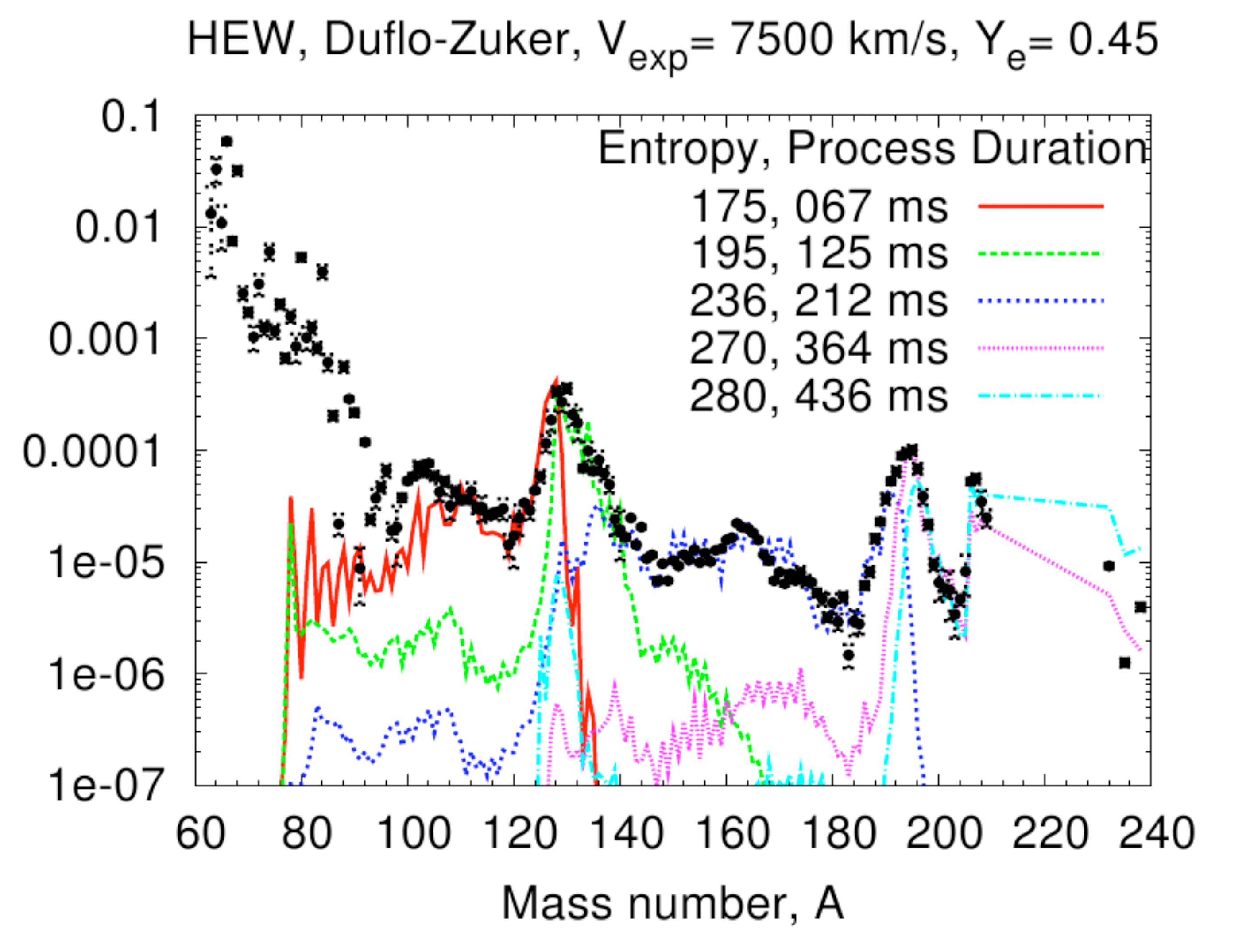}
\caption{\label{label}High entropy neutrino wind results for the mass model 
by \citet{1995PhRvC..52...23D}, 
expansion parameters and proton/nucleon ratio
$Y_e$ as given in the label, for a variation in extropies per baryon and $k_B$.
\citep{2010ApJ...712.1359F}}
\end{centering}
\label{4:rprocess}
\end{figure} 

The calculations presented here are based on trajectories 
for densities and temperatures originating from expansions with a complete
parameter study in terms of entropy $S$, electron fraction $Y_{e}$ and 
expansion velocity $V_{exp}$, the latter being related to 
the expansion timescale $\tau_{exp}$ 
\citep{1999ApJ...516..381F,2010ApJ...712.1359F}.
Here we only show the results utilizing the Duflo-Zuker mass model for a 
range of entropies. It is assumed
that in the late phases of the neutrino wind of a deleptonized neutron star
conditions with $Y_e<0.5$ prevail (but see the discussion above).

Either higher entropies than obtained by the simulations discussed above or
conditions with intrinsically high neutron densities (like expanding
neutron star matter with $Y_e\approx 0.1-0.2$) can lead to neutron/seed ratios
which are sufficiently high to reach fissionable nuclei in the $r$~process.
The fission fragments can again capture neutrons and produce fissionable
nuclei, leading to an $r$~process with fission recycling 
\citep{1994ApJ...429..499R,2007PrPNP..59..199M}.
This requires reliable fission barriers (and
fission fragment distributions) to test the possibility for the production
of superheavy elements. It was shown recently that neutron-induced 
fission is more important in $r$-process nucleosynthesis than beta-delayed 
fission \citep{2003NuPhA.718..647P,2007PrPNP..59..199M}. 
Thus, the need to provide a compilation
of neutron-induced fission rates is obvious and has been performed 
recently \citep{2005NuPhA.747..633P,2009PhRvC..79b4612G,2009arXiv0911.2181P}.
Comparison of rates obtained with different
sets for mass and fission barrier predictions give a
measure of the uncertainties involved. 

\subsection{Exotic SN Types: Hypernovae and GRBs}

As was outlined in various parts of the preceeding sections, massive stars in 
the range of 8 to $\sim$ 130$M_\odot$ undergo
core-collapse at the end of their evolution and become Type II and
Ib/c supernovae unless the entire star collapses into a black hole
with no mass ejection \citep{2003ApJ...591..288H}. 
\index{supernova!hypernova} \index{gamma ray burst} \index{supernova!type Ibc} \index{supernova!type II}
These Type II and Ib/c 
supernovae (as well as
Type Ia supernovae, see Ch.~5) release large explosion energies and eject
matter which underwent explosive nucleosynthesis, thus having strong dynamical,
thermal, and chemical influences on the evolution of interstellar
matter and galaxies. They have been the main focus of the present chapter
up to now. The explosion energies of
core-collapse supernovae are fundamentally important quantities, and
an estimate of $E \sim 1\times 10^{51}$ erg has often been used in
calculating nucleosynthesis and the impact on the interstellar medium.
(Here we use the explosion energy $E$ for the final
kinetic energy of the explosion.)  A good example is SN1987A in the Large
Magellanic Cloud, whose energy is estimated to be $E = (1.0$ - 1.5)
$\times$ $10^{51}$ ergs from its early light curve

One of the most interesting recent developments in the study of
supernovae (SNe) is the discovery of some very energetic supernovae
(see e.g. \citet{2006NuPhA.777..424N},
whose kinetic energy (KE) (in spherically symmetric analysis, see also
\citet{2004RvMP...76.1143P}) 
exceeds $10^{52}$ erg, about 10 times the
KE of normal core-collapse SNe (hereafter $E_{51} = E/10^{51}$ erg).
The most luminous and powerful of these objects, the Type Ic supernova
(SN~Ic) 1998bw, was probably linked to the gamma-ray burst GRB 980425,
thus establishing for the first time a connection
between gamma-ray bursts (GRBs) and the well-studied phenomenon of
core-collapse SNe.  However, SN~1998bw was exceptional for a SN~Ic: it
was as luminous at peak as a SN~Ia, indicating that it synthesized
$\sim 0.5 M_\odot$ of $^{56}$Ni, and its KE was estimated at $E \sim 3
\times 10^{52}$ erg.

There is another class of supernovae which appears to be rather faint
with apparently almost vanishing $^{56}$Ni ejection. Thus, the question
emerges how these different objects are related, whether they correspond
to different initial masses and how the explosion mechanism changes.
The questions to be answered are the following:

\begin{itemize}
\item {do 8-10~M$_\odot$ stars which produce an Fe-core in a collapse
initiated via electron capture after core He-burning (electron capture
supernovae) have a different explosion mechanism after core collapse than
more massive stars? Is here only a small amount of material involved
outside the collapsing C-core and litte Ni-ejection occurring?}
\item {for which stellar progenitor masses do we have a transition from the
formation of neutron stars to the formation of black holes after collapse?}
\item {to which extent is this transition region shifted by the nuclear
equation of state?}
\item {for which transition region are initially neutron stars formed, causing
a regular supernova explosion, and only fall back by the reverse shock swallows
inner matter, leading to a small final Ni-ejection and faint light curves?}
\item{for which progenitor masses are black holes formed directly during
collapse and how can this be observed?}
\item{what is the role of rotation and magnetic fields to cause gamma-ray
bursts?}
\item {can we give reliable nucleosynthesis yields for such events?}
\end{itemize}

Before going into a too involved discussion of the causes of these events,
let us first consider the possible effect which higher energy explosions
have on the ejecta, i.e. nucleosynthesis products.
Here we use the term 'hypernova' to describe an
extremely energetic supernova with $E \ge 10^{52}$ erg without
specifying the explosion mechanism 
\citep{2001sgrb.conf..144N}.  
Following SN
1998bw, other \emph{hypernovae} of Type Ic have been discovered or
recognised.
Nucleosynthesis features in such hyper-energetic supernovae must show
some important differences in comparison to normal supernova explosions.  
The higher explosion energies could lead to larger ejected $^{56}$Ni masses, as
observed in such explosions. They also cause higher entropies in the innermost 
ejecta,
which result in a more extreme  $\alpha$-rich freeze-out from explosive Si-burning.
Such conditions permit the sizable production of Fe-group nuclei beyond
$^{56}$Ni, up to $^{64}$Ge which decays to $^{64}$Zn 
\citep{2001ApJ...555..880N}.
This feature could have an influence on abundance patterns observed in 
extremely metal-poor halo stars.  
In fact, the observational finding that Zn behaves like an Fe-group element
in galactic evolution - and was underproduced in existing supernova models 
(which were not including the $\nu p$~process) -
was used as a strong argument that a large fraction of massive stars
explode as hypernovae 
\citep{2006NuPhA.777..424N,2006ApJ...653.1145K}. 

The observed frequency of type Ib supernovae is about 20\% of SNe II
at solar metallicity \citep{1999A&A...351..459C}. 
In four cases the typical 
spectrum of type Ic supernovae has been observed, associated with long soft
gamma-ray bursts \citep{2006ARA&A..44..507W}, 
indicating a link between SNe Ic and long 
soft GRBs. 
\citet{2003A&A...406..259P} 
found an increase
in the ratio of SNe Ibc / SNe II with metallicity. In order to understand
this trend one has to understand stellar models as a function of metallicity,
from the first stars in the Universe, i.e., metal-free, Population III (Pop
III) stars which were born in a primordial hydrogen-helium gas cloud to
present metallicities.
This is one of the important challenges of current
astronomy and relates to Sect.~\ref{sec:4-2}, where we have discussed the
evolution of massive stars from lowest metallicities to solar values,
including rotation and metallicity-dependent winds.
In fact 
\citet{2003A&A...404..975M} 
could reproduce this observed trend 
metallicity trend. On the other hand different groups 
\citep{1992ApJ...391..246P,2007ApJ...662L.107V,2008MNRAS.384.1109E} 
provide models from binary evolution with mass transfer
(removing the H-rich envelope) which seems to reproduce this trend as 
well. Recent observations 
\citep{2008ApJ...673..999P} 
provide for the first time
the individual ratios of SNe Ib / SNe II and SNe Ic / SNe II, rather than
only the combined SNe Ibc / SNe II ratio. 
\citet{2009A&A...502..611G} 
have 
studied stellar evolution in detail as a function of initial mass, metallicity 
and rotation, based on the Geneva evolution models. As a full
understanding of the GRB mechanism is pending, two options have been 
considered for the cases where the Fe-core is massive enough that the
formation of a black hole in the collapse is expected: (i) nevertheless
a supernova-type explosion is assumed, (ii) a black hole forms without a 
supernovae. They find that current models of stellar evolution can account
for the observed number ratios of SNe Ib / SNe II and SNe Ic / SNe II and
their variation with metallicity. In case (ii), i.e. when no supernova
occurs after black holes are formed, single-star models can still account for
more than one half of the combined SNe Ibc / SNe II ratio for metallicities
above solar, however, low metallicity SNe Ic events have to come from
binary evolution. If black hole
formation is identified with the occurrance of GRBs, the resulting number is 
too large, indicating that only a fraction of such events, most
probably very rapid rotators, result in GRBS after collapse
\citep{1999ApJ...524..262M}.

As mentioned earlier in this subsection, the explanation of SN Ic SN1998bw
\index{supernova!SN1998bw} \index{gamma-ray burst!GRB980425} \index{gamma ray burst!GRB030329}
\index{supernova!SN2003dh}
is based on a very large progenitor mass $M$ and explosion energy $E$.
The type Ic hypernovae 1998bw and 2003dh were clearly linked to the
gamma-ray bursts GRB 980425 and GRB 030329, thus establishing the connection
between long GRBs and core-collapse supernovae (SNe).  SNe~1998bw and 2003dh
were exceptional for SNe~Ic: they were as luminous at peak as a SN~Ia,
indicating that they synthesized 0.3 - 0.5 $M_\odot$ of $^{56}$Ni, and
their kinetic energies (KE) were estimated in the
range $E_{51} = E/10^{51}$\,erg $\sim$ 30 - 50.
Other \emph{hypernovae} have been recognized, such as SN~1997ef
and SN~2002ap. \index{supernova!SN2002ap}
These hypernovae span a wide range of properties, although they all
appear to be highly energetic compared to normal core-collapse SNe.
The mass estimates, obtained from fitting the optical light curves and
spectra, place hypernovae at the high-mass end of SN progenitors.

In contrast, SNe II 1997D and 1999br were very faint SNe with very low
kinetic energy. 
\index{supernova!SN1997D}
 This leads to a diagram with the
explosion energy $E$ or the ejected $^{56}$Ni mass $M(^{56}$Ni) as a
function of the main-sequence mass $M_{\rm ms}$ of the progenitor star
which shows two branches. Therefore, one is led to the conclusion that
SNe from stars with
$M_{\rm ms} \mathrel{\rlap{\lower 4pt \hbox{\hskip 1pt $\sim$}}\raise
1pt \hbox {$>$}}$ 20-25 $M_\odot$ have different $E$ and $M(^{56}$Ni),
show a bright, energetic \emph{hypernova branch} at one extreme and a
faint, low-energy SN branch at the other extreme.  For the
faint SNe, the explosion energy was so small that most $^{56}$Ni fell
back onto the compact remnant.  Thus the faint SN branch may become a
\emph{failed} SN branch at larger $M_{\rm ms}$.  Between the two
branches, there may be a variety of SNe.

This trend could be interpreted as follows.  Stars more massive than
$\sim$ 25 $M_\odot$ form a black hole at the end of their evolution.
Stars with non-rotating black holes are likely to collapse \emph{quietly}
ejecting a small amount of heavy elements (Faint supernovae). A preceding
stage could be the temporary formation of a neutron star and a supernova
explosion, but fallback of matter leads to an increase of the neutron
star mass beyond its maximum stable value. (The combination of mixing processes 
in the innermost ejecta and fallback can influence the ejecta composition.) In
contrast, stars which formed rotating black holes are likely to give rise to
hypernovae. Here disk and jet formation seems to be a necessary ingredient
to understand the explosion. (An option is that hypernova progenitors might 
form from the rapidly rotating cores after spiraling-in of a companion star 
in a binary system).

\section{The Aftermath of Explosions}

\label{sec:4-5}
In the preceding sections we have given an overview of hydrostatic and
explosive burning processes in massive stars, the individual phases of 
stellar evolution, the endstages like core collapse,  \index{supernova!remnant}
explosive nucleosynthesis products from supernovae explosions
and possible variations in outcome if core collapse ends in black hole 
formation, related possibly to hypernovae or gamma-ray bursts.
What remains to be done is to (i) get a complete
picture from stellar models and simulations how hydrostatic/wind and
explosive contributions add up to the complete yields observed in
such events, (ii) verify such models with individual 
observations, e.g. from lightcurves and from remnants, 
(iii) finally to integrate
all these events/stellar yields over a mass distribution and metallicity
evolution of galaxies, in order to make comparisons with overall galactic
oberservations of very long-lived radioisotopes which average over several
stellar generations.

\subsection{Massive Stars and Their Complete Yields}

In section \ref{sec:4-4} we have introduced in Eq. (\ref{4:totale}) a 
simplified rule which determined at which radius certain temperatures are 
attained in the explosion, assuming that the explosion energy is distributed 
at all times in a homogenous bubble within the radius of the present shock
front position. If one knows the radial mass distribution $M(r)$ in 
pre-explosion models through which the shock front passes, one knows the
amount of matter which encountered certain burning conditions. In table
\ref{4:yields} we provide this information for different initial stellar masses
(still based on models from 
\citet{1988PhR...163...13N}), 
at (up to) which radial
mass position explosive (complete and incomplete) Si-burning, O-burning, 
Ne/C-burning are occurring (upper portion) and the size of these regions
(in M$_\odot$) involved (lower portion). In addition, we give the size of
the CO-core of prior He-burning in stellar evolution. To first order matter
between explosive C/Ne-burning and the stellar surface is ejected unchanged.
As this simplified treatment does not know anything of the explosion mechanism
which produced this explosion energy, the position of the mass cut is not
known and therefore also not the total amount of complete Si-burning material.
The core sizes given (e.g. CO-core after He-burning) also make no difference
whether this matter resulted from initial core burning of this burning stage
or subsequent outward propagating shell burning (e.g. shell He-burning of
shell C-burning) which produce specific isotopes of interest (e.g.
$^{26}$Al, $^{60}$Fe as discussed in Sect.~\ref{sec:4-3}).

Initially we want to focus here on the explosive burning phases. We also
want to add that table \ref{4:yields} includes always two values for the
radial masses involved, (a) from the simplified Eq.(\ref{4:totale}) applied
to the appropriate stellar model ($M(r)$) and (b) resulting from an actual
explosion calculation (initiated via a thermal bomb), as obtained in 
\citet{1996ApJ...460..408T}. 
When comparing these numbers, we see a quite
close agreement, except for the most massive star where non-negligible
deviations are encountered.

\begin{table}
\label{4:yields} \index{nucleosynthesis!yield}
\caption{Masses and products in explosive and hydrostatic burning}
\begin{center}
\begin{tabular}{c c c c c c}
\hline
\hline
 $M(r)$ & Burning site & 13M$_\odot$ & 
15M$_\odot$ & 20M$_\odot$ & 25M$_\odot$ \\
\hline
 Fe-core & hydr. Si-burning       &  1.18 &  1.28 &  1.40 &  1.61 \\
 mass cut &  (expl. mechanism)      &    ?  &    ?  &  ?  &    ?  \\
 ex Si-c &  expl. compl. Si-burning  &  1.42\ 1.40 &  1.46\ 1.44 &  1.70\ 1.69 &  1.79\ 1.80 \\
 ex Si-i & expl. incompl. Si-burning   &  1.48\ 1.47 &  1.52\ 1.51 &  1.75\ 1.75 &  1.85\ 1.89 \\
 ex O & expl. O-burning &  1.54\ 1.54 &  1.57\ 1.57 &  1.81\ 1.81 &  1.92\ 2.00 \\
 ex C/Ne & expl. Ne-burning &  1.66\ 1.65 &  1.73\ 1.70 &  2.05\ 2.05 &  2.26\ 2.40 \\
 CO-core   & hydr. He-burning &  1.75 &  2.02 &  3.70 &  5.75 \\
\hline
$\Delta M$  & (Main) products, major radioactivities & & & &  \\
\hline
ex. Si-c & "Fe", He; $^{56,57}$Ni, $^{61,62}$Zn, $^{59}$Cu, $^{52}$Fe, $^{48}$Cr  &   ?   &   ?   &  ?  &   ?   \\
ex. Si-c & $^{44}$Ti, $\nu p$~process, $r$~process?   &   ?   &   ?   &  ?  &   ?   \\
ex. Si-i & Si, S, Fe, Ar, Ca; $^{55}$Co, $^{52}$Fe, $^{48}$Cr &  0.06\ 0.07 &  0.06\ 0.07 &  0.05\ 0.06 &  0.06\ 0.09 \\
ex. O & O, Si, S, Ar, Ca  &  0.06\ 0.07 &  0.05\ 0.06 &  0.06\ 0.06 &  0.07\ 0.11 \\
ex. C/Ne & O, Mg, Si, Ne; $^{26}$Al,  $p$~process &  0.12\ 0.11 &  0.16\ 0.13  &  0.24\ 0.24 &  0.34\ 0.40 \\
hydr. He& O, Ne, Mg, Si, $s$~process &  0.09\ 0.10 &  0.29\ 0.32 &  1.65\ 1.65 &  3.49\ 3.35 \\
\hline
\end{tabular}
\end{center}
\end{table}

Based on this information we want to discuss
complete nucleosynthesis yields, including explosive processing (also the
$\nu p$~process, affected by neutrinos in the innermost ejecta, as well as
the  $p$~process in explosive Ne/O-burning), hydrostatic yields from the outer
layers (including $s$~process) which are ejected unaltered, and prior wind
yields lost during stellar evolution. Then we concentrate on
the long-lived radioactivities $^{26}$Al, $^{60}$Fe, 
$^{44}$Ti, other Fe-group and lightcurve-determining nuclei, including
their origin which is e.g. important for $^{26}$Al and $^{60}$Fe,
which have hydrostatic burning as well as explosive origins.
The $r$~process in the neutrino wind or possibly polar jets has been
presented qualitatively with entropy, $Y_e$ and expansion timescale as free
parameters or expansion timescale of neutron star matter as a free parameter.
Presently no realistic explosion models are available to discuss this
matter in a real stellar context, but a short discussion of long-lived radioactive
chronometers is presented.

Table \ref{4:yields} provides the following conclusions: The amount of 
ejected mass from the 
unaltered (essentially only hydrostatically processed) CO-core varies strongly 
over the progenitor mass range. The variation is still large for the matter 
from explosive Ne/C-burning, while the amount of mass from explosive O- and 
Si-burning is almost the same for all massive stars.
Therefore, the amount of ejected mass from the unaltered (essentially
only hydrostatically processed) CO-core and from explosive Ne/C-burning
(C, O, Ne, Mg)
varies strongly over the progenitor mass range, while the amount of mass
from explosive O- and Si-burning (S, Ar, and Ca) is almost the
same for all massive stars. Si has some contribution from hydrostatic
burning and varies by a factor of 2-3.
The amount of Fe-group nuclei ejected
depends directly on the explosion mechanism which also affects the $Y_e$
in these inner zones. Thus, we have essentially
three types of elements, which test different aspects of supernovae,
when comparing with individual observations. The first set (C, O, Ne,
Mg) tests the stellar progenitor models, the second (Si, S, Ar, Ca) the
progenitor models and the explosion energy in the shock wave, while the
Fe-group (beyond Ti) probes clearly in addition the actual supernova
mechanism. Thus, we require that all three aspects of the predicted abundance
yields are based on secure modeling (stellar evolution, explosion energy,
and explosion mechanism) in order to be
secure for their application in lightcurve modeling, radioactivities in 
remnants as well as the the in chemical evolution of galaxies.\\

\noindent{\bf $r$-process ejecta}\\
The biggest uncertainty exists for the absolutely innermost ejecta, i.e. the
possible $r$~process ejection in the neutrino wind. This matter escapes after the 
supernova explosion shock wave was launched and the continuing neutrino escape from
the remaining neutron star leads to its \emph{surface erosion/evaporation}, i.e.
a neutron-rich wind which could trigger an $r$~process  \index{process!r process}
\citep{1996ApJ...471..331Q}.
Early modelling seemed to lead to a full $r$-process abundance distribution
\citep{1994ApJ...433..229W}, 
which was, however, already then questioned by
other investigations 
\citep{1994A&A...286..857T}, 
when utilizing the entropies
obtained from their calculations. Results with present neutrino physics
and detailed transport modeling seem to find the opposite behavior, i.e.
proton-rich conditions for more than the first 10s after the explosion 
\citep{2009arXiv0908.1871F,2009ApJ...694..664M}, 
as noticed first by 
\citet{2003NuPhA.719..144L} 
and discussed in Sect.~\ref{sec:4-4} together with 
the $\nu p$~process. A major question is if and how this turns
to be neutron-rich in later phases, what physics causes this change (the
nuclear EoS or neutrino properties?), and how very high entropies can be 
attained to produce also the heaviest nuclei. Present observations of low
metallicity stars show huge variations in heavy $r$-process content and 
indicate that in most explosions the latter is not taking place, making the
\emph{full} $r$~process a rare event. Then typical supernovae would only provide
a weak $r$-process environment.
Whether either high entropies are only attained in exceptional cases or
other origins of low entropy, highly neutron-rich matter (neutron star mergers
or neutron-rich jets from rotating core collapse supernovae, 
\citet{1999ApJ...525L.121F,2001ApJ...562..456C,2003ApJ...587..327C})
cause the main $r$~process has to be explored, in parallel with
the still remaining challenges of nuclear physics far from stability.

In the preceding sections we have shown that there exists a principle understanding
about the nuclear working of the $r$~process and that it is possible to reproduce
solar system $r$-process abundances by superpositions of components with varying
environment conditions. What seems not possible, yet, is to clearly identify, without
doubt, the responsible astrophysical site. Taking, however, such $r$~process 
superposition fits as \emph{zero-age} abundances with e.g. production ratios for
$^{232}$Th/$^{238}$U or other actinide (chronometer) nuclei, one can predict such ratios also
as a function of decay time (present-age abundances) and identify ages of very 
metal-poor stars which were
born with a fresh pollution of an $r$~process pattern
(see e.g. \citet{1991PhR...208..267C,1999ApJ...521..194C,2002SSRv..100..277T,2007ApJ...662...39K} and references therein).
The typical result is that these chronometers indicate an age of the oldest
\index{nucleocosmochronology} \index{Galaxy!age}
stars in our Galaxy in the range of 14-15~Gyr with an uncertainty of about 
3-4~Gyr. A more complex utilization of long-lived $r$-process radioactivities
in terms of a continuing enrichment in galactic evolution is given in 
Ch.~2.\\

\noindent{\bf Fe-group and beyond}\\
The neutrino wind works via evaporation of the neutron star surface after
the supernova shock wave emerged and caused the explosion. The innermost
matter which experienced the shock is thus ejected earlier and
a typical example for its composition is seen in 
Fig. \ref{4:alphar}, which shows the zones of complete Si-burning with  $\alpha$-rich
freeze-out. The change in abundances at $M=1.63$M$_\odot$ is due to a
change in $Y_e$ in the original stellar model, which was utilzed for explosive
nucleosynthesis predictions just by introducing a thermal bomb of $10^{51}$erg.
If one accounts correctly for the neutrino interactions during the collapse
and explosion, this turns matter even slightly proton-rich ($Y_e>0.5$),
see the discussion of the $\nu p$~process in Sect.~\ref{sec:4-4} and
Figs. \ref{4:yeveut}, \ref{4:nupi1} and \ref{4:nupabundance}
\citep{2006ApJ...637..415F,2006PhRvL..96n2502F,2005ApJ...623..325P,2006ApJ...644.1028P,2006ApJ...647.1323W}. 
This results
also in substantial fractions of $^{64}$Ge (decaying to $^{64}$Zn via $^{64}$Ga,
both of short half-life in the minute to second regime and therefore not
of interest in terms of radioactivities with long half-lifes), but within
the $\nu p$~process also to the production of Sr and heavier nuclei.
The isotopic ratios $^{58}$Ni, $^{60,61,62}$Zn change strongly. Alpha-nuclei
such as $^{40}$Ca, $^{44}$Ti, $^{48}$Cr, and $^{52}$Fe are affected as well.
Higher entropies and $Y_e$-values close to 0.5 increase the fraction of these
 $\alpha$-nuclei (and would in hypernovae also cause a substantial production
of $^{64}$Ge as discussed in Sect.~\ref{sec:4-4} 
in the context of the $\nu p$~process \citep[see also][]{2010arXiv1007.1275A,2010arXiv1004.4916R}. 
It was also discussed there
that probably only a small fraction of black-hole producing events actually
lead to hypernovae (not 50\% as assumed in some cases 
\citep{2006NuPhA.777..424N,2006ApJ...653.1145K}). 
However, this reduction in the $^{64}$Ge production can well
be balanced by the larger $^{64}$Ge production in regular supernovae
due to the correct inclusion of neutrino-interactions and their effect on 
increasing $Y_e$ to values larger than 0.5.\\

\noindent{\bf Other explosive burning zones, $^{44}$Ti, $^{26}$Al}\\
While such a discussion is of general interest for the composition of the
Fe-group and filling in of lighter $p$-process nuclei, which are underproduced
in the classical picture (see Sect.~\ref{sec:4-4}), we would like to 
concentrate here on long-lived radioactivities. The nucleus which is mainly
produced in the complete Si-burning regime with  $\alpha$-rich freeze-out is
$^{44}$Ti. As discussed above, its total amount depends on the matter
experiencing this burning outside the mass cut (which is in principle 
unknown without successful explosion calculations). Several authors have made
assumptions on its position, either based on total $^{56}$Ni ejecta
in thermal bombs 
\citep{1996ApJ...460..408T,2006NuPhA.777..424N} 
or entropy jumps in the pre-collapse models in piston-induced explosions 
\citep{1995ApJS..101..181W,1995ApJS..101..181W}. 
Based on these assumptions
different authors find somewhat different predictions, which are, however, a 
relatively flat function of the stellar progenitor mass. 
The values of \citet{1996ApJ...460..408T}
range between $2\times 10^{-5}$ and $1.5\times 10^{-4}$~M$_\odot$. The
interesting point here is that variations in the $Y_e$-structure can lead
to changes up to a factor of 2. Here either the initial pre-collapse 
distribution was assumed (with possible $Y_e$-dips in the innermost parts 
from shell O-burning) or a smoothed flat $Y_e$-distributions closer to 0.5 
(which also reproduces the solar $^{56}$Fe/$^{57}$Fe-ratio). This latter
case leads to the higher values and will also be closer to reality when
neutrino interactions are taken into account during the explosion
(see discussion above on the $\nu p$~process and $Y_e$). 
\citet{2002ApJ...576..323R} and \citet{2009arXiv0908.4283T}
find smaller values (either $1.5-5\times 10^{-5}$ and $3.5-6\times 
10^{-5}$~M$_\odot$). The latter is based on a readjustment of the  $3\alpha$-
and $^{12}$C($\alpha,\gamma)^{16}$O-rates to most recent experimental values,
which does not have a drastic influence, however. Indirectly, core sizes and other
stellar properties, including explosion energies, 
can enter. \index{isotopes!44Ti} 

$^{44}$Ti is only made in the explosive phase of complete Si-burning
with  $\alpha$-rich freeze-out from charged particle equilibrium. There have been
investigations on the reactions producing ($^{40}$Ca($\alpha,\gamma)^{44}$Ti)
and distroying ($^{44}$Ti($\alpha,p)^{47}$V) reactions as well as the half-life
of $^{44}$Ti. Due to the fact, that this is a freeze-out from equilibrium, 
\citet{1999ApJ...521..735H} 
found that even rate changes by a factor of 6 
change the $^{44}$Ti production only by a factor of 1.3
What is different between both approaches, leading to productions either
larger or smaller than $5\times 10^{-5}$~M$_\odot$, is the introduction
of the explosion (i) a thermal bomb, (ii) a piston. This apparently leads to
higher entropies in the first case and a more intense  $\alpha$-rich freeze-out.
It should also be mentioned that non-spherical explosions can lead to
larger $^{44}$Ti-production than spherical models 
\citep{1998ApJ...492L..45N,2000ApJS..127..141N}.
A comparison to observations from either SN 1987A or the SNR Cas A is discussed
in the following subsection.

Complete explosive nucleosynthesis predictions for a range of progenitor stars 
with induced explosions have been given by a number of authors in recent years 
\citep{2002ApJ...576..323R,2002RvMP...74.1015W,2006NuPhA.777..424N,2006ApJ...647..483L,2007PhR...442..269W,2008ApJ...673.1014U}, 
updating some 
of the discussions made above, based on earlier models
\citep{1995ApJS..101..181W,1996ApJ...460..408T}.
Also specific investigations were undertaken for 
Pop III low metallicity stars 
\citep{2005ApJ...619..427U,2007ApJ...660..516T}. 
We limit the present discussion to stars below
130~M$_\odot$ which still undergo core collapse and do not explode via 
explosive O-burning like the so-called pair-creation supernovae 
\citep{2003ApJ...591..288H}. 
While such explosions seem theoretically possible, provided that these
massive cores can result from stellar evolution, the apparent absence
of predicted abundance patterns in low metallicity stars plus our present
understanding of massive stars with rotation 
\citep{2010RvMP...??..???B}, 
seem to exclude this outcome.

Then, the basic pattern given in table \ref{4:yields} always applies. The
abundances from incomplete Si-burning and explosive O-burning can explain
Galactic chemical evolution. As mention in Sect.~\ref{sec:4-4}, the 
classical  $p$~process takes place in explosive Ne-burning via photodisintegrations
of pre-existing heavy nuclei, but even with the best nuclear input
the underproduction of light p-nuclei cannot be solved. The solution
can be obtained by adding a \emph{light (heavy) element primary process} (LEPP,
\citep{2004ApJ...601..864T} 
\index{process!LEPP} \index{process!$\nu p$ process} \index{process!p process}
where the best candidate is the $\nu p$~process.
Thus, the classical $p$-process isotopes have to be explained by a superposition
of the innermost proton-rich complete Si-burning ejecta with those of
explosive Ne-burning in outer zones.

The scheme indicated in table \ref{4:yields} is a bit too simplified when
considering the ejecta of outer layers, whose composition was produced 
during stellar evolution and ejected essentially unaltered during the
explosion. However, the \emph{CO-core} scheme is not sufficient to describe massive-star yields.
While it includes all matter which underwent He-burning, it does not 
differentiate between core He-burning and shell He-burning. The latter occurs
at higher temperatures and has specific features different from
core He-burning. In a similar way, the NeO-core contains all matter which
underwent C-burning during stellar evolution, but also here, no difference
is made between core C-burning and higher temperature shell C-burning.
The same is true for Ne-burning.


\begin{figure}  
\centering 
\includegraphics[width=\textwidth]{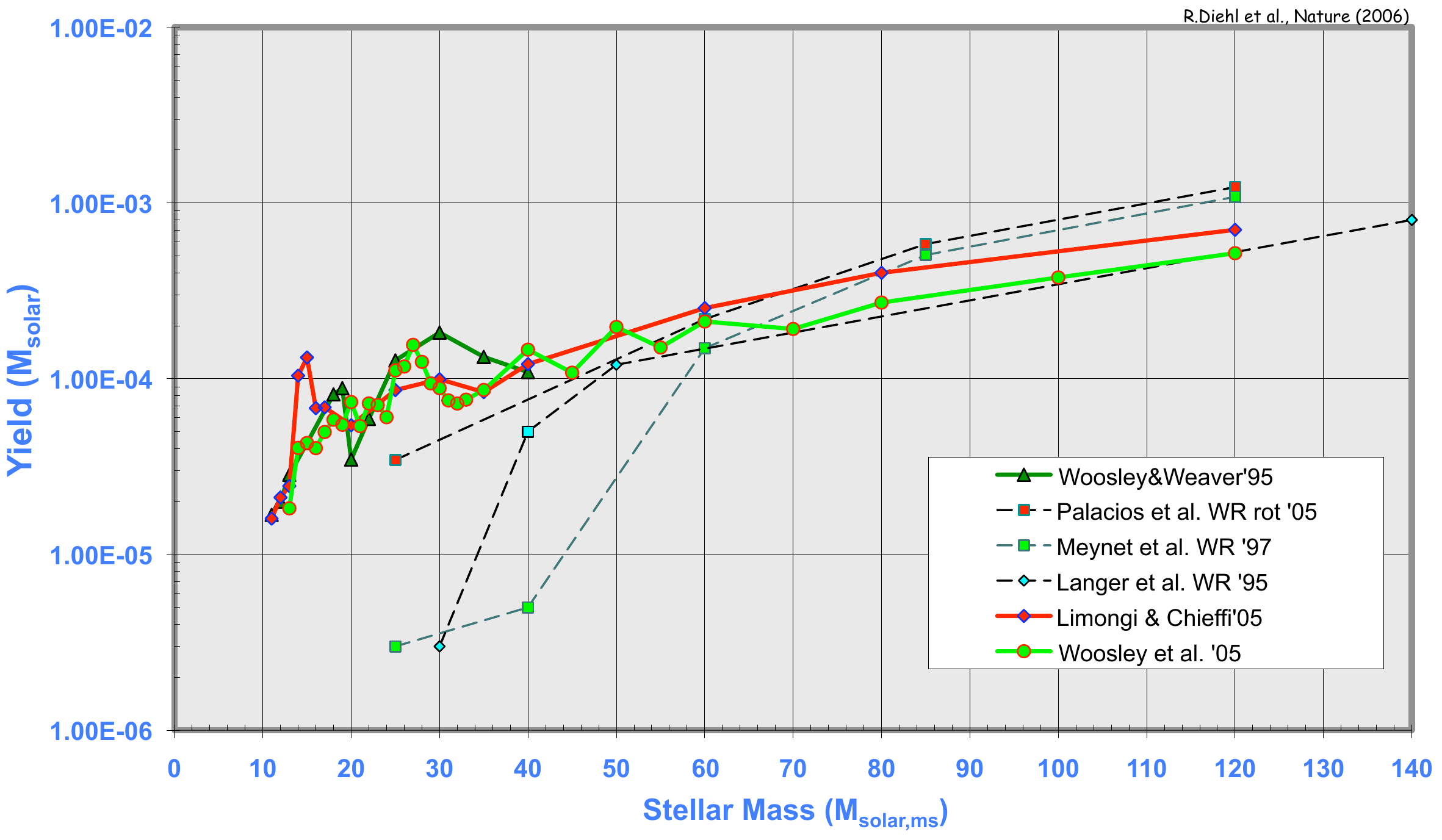}
\caption{The \Al yields from wind ejections and from the explosive release in the 
supernova, as a function of the initial mass of the star \citep[as assembled by][(in their appenix]{2006Natur.439...45D}.}

\label{fig4:al_yields} 
\end{figure}   

\index{nucleosynthesis!yield} \index{isotopes!26Al} 
$^{26}$Al production during stellar evolution was discussed in Sect.~\ref{sec:4-3}; now we include
also the explosive production of $^{26}$Al. It occurs in the regions
of explosive Ne/C-burning. Under these conditions $^{25}$Mg is produced
via $^{24}$Mg($n,\gamma)^{25}$Mg and the protons arise from 
$^{23}$Na($\alpha,p)^{26}$Mg, similar to the reaction pattern shown in 
table \ref{4:neon} for hydrostatic Ne-burning (and partially also C-burning).
Under explosive conditions at temperatures of the order $2.3\times 10^9$K,
these burning stages act explosively in a combined way, and the temperatures
are also suffuciently high to utilize the released protons for the
$^{25}$Mg($p,\gamma)^{26}$Al reaction. However, as also seen from 
table \ref{4:neon}, neutrons are abundantly produced. They 
act as the main destructive species via $(n,p)$ and $(n,\alpha)$ reactions.
As can be seen from table \ref{4:yields}, the mass involved in
explosive Ne/C burning is strongly dependent on the progenitor mass.
Thus, we expect a dramatic increase with increasing initial stellar masses, which is
exactly what we see. 
\citet{2006ApJ...647..483L} 
have analyzed in detail
the contributions from (i) wind ejecta during stellar evolution, (ii) 
hydrostatic burning products ejected during the explosion, and (iii) 
explosive Ne/C-burning. The latter dominates up to about 60~M$_\odot$
and increases from initially about $2\times 10^{-5}$~M$_\odot$ per
event to $2-3\times 10^{-4}$~M$_\odot$. Then wind ejecta start to take over
and flatten out close to $10^{-3}$~M$_\odot$ at initial stellar masses of 
120-140~M$_\odot$. The latter are subject to rotational effects
\citep{1995Ap&SS.224..275L,1997A&A...320..460M,2005A&A...429..613P} 
and increase in fact with higher rotation rates (see Sect.~\ref{sec:4-3}).
\citet{2009arXiv0908.4283T} 
have reanalyzed this behavior in the lower mass range
from 15-25~M$_\odot$ and confirmed this trend. They also did not find
a strong dependence of the result on the He-burning reactions
 $3\alpha$- and $^{12}$C($\alpha,\gamma)^{16}$O. They show nicely how
$^{26}$Al is produced starting in H-burning, but the final explosion produces
close to a factor of 10 more of it.
Yields from different studies have been assembled in Fig.~\ref{fig4:al_yields}.\\

\noindent{\bf $^{60}$Fe}\index{isotopes!60Fe} \\

\begin{figure}  
\centering 
\includegraphics[width=\textwidth]{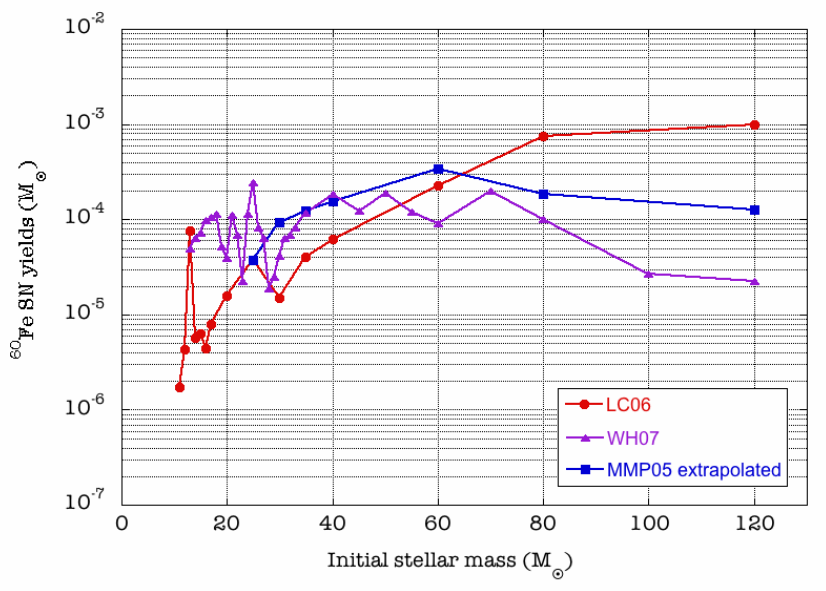}
\caption{The \Fe yields from the explosive release in the supernova, as a function of the initial mass of the star \citep[as assembled by][]{2010A&A...511A..86M}.}

\label{fig4:60fe_yields} 
\end{figure}   

Massive-star yields for $^{60}$Fe are summarized in Fig.~\ref{fig4:60fe_yields} and 
 should only be mentioned here for completeness, although not produced in explosive nucleosythesis. It is entirely
produced in the $s$~process during shell He-burning and thus a pure product
of stellar evolution. The explosion
only acts to eject the corresponding layers. As $^{60}$Fe is 
produced via neutron capture of beta-unstable $^{59}$Fe, 
a relatively high neutron density of about $3\times 10^{10}$cm$^{-3}$ is 
required in order for its efficient production. This is only attained in
shell He-burning during late evolution stages after core C-burning. The 
production ranges from $2\times 10^{-6}$ to  $8\times 10^{-5}$~M$_\odot$
for initial stellar masses between 10 and 40 ~M$_\odot$. This result
is dependent on the He-burning reactions
 $3\alpha$- and $^{12}$C($\alpha,\gamma)^{16}$O, as they compete with
the neutron producing reaction $^{22}$Ne($\alpha,n)^{25}$Mg. There exist
also uncertainties in $^{59}$Fe($n,\gamma)^{60}$Fe and 
$^{60}$Fe($n,\gamma)^{61}$Fe, which cause yield uncertainties by a factor
of up to 5. If the star experiences strong mass loss, the He-burning shell
does not encounter the higher density conditions required for the
high neutron density of  $3\times 10^{10}$cm$^{-3}$. Thus for initial
stellar masses in excess of 40~M$_\odot$, the mass loss treatment can also
lead to variations in predicted yields of more than a factor of 10. Apparently a high mass loss
rate is required to not overproduce $^{60}$Fe in high mass stars 
$M>40$~M$_\odot$
\citep{2006ApJ...647..483L} with respect to $\gamma$-ray line constraints \citep[see][and Ch.~7]{2007A&A...469.1005W}.

\subsection{Observational Diagnostics: Lightcurves, Spectra and SNR}

\subsubsection*{Lightcurves}

Supernova light curves are powered by radioactive decays.  
Very early interpretations of supernova lightcurves related them to the
radioactive decay of $^{254}$Cf \citep{1957RvMP...29..547B}. 
In fact, a strong $r$~process (with fission-cycling) would cause observable 
features based on the decay of heavy radioactive nuclei. 
\index{supernova!light curve} \index{isotopes!56Ni} 
(This question was addressed very recently with respect to $r$-process ejecta
from neutron star mergers \citep{2010MNRAS.406.2650M}).
Supernova lightcurves, however, are dominated by Fe-group ejecta.
In addition to abundant $^{56}$Ni, there are a number of radioactive nuclei which will
decay on time scales of ms up to $10^7$y.
Here we only want to concentrate on a few
nuclei, which by a combination of their abundances and half-lives, can
be of importance.
These nuclei are $^{56}$Co ($^{56}$Ni), $^{57}$Co ($^{57}$Ni), $^{55}$Fe
($^{55}$Co), $^{44}$Ti, and $^{22}$Na. For a 20 M$_\odot$ star like
SN 1987A they were predicted with total masses of
0.07, 3.12$\times 10^{-3}$, 3.03$\times 10^{-4}$, 1.53$\times 10^{-4}$,
and 1.33$\times 10^{-7}$M$_\odot$ 
\citep{1990ApJ...349..222T,1996ApJ...460..408T}.

Observations of light curves in radiation which reflects the thermalized energy of this radioactivity 
constrained these values  to 
$M (^{56}$ Ni)$\approx 0.071$ M$_\odot$ 
(e.g. \citet{1990AJ.....99..650S}) \index{isotopes!57Ni} 
and $M (^{57}$ Ni)$\approx 3.3 \times 10^{−3}$ M$_\odot$ 
(\citet{1993ApJ...408L..25F} and references therein). 
Only more recently \index{isotopes!44Ti} 
a very careful analysis confirmed an upper limit on $^{44}$Ti of the
order 1.1$\times 10^{-4}$M$_\odot$ 
\citep{2001A&A...374..629L}.

\begin{figure}[!tbp] 
\begin{centering}
\includegraphics[width=\textwidth]{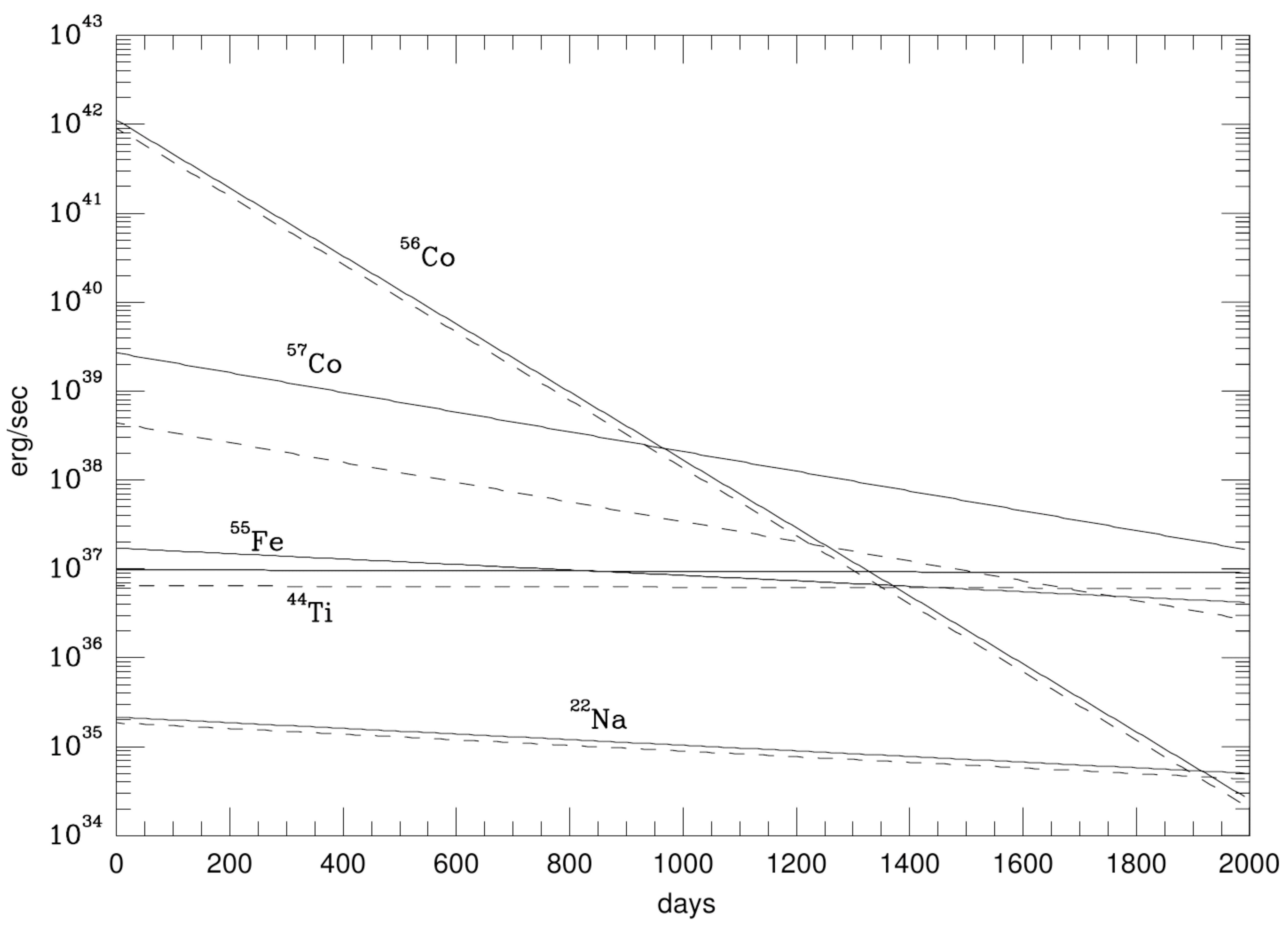}
\caption{Total energy release due to the decay of long-lived radioactive 
species (dashed lines) and due to the release in terms of thermalized
decay photons (solid lines). The ejected masses of radiactive species
are takem from a 20~M$_\odot$ model for SN 1987A 
\citep{1990ApJ...349..222T,1996ApJ...460..408T}
.}
\end{centering}
\label{4:lightcurve}
\end{figure}

Generally, after
beta-decay or electron capture, a daughter nucleus is produced in an excited state ( $^{55}$Fe is a notable exception, see below). 
The ground state is reached by one or several gamma transitions, \index{decay!gamma decay}
observable by current gamma-ray detectors for nearby sources (see Sect.~10.1).
Photons, positron-electron annihilations following $\beta ^+$-decays, and the
kinetic energy given to the decay products can contribute to the light curve
at later times.

The number of photons released for each of the transitions, occuring in the
daughter nucleus after beta-decay, is equal to the number of decays
$N_d$, multiplied with the appropriate percentage of the occurrance (\emph{branching ratio}) for
the specific transition.
The total energy released corresponds to the product of the number of
decays with the decay Q-value:

\begin{equation}
N_d (t)=- {dN \over dt}(t) = \lambda N_o {\rm exp}(-\lambda t) \ \ \ \ 
{d E \over dt}(t) =  Q  N_d (t) = Q \lambda N_o {\rm exp}(-\lambda t),
\end{equation}

where $\lambda= \ln2/t_{1/2}$ is the decay rate of the nucleus. The
initial number of radioactive nuclei can be calculated from their total
mass by $N_o=M/Am_u$, with $A$ being the nucleon number
of the nucleus, $m_u$ the atomic mass unit, and $M$ the mass given above.
When using the radioactivity half-lives of relevant isotopes expected in supernova ejecta
(i.e., 78.76d, 271.3d, 2.7y, 54.2y, and 2.602y, and atomic Q-values of 4.566, 0.835,
0.232, 3.919, and 2.842~MeV)  we can estimate radioactive-energy generation
rates in erg~s$^{-1}$ and the total number of decays per sec.
The Q-value used for $^{44}$Ti combines the subsequent decays
of $^{44}$Ti and $^{44}$Sc.
Q-values include all available energies, i.e. the kinetic energy of the decay products, the energy
in photons, the annihilation energy of positron-electron pairs in
$\beta ^+$-decays, and the neutrino energy. At densities prevailing in
the expanding remnant, neutrinos will escape freely and their energy has to
be subtracted, which leaves corrected values for the appropriate energy
deposits of 3.695, 0.136, 0.0, 2.966, and 2.444~MeV.
\index{gamma-ray lines}
Because the electron
capture on $^{55}$Fe does only lead to an energetic neutrino, there is
no local energy deposition from this isotope\footnote{This situation was recently re-evaluated by 
\citet{2009MNRAS.400..531S}.
The electron capture occurs from an electron
in an atomic orbit, leaving a hole which can be filled by other
electrons cascading down to fill this hole, thus emitting photons  - X-rays - or
depositing the energy in ejecting outer electrons - Auger electrons. Thus,
in cases where only ground-state to ground-state electron capture occurs
and the energy is emitted in an escaping neutrino only Auger electrons or
X-rays can contribute to local energy deposition.}.
Gamma transitions following the decays of the other isotopes under consideration obtain candidate $\gamma$-rays at (rounded to
full percent values): $^{56}$Co, 847~keV (100\%), 1038~keV (14\%), 1238~keV
(68\%), 1772~keV (16\%), 2599~keV (17\%); $^{57}$Co, 122~keV (86\%), 136~keV
(11\%); $^{44}$Ti, 78~keV (93\%), 68~keV (88\%), 147~keV (9\%), 1157~keV (100\%);
$^{22}$Na, 1275~keV (100\%; branching ratios given as percentages per dcay).
If positrons from $\beta^+$-decay slow down and annihilate with electrons locally within the supernova envelope, the full neutrino-loss corrected energy corresponding to the reaction Q-value \index{positron} \index{process!beta decay} \index{process!gamma decay}
will be deposited in the envelope. Observable signatures include
high energy photons such as the ones from the gamma transitions,
and their Compton scattered and completely thermalized descendants\footnote{
Deposition of energy from radioactive decay involves absorption of high-energy photons, slowing down of $\sim$MeV-type energy electrons and positrons, and proper treatment of temporary energy reservoirs such as ionization and inhibited radioactive decay from completely-ionized nuclei \citep[see, e.g.,][]{2009arXiv0911.1549S,1999A&A...346..831M,2007ApJ...662..487W}
}. 

\begin{figure}[!tbp] 
\includegraphics[width=0.6\textwidth]{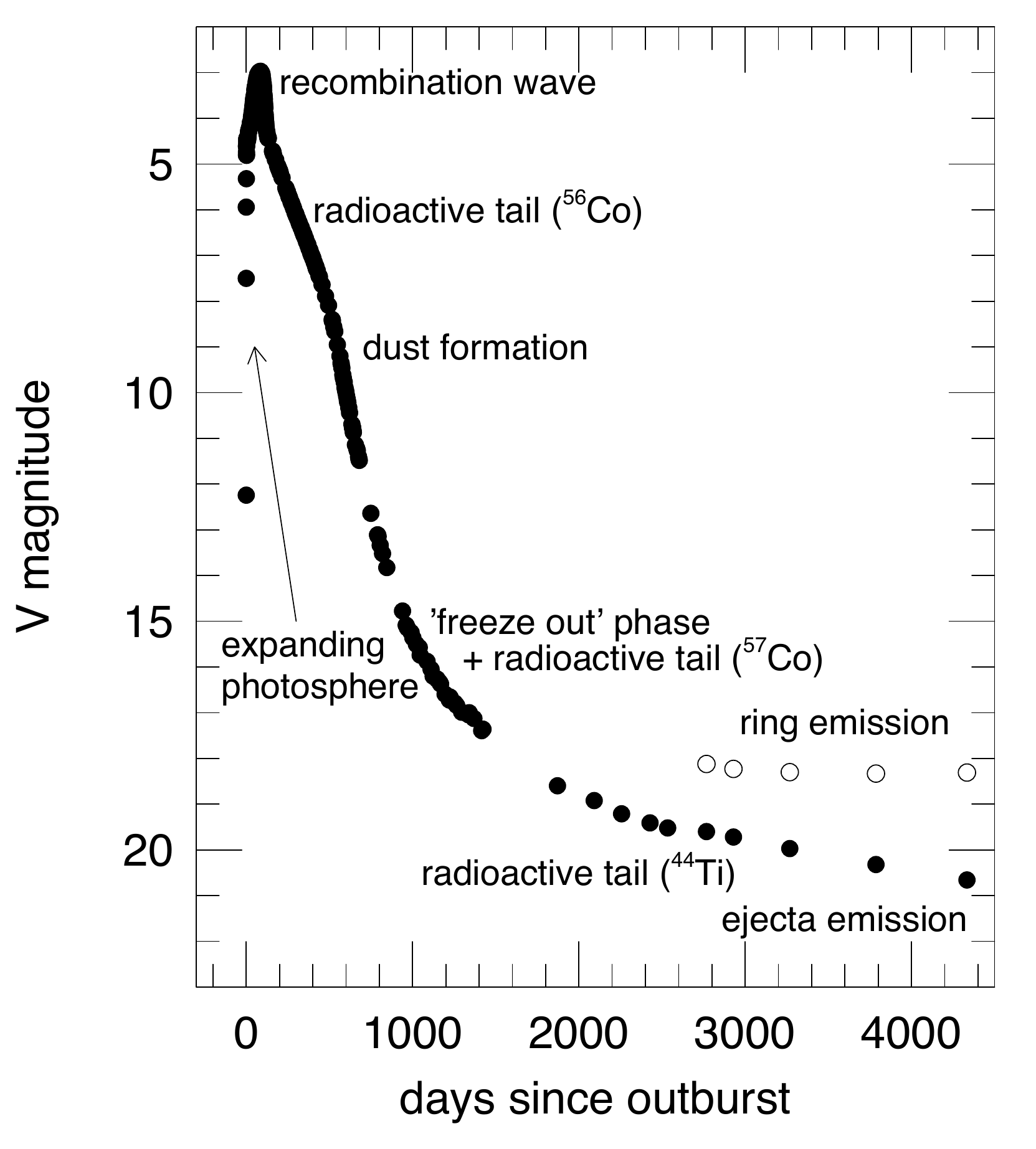}
\sidecaption
\caption{Observed (visual) lightcurve of SN1987A for the first
1500 days from 
\citet{2003LNP...598...77L}.
}
\label{4:bruno}
\end{figure}

Then the sum of all
individual contributions discussed above
would make up the bolometric lightcurve of the supernova
(see Fig.\ref{4:lightcurve}).
The \emph{light curve}\index{supernova!light curve}, i.e. the brightness as a function of time,
will be dominated first by the decay of $^{56}$Co, and then
$^{57}$Co and $^{44}$Ti, if one neglects possible radiation
from a pulsar. $^{22}$Na never plays a dominant role
for the lightcurve. 
At lower densities
(and later times),  escaping high energy photons or positrons lead to a reduction of the brightness of \emph{bolometric} emission. This
can be seen in late time observations as shown e.g. in 
\citet{2003LNP...598...77L} 
(see Fig.\ref{4:bruno}). 
An important consistency check would be to compare this bolometric light curve  (which includes only optical,
UV and IR emission, hence thermalized gas and dust components) to the high-energy photons more directly reflecting radioactive decays.
At late times, those high energy photons escape
freely.

The $\gamma$-ray detections of SN1987A were 
the first to identify $\gamma$-rays from $^{56}$Co decay. Later improved
observations by CGRO for $^{56}$Co and $^{57}$Co were a direct proof of
these unstable species in the right amounts. \index{Solar Maximum Mission}
It turned out that Ni decay $\gamma$-ray lines were seen with the Gamma-Ray Spectrometer on the Solar Maximum Mission \citep{1988Natur.331..416M,1990ApJ...357..638L} significantly earlier than expected from a spherically stratified
distribution of elements, where the Fe-group nuclei are produced in the
center. This is understood from deviations from spherical symmetry in the expanding remnant, bringing Ni-rich clumps to the surface earlier by convective instabilities, mixing 
$^{56}$Ni/$^{56}$Co to outer layers at early times. Gamma-ray line profiles measured with high spectral resolution indicated Doppler broadening of the lines from their ejecta motion \citep{1990ApJ...351L..41T}.
 
Many supernova remnants (such as the 300-year old Cas A remnant) \index{supernova!SN1987A} \index{supernova!Cas A}
show mixing in their ejecta \citep{2005AdSpR..35..976V}. While there existed some theoretical indications
that this is due to instabilities of the propagating shock wave,
it is generally and more plausibly associated with the expansion of the supernova
into an inhomogeneous medium. SN1987A observations  showed clearly that
mixing is part of the supernova explosion itself. Several
independent reasons lead to such a conclusion. The supernova produces large
amounts of unstable long-lived nuclei, the dominant abundance is found
in the doubly-magic nucleus $^{56}$Ni, which is produced in the innermost
part of the ejecta. $^{56}$Ni decays with half-lives of 6.1 days to $^{56}$Co
and 77.8 days to $^{56}$Fe. After the beta-transition, deexcitations to the
ground state of the daughter nucleus lead to the emission of high energy
gamma-rays. While these gamma-rays would escape freely at low densities,
Compton scattering will reduce their energies into the X-ray and even
thermal regime at higher densities.
With decreasing densities during the expansion, initially only thermalized
photons will escape, then X-rays and finally gamma-rays.
For SN1987A X-ray observations  with GINGA, HEXE, and balloons
and gamma-ray observations
with the SMM-satellite and balloons
actually agreed with this general behavior \citep[e.g.][]{1990PAZh...16..403S,1990ApJ...357..638L}. 
The main problem was that the
predicted time scales did not agree with
the observations, where X-rays and gamma-rays appeared earlier than
predicted.
An agreement could only be obtained when part of the $^{56}$Ni,
being produced initially in the inner parts of the ejecta, was mixed out
to larger distances 
\citep{1989ApJ...340..414F}.
Mixing is also required to explain
the spread of expansion velocities seen in line widths of infrared
observations for various elements
and in the gamma ray lines of $^{56}$Co.
The inferred velocities differ strongly from the much smaller ones, expected
from an expanding remnant, which maintains the stratified composition from
explosive and hydrostatic nuclear burning.
Other indications came from the modeling of the optical light curve.
The best agreement between
calculated and observed light curves were obtained for a composition which
mixed a small fraction of Ni all the way into the 10M$_\odot$ hydrogen
envelope and hydrogen into the deeper layers, containing mostly heavy elements
(see e.g. \citet{1990ApJ...348L..17B}).

The  lightcurve from SN1987A could be reproduced with
theoretical modelling, including the effects of
X-ray and $\gamma$-ray escape, as well as mixing of $^{56}$Ni.
SNe Ib and Ic events, believed to be core collapse events without an
overlying hydrogen envelope have to be treated accordingly. The combination of
small masses involved
(only He-cores or C-cores without H-envelope) and the assumption of mixing
can reproduce the steeper decline than found in massive SNe II.
A typical case of a type Ic supernova is SN 1998bw, associated with 
GRB 980425. The straight-forward modeling of the observed lightcurve 
\citep{2002A&A...386..944S}, 
similar to the discussion in the beginning of this 
subsection \citep{2001ApJ...555..880N}, 
led to interpretations of a largely 
non-solar $^{56}$Ni/Fe to $^{56}$Ni/Fe ratio. The inclusion of internal
conversion and Auger electrons, as suggested by 
\citet{2009MNRAS.400..531S}
could naturally explain the observed slowdown of the lightcurve without 
invoking such extreme abundance ratios.

We note that in recent years photon transport calculations have reached major improvements,
and are now able to consistently reproduce both light curves and spectra from SNIa, and also from core collapse supernovae \citep[e.g.][for a description o the method]{2009arXiv0911.1549S}. Presently, systematic uncertainties of the method are being investigated, and appear rather well understood (at least for SNIa \citep[see, e.g., ][]{2007ApJ...662..487W}. As optical-to-IR light curves and spectra will be collected in abundance through large telescope survey programs for cosmological studies, it is likely that those (more indirect) measurements of core-collapse supernova nucleosynthesis will generate the tightest constraints to learn more about these events and their internal nuclear processes.

\subsubsection*{Optical/IR Spectra and Dust Formation}

In the preceding subsection we laid out the framework of \index{stardust}
understanding supernova lightcurves. What is missing here, is the evolution
of supernova spectra. The receeding photosphere in terms of radial mass
in an expanding, radiation-filled bubble can give clear indications of the
element composition (as a function of time equivalent to declining Lagrangian
mass). The problem of type II supernovae is that the huge H-envelope does
not really contain much information in terms of nucleosynthesis. Type Ib and
Ic supernovae, which lost their H- and possibly He-envelope, reveal much
more information of the compact inner part, which experienced
explosive processing 
\citep{2001AJ....121.1648M,2002ApJ...566.1005B,2006MNRAS.369.1939S}.
This is similar to type Ia supernovae, originating
from exploding white dwarfs, which have been extensively utilized for
abundance diagnostics.


\begin{figure}  
\centering 
\includegraphics[width=1.0\textwidth]{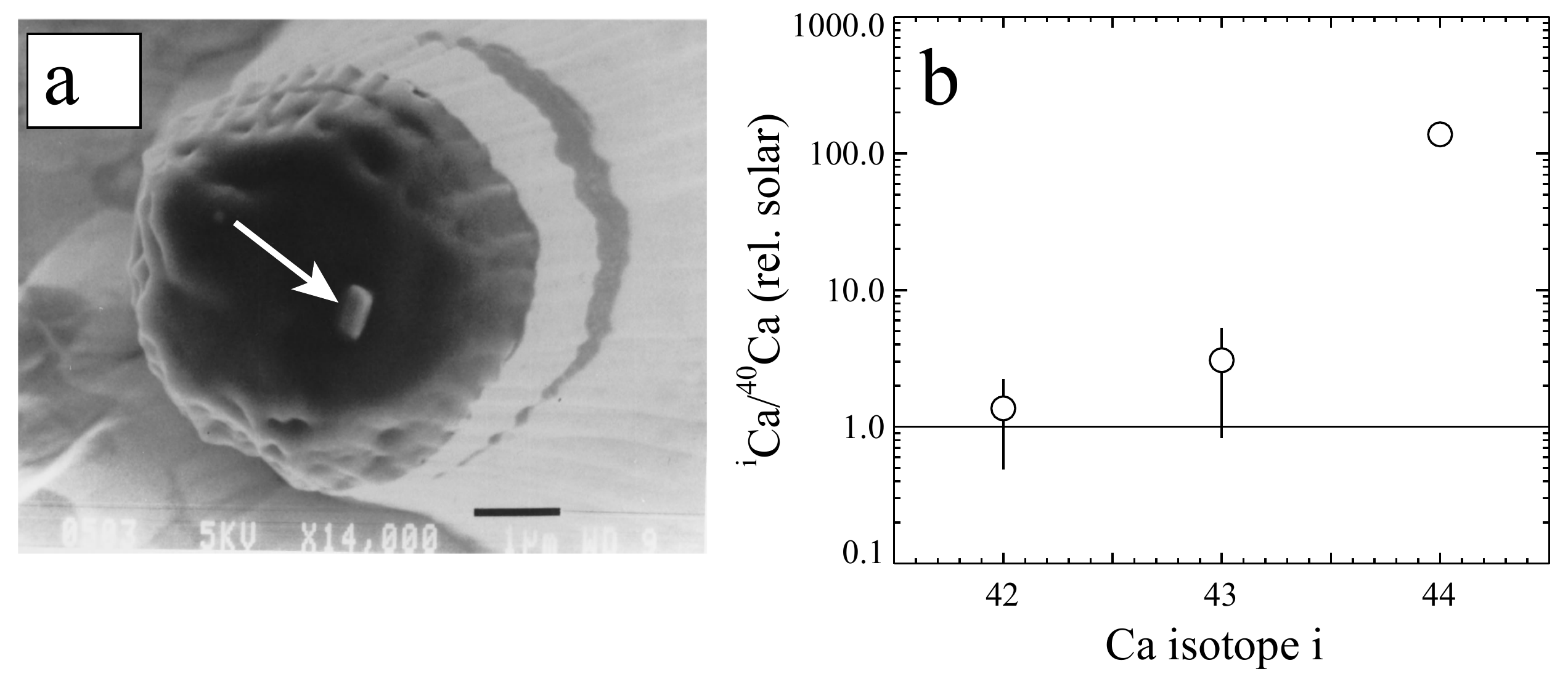}
\caption{ {\it (a)} Scanning electron microscope image of a presolar graphite grain following isotopic measurement with an ion microprobe \citep[from][]{1996ApJ...462L..31N}. The arrow indicates TiC sub-grain originally enclosed in graphite but revealed 
by the ion-probe sputtering. {\it (b)} Calcium isotopic composition of graphite 
grain shown in a. The $^{44}$Ca/$^{40}$Ca ratio is 137 times the solar ratio, 
whereas the other Ca-isotopic ratios are normal within $2-\sigma$ errors (the error 
bar in the isotope 44 abundance is smaller than the symbol size). This is a clear 
signature of in situ decay of live \Ti, originally condensed in the TiC sub-grain, 
and demonstrates that the grain formed in the ejecta of a supernova.   }

\label{fig4:ti-grain} 
\end{figure}   

Another issue, which has been discussed already in the subsection on supernova
lightcurves is  related to
convective instabilities and mixing of nucleosynthesis products. The extent
of mixing is responsible for the element mixture in the expanding and cooling
remnant when chemical reactions and dust formation set in. For which
compositions and conditions does this environment lead to presolar grains
with high melting temperatures which can survive the interstellar medium and
formation of the solar system in order to be detected today in meteoritic
inclusions? When do we form oxides, diamonds, hybonites, carbides ... and
how selectively do they include matter from the regions where they form?
How can we relate $^{26}$Mg and $^{44}$Ca excesses in presolar grains
to initially embedded $^{26}$Al and $^{44}$Ti? \index{isotopes!44Ti} 
Fig.~\ref{fig4:ti-grain} shows an example of a grain attributed to supernova condensation.
A discussion of these questions can be found in Sect.~10.3.

\subsubsection*{$^{26}$Al in the Vela Region}    
\label{sec:7-4-8}
\index{supernova!remnant} \index{isotopes!26Al} 
The Vela region appears prominent in several astronomical images of our Galaxy: It includes the Gum nebula and the Vela Supernova Remnant, both prominent agents to form nearby structures of the interstellar gas and bright in X- and radio emissions, and it includes the Vela pulsar where bright gamma-ray pulsations teach us about particle acceleration in neutron star magnetospheres, and furthermore with Vela~X-1 a remarkable X-ray source and prototype of a binary system where a neutron star accretes wind material from a high-mass companion star. All those objects are relatively nearby, mostly in the foreground of the Vela molecular ridge which is one of the nearest star-forming regions and located in about 700 ($\pm$200)~kpc distance \citep{2007A&A...466.1013M}. 

Three prominent sources have been discussed in the context of measuring \Al production for individual objects -- all related to massive star and explosive nucleosynthesis, respectively (\Al observations are discussed in Chap.~7 otherwise): The Vela supernova remnant, a recently-discovered supernova remnant called \emph{Vela Junior}, and a Wolf-Rayet binary system \emph{$\gamma^2Velorum$}.

\begin{figure}  
\centering 
\includegraphics[width=0.8\textwidth]{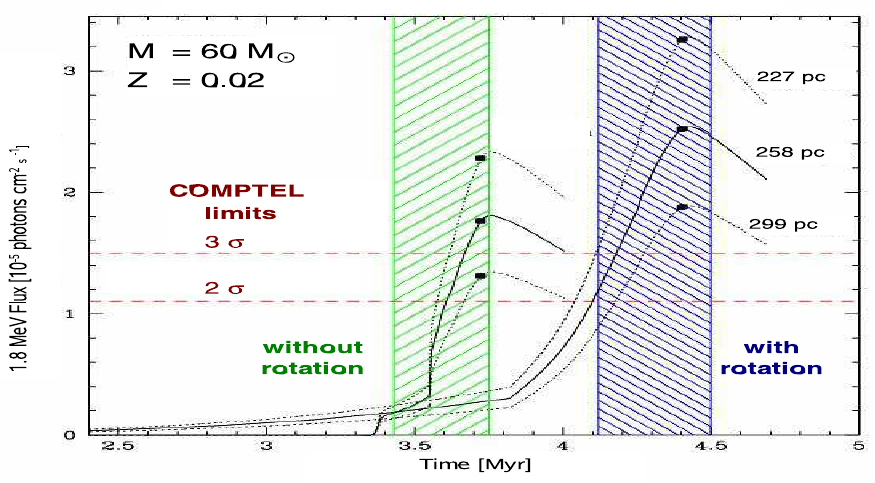}
\caption{The \Al gamma-ray flux estimated from ejection of \Al-rich envelope material during the Wolf-Rayet phase  \citep{2005NuPhA.758..320M} . Shown are two different stellar models, with and without stellar rotation. In addition to delaying the WR phase, rotation also increases the \Al production. Shown are gamma-ray fluxes for different distances, not including the more recent and larger distance of 330--390~pc.  \index{stars!$\gamma^2$ Velorum} }
\label{fig7:WR11_26Al} 
\end{figure}   

The Vela supernova remnant is relatively nearby at 250~pc, about 10,000~y old, and spans an area of about 8\degree diameter on the southern sky. It hosts the Vela pulsar and a plerionic pulsar nebula at its center. Such a nearby supernova explosion in the recent past seems a unique opportunity to calibrate the \Al yield of a core-collapse supernova. With COMPTEL, diffuse and extended emission had been recognized from this direction \citep{1995A&A...298L..25D}. But possibly-underlying extended \Al emission limits the precision of this measurement, the \Al gamma-ray flux attributed to this supernova remnant is 0.5--2.7~10$^{-5}$ph~cm$^{-2}$s$^{-1}$. This is well within expectations (an \Al yield of 10$^{-4}$~\Msol would result in a gamma-ray flux of $\sim$2~10$^{-5}$ph~cm$^{-2}$s$^{-1}$). INTEGRAL could not detect the supernova remnant, however, despite sufficient exposure. This may be due to the low surface brightness of this extended source and line broadening from the remaining ejecta motion of this young supernova remnant.

In refined X-ray imaging analysis of the Vela supernova remnant, a new supernova remnant was discovered being superimposed with a diameter of $\sim$2\degree and named RXJ0852.0-46.22 or \emph{Vela Junior} for short \citep{1998Natur.396..141A}. Early speculations about its \Al emission were stimulated from hints for \Ti emission, \index{isotopes!44Ti} which would have implied that this supernova remnant would be both young and nearby; these could not be substantiated by additional measurements and studies. Now absence of \Al emission from Vela Jr. appears plausible, as the supernova remnant is probably older than 1000~y and more distant than 740~pc \citep{2009AdSpR..43..895K}.

The \emph{$\gamma^2Velorum$} binary system is yet another tantalizing \Al source: It consists of a Wolf Rayet star (\emph{WR11}) and an O-star companion. If the binary interactions can be ignored for \Al production of the WR11 star, this would be \emph{the} opportunity to calibrate the \Al ejection during the Wolf-Rayet phase of a massive star, as the distance to this object has been derived from Hipparcos measurements as only 258~pc. The Wolf-Rayet star currently has a mass of $\sim$9~\Msol, with a 30~\Msol O star companion \citep{2000A&A...358..187D}.  Modeling the Wolf-Rayet star evolution in detail, and also accounting for possibly rapid rotation as it delays the wind ejection phase,  \citet{2005NuPhA.758..320M} show that the upper limit on \Al emission from WR11 would be hard to understand (Fig.~\ref{fig7:WR11_26Al}). But doubts have appeared on the Hipparcos distance measurement, and current belief is that the system is part of the Vela OB2 association and rather at a distance of 330--390~pc \citep[see][and references therein]{2009MNRAS.400L..20E}. Additionally, the system's age may also be somewhat higher than estimated earlier (beyond 5~My, rather than 3--5~My) \citep{2009MNRAS.400L..20E}, and part of the \Al ejected in the earlier wind phase may now be spread over a shell extended by up to 6\degree \citep{2006NewAR..50..484M}, hence of lower surface brightness and still consistent with the non-detection by COMPTEL and INTEGRAL gamma-ray telescopes. Note, however, that binary mass transfer may have altered the evolution of the Wolf-Rayet star substantially; this mutual impact on stellar evolution in close binaries is still very uncertain, but could lead to orders of magnitude increases of \Al production in rare cases \citep{2009A&A...507L...1D,1998nuas.conf...18L}.

\subsubsection*{Supernova Remnants}

An expanding supernova as it interacts with circumstellar matter gives rise to thermal X-ray line emission, for remnants of ages between $\sim$hundreds to several 10000 years. Although in principle circumstellar and ejecta materials are both contributing to such emission, and such atomic-shell emission carris uncertainties of ionization degrees and local temperature (to define a local thermodynamic equilibrium), abundance determinations for specific ejecta species can be made, and also have been explored for determinations both of supernova types and of nucleosynthesis yields \citep{1995ApJ...444L..81H,2006A&A...449..171N}. 
Even an (unsuccessful) search for X-ray line emission of radioactive $^{44}$Sc from the decay of $^{44}$Ti had been performed for the presumably young and nearby Vela Junior SNR  \citep[][and references herein]{2009PASJ...61..275H}.
More directly, 
the very late phases of the expansion into the interstellar medium permits the detection
of long-lived radioactivities by detecting its decay photons of specific $\gamma$-ray
energies, when
these can escape freely from the expanding debris. Such detections can then be identified with the amount of matter
existing in radioactive isotopes.

The discovery of the 1157 keV $^{44}$Ti $\gamma$-ray line emission from the 
youngest Galactic SNR Cas A with COMPTEL  \index{isotopes!44Ti} \index{supernova!Cas A} \index{Compton observatory!COMPTEL}
\citep{1994A&A...284L...1I} 
was the first 
direct proof that this isotope is indeed produced in SNe. This has been 
strengthened by the BeppoSAX/PDS detection of the two low-energy $^{44}$Ti 
lines 
\citep{2001ApJ...560L..79V}. 
By combining both observations, 
\citet{2001ApJ...560L..79V} 
deduced a $^{44}$Ti yield of $(1.5\pm1.0) \times 10^{-4}$~M$_\odot$. This 
value seemed higher than the predictions of most models (see the
previous subsection), although it is not outside the error bars. 
Several aspects have been considered to explain this large value: a large 
energy of the explosion ($\approx 2 \times 10^{51}$ 
erg), asymmetries 
\citep{1998ApJ...492L..45N} 
currently observed in the ejecta 
expansion, and a strong mass loss of the progenitor consistent with the 
scenario of a Type Ib SN 
\citep{2004ApJ...604..693V}. 

If \Ti ejection as seen in the Cas A event was typical for core-collapse events, 
the $gamma$-ray surveys made with COMPTEL\citep{1997A&A...324..683D,1999ApL&C..38..383I} and with INTEGRAL\citep{2006NewAR..50..540R} should have 
seen several objects along the plane of the Galaxy through their \Ti decay emission (see Sect.~7.6 for a detailed discussion of Galactic supernovae and \Ti).
From this, it had been concluded that \Ti ejection is rather a characteristic of a rare subclass of core-collapse supernovae \citep{2006A&A...450.1037T}.

From the three different $\gamma$-ray lines resulting from the \Ti decay chain, constraints for kinematic Doppler broadening can be derived: The Doppler broadening being a linear function of energy, it would broaden the 1157~keV line to values in the few to tens of keV range, which can be measured with Ge spectrometers; the lower-energy lines at 68 and 78~keV would not show significant kinematic broadening.  \citet{2009A&A...502..131M} have exploited INTEGRAL/SPI spectrometer data to show that \Ti ejecta as seen by above measurements need to be faster than 500~km~s$^{-1}$, as the 1157~keV line is not found with SPI and assumed to be broadened such as to disappear in instrumental background.

\subsubsection*{Ejecta from a Supernova Remnant on Earth} %


\begin{figure}  
\centering 
\includegraphics[width=1.0\textwidth]{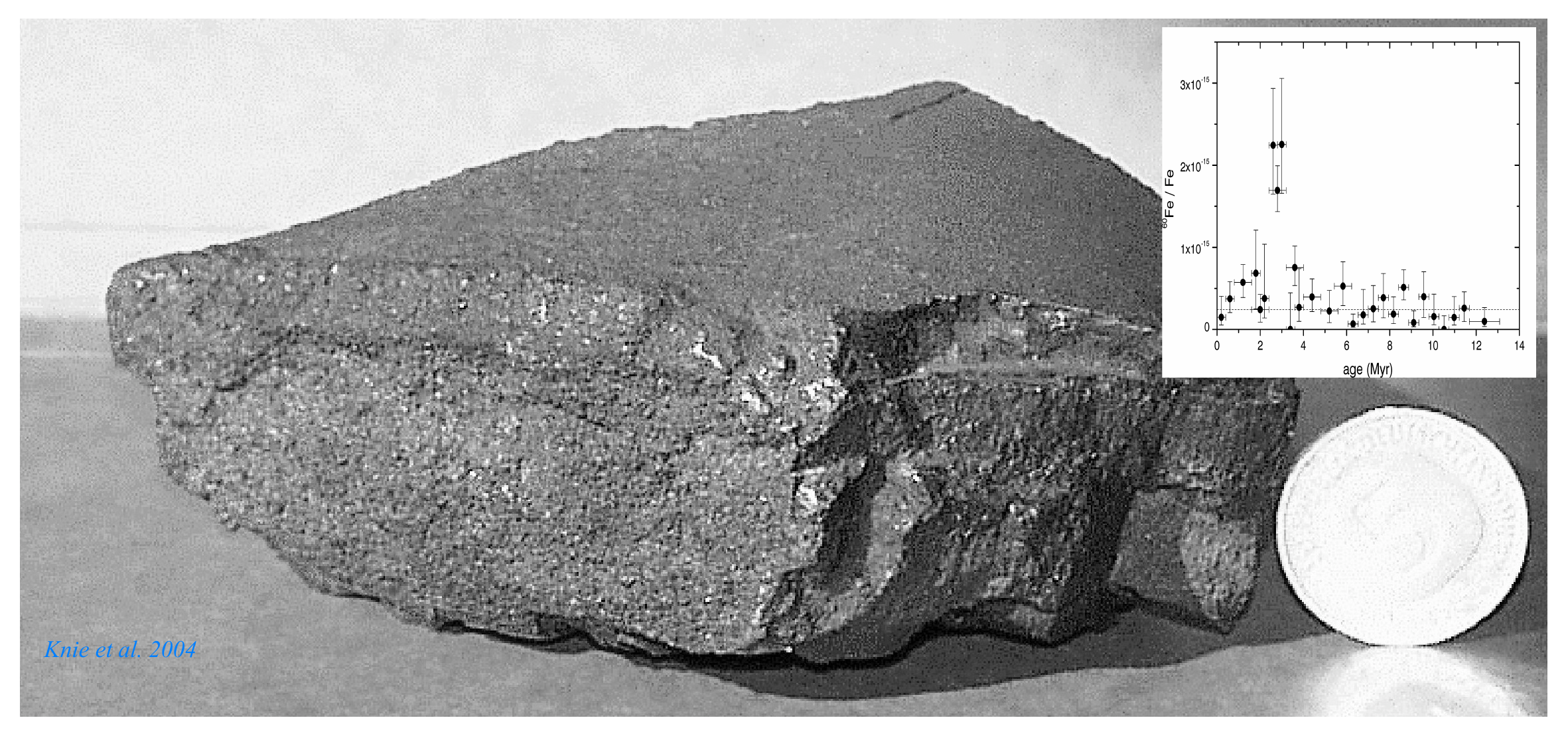}
\caption{ Ocean crust sample as analyzed by   \citet{2004PhRvL..93q1103K}  for \Fe content. The insert graph shows their result, i.e. the number of \Fe nuclei versus age as determined from cosmic-ray produced Be radioactivity. }

\label{fig4:60Fe_crust} 
\end{figure}   

\Fe has been discovered through accelerator-mass spectroscopy (AMS) \index{accelerator mass spectrometry} \index{isotops!60Fe} \index{ocean crust} analyses of ocean crust material \citep{2004PhRvL..93q1103K} (Fig.~\ref{fig4:60Fe_crust}).  If taken from places on Earth which are remote from any antropogeneous contamination, such as in deep parts of the Pacific ocean, they provide a record of past composition of ocean water. Manganese crusts grow very slowly from sedimentation. Therefore, a rather small sample will cover tens of My of sedimentation history within a few cm of depth. \Fe production from cosmic ray irradiation in the atmosphere is unlikely, other systematic contaminations also seem low. The age of each depth layer can be determined from Be isotopes produced by cosmic rays in the atmosphere of the Earth, also ingested into ocean water with other atmospheric gas and dust. The AMS method is one of the most-sensitive techniques to detect small amounts of specific isotopes, reaching a sensitivity of 10$^{-16}$. Evaporization of chemically-prepared Fe-enriched crust samples and successive ion acceleration and mass spectrometry obtained the result shown in Fig.~\ref{fig4:60Fe_crust}. This discovery was taken as evidence that debris from a very nearby supernova event must have been deposited on Earth about 3~million years ago. Unclear remain the deposition and crust uptake efficiencies, such that the quantitative estimation of interstellar \Fe flux or supernova distance is uncertain; distances in the 10--30~pc range have been discussed. Studies of other sediment samples are underway to estimate these effects, and to confirm this exciting record of nearby supernova activity. 
Supernova nucleosynthesis of radioactivities appears close to our lives, indeed.

\section*{Acknowledgements}

Many colleagues have contributed to the subject of this chapter. We would like to acknowledge specifically our discussions with 
Alessandro Chieffi,
John Cowan, 
Iris Dillmann,
Khalil Farouqi,
Claes	Fransson,
Carla Fr\"ohlich,
Alexander	 Heger,
Wolfgang	Hillebrandt,
Thomas Janka, 
Gunther	Korschinek,
Karl-Ludwig	Kratz,
Karlheinz Langanke, 
Bruno Leibundgut, 
Marco	Limongi,
Gabriel Martinez-Pinedo,
Bradley	Meyer,
Georges	Meynet,
Yuko	 Mochizuki,
Ewald M\"uller,
Ken'ichi Nomoto,
Igor Panov, 
Thomas Rauscher,
Lih-Sin	The,
James Truran,
Jacco	Vink,
Stan	Woosley,
and Hans	Zinnecker.


	\bibliographystyle{spbasic}






\backmatter

\end{document}